\documentclass[12pt, reqno, oneside,amsart]{article}
\usepackage[
left=1.2in,
right=1.2in,
top=1.in,
bottom=1.in,
includeheadfoot,
footskip=0.7in
						]{geometry} 
\usepackage[T1]{fontenc}

\usepackage{mathrsfs}  

\usepackage[tbtags]{amsmath}	

\usepackage{cite}
\usepackage{graphicx}
\usepackage{amsfonts}
\usepackage{amssymb}
\usepackage{dsfont}		

\input epsf

\usepackage[shortlabels]{enumitem}

\usepackage{framed}

\usepackage{braket}

\usepackage{lipsum}

\usepackage[titletoc]{appendix}

\usepackage{enumitem}

\usepackage[dvipsnames]{xcolor}

\usepackage{hyperref}
\hypersetup{
    colorlinks=true,
    anchorcolor=red,
    urlcolor=magenta,
    citecolor=red,
    linkcolor=cyan}
        
\usepackage{simplewick}    

\usepackage{empheq}

\usepackage{stmaryrd}

\usepackage{subfigure}

\usepackage{slashed}

\numberwithin{equation}{section}

\usepackage{etoolbox}
\usepackage[noindentafter]{titlesec}	
\usepackage{titletoc}

\titleformat{\section} 
	{
	\normalsize
	\large
	\bfseries
	} 
{\thesection.}       
{0.3em}                  
{}

\titlespacing{\section}
{0pt}                  
{15pt}                  
{10pt}                  

\titleformat{\subsection} 
[hang]                 
{
\bfseries
} 
{\thesubsection.}       
{0.3em}                  
{}                     

\titlespacing{\subsection}
{0pt}                  
{15pt}                  
{10pt}                  

\titleformat{\subsubsection} 
[hang]
{
\normalsize
\bfseries
} 
{\thesubsubsection.}    
{0.3em}                
{}                     

\titlespacing{\subsubsection}
{0pt}                  
{10pt}		        
{2pt}                  

\definecolor{PastelBlue}		  	    {RGB}{81, 167, 249}    

\definecolor{PastelGreen}		  	    {RGB}{112, 191, 65}    

\definecolor{PastelRed}		  	    {RGB}{236, 93, 87}    

\definecolor{PastelPurple}		  	    {RGB}{179, 106, 226}    

\definecolor{PastelOrange}		 	{RGB}{243, 144, 25}    

\def\a{\alpha}
\def\b{\beta}
\def\s{\sigma}

\def\la{\lambda}

\def\vro{\varrho}

\def\e{\epsilon}
\def\vro{\varrho}

\def\rm{\mathrm}
\def\cal{\mathcal}
\def\scr{\mathscr}
\def\bs{\boldsymbol}

\def\pa{\partial}

\def\dd{{\mathrm d}}

\def\be{\begin{equation}}
\def\ee{\end{equation}}
\def\bsub{\begin{subequations}}
\def\esub{\end{subequations}}

\def\R{\mathrm{R}}

\def\Oint{O^{(\rm{int})}_{[2]}}

\def\Oincov{O^{(\rm{int})}}

\def\inter{\rm{int}}

\def\cover{\rm{cover}}

\def\bbS{{\mathbb S}}
\def\bbT{{\mathbb T}}
\def\AdS{{\mathrm{AdS}}}
\def\CFT{{\mathrm{CFT}}}

\def\id{\rm{id}}

\def\Cent{{\mathrm{Cent}}}

\def\Class{{\mathrm{Cl}}}

\def\Symm{\rm{Sym}}

\def\c{\check}
\def\h{\hat}

\def\Nzn{{N^\zeta_n}}

\newcommand{\AmZ}[1]{A\substack{\alpha\beta | \hat\s \hat\vro |\check\s\check\vro \\ n_1,n_2}(#1)} 

\newcommand{\Amint}[1]{A\substack{\inter | \hat\s \hat\vro |\check\s\check\vro \\ n_1,n_2 }(#1)} 

\newcommand{\AmZnn}[1]{A\substack{\alpha\beta | \hat\s \hat\vro |\check\s\check\vro \\ n,n}(#1)} 

\newcommand{\Amintnn}[1]{A\substack{\inter | \hat\s \hat\vro |\check\s\check\vro \\ n,n }(#1)}

\def\hZ{h_Z}

\def\hX{h_{XX}}

\title{Four-point functions with multi-cycle fields in symmetric orbifolds and the D1-D5 CFT}

\date{\today} 

\usepackage{authblk}

\author[1]{\normalsize Andre Alves Lima\thanks{andrealves.fis@gmail.com}}
\author[1]{\normalsize G.~M. Sotkov\thanks{gsotkov@gmail.com}}
\author[2]{\normalsize M. Stanishkov\thanks{marian@inrne.bas.bg}}

\affil[1]{\textit{\footnotesize Department of Physics, Federal University of Esp\'irito Santo, 29075-900, Vit\'oria, Brazil}}
\affil[2]{\textit{\footnotesize Institute for Nuclear Research and Nuclear Energy, Bulgarian Academy of Sciences, 1784 Sofia, Bulgaria}}

\begin{document}
\begin{titlepage}

\maketitle

\begin{abstract}

We study $S_N$-invariant four-point functions with two generic multi-cycle fields and two twist-2 fields, at the free orbifold point of the D1-D5 CFT. We derive the explicit factorization of these functions following from the action of the symmetric group on the composite multi-cycle fields. Apart from non-trivial symmetry factors that we compute, the function with multi-cycle operators is reduced to a sum of connected correlators in which the composite fields have, at most, two cycles. The correlators with two double-cycle and two single-cycle fields give the leading order contribution in the large-$N$ limit. We derive explicit formulas for these functions, encompassing a large class of choices for the single- and the double-cycle fields, including generic Ramond ground states, NS chiral fields and the marginal deformation operator. We are thus able to extract important dynamical information from the short-distance OPEs: conformal dimensions, R-charges and structure constants of families of BPS and non-BPS fields present in the corresponding light-light and heavy-light channels.  We also discuss properties of generic multi-cycle $Q$-point functions in $M^N/S_N$ orbifolds, using a technology due to Pakman, Rastelli and Razamat.

{\footnotesize 
\bigskip
\noindent
\textbf{Keywords:}
\noindent
Symmetric product orbifold of $\mathcal {N}=4$ SCFT;
multi-cycle Ramond and NS fields; four-point correlation functions at large $N$ limit, OPEs and fusion rules.
}

\end{abstract}

\pagenumbering{gobble}

\end{titlepage}

\thispagestyle{empty}

\pagenumbering{roman}

\tableofcontents

\newpage

\pagenumbering{arabic}

\section{Introduction and summary}

The AdS/CFT correspondence has been a two-way lane leading to insights both into quantum gravity and into aspects of the strong coupling structure of quantum field theories. 
Significant computational (and conceptual) developments 
have sprung from modern technologies devised to compute 
correlation functions using the full resource of the symmetries of holographic CFTs.
Progress has been notable, in particular, in the context of four-dimensional ${\cal N} = 4$ SYM dual to $\AdS_5 \times \bbS^5$, with some results extending to other instances of $\AdS_{d+1}/\CFT_d$ with $d > 2$.
Meanwhile, the study of correlation functions in $\AdS_3/\CFT_2$ has progressed at a somewhat different pace. 
Methods such as Witten diagrams and Mellin transforms
meet some technical difficulties when faced with the idiosyncrasies of two-dimensional CFTs 
\cite{Giusto:2018ovt,Giusto:2019pxc,Rastelli:2019gtj}.   
On the other hand, it is precisely the exceptional nature of $\CFT_2$ and of $\AdS_3 \times \bbS^3$ that makes $\AdS_3/\CFT_2$ special \cite{Eberhardt:2019qcl,Eberhardt:2018ouy,Eberhardt:2019ywk,Gaberdiel:2020ycd}, and correlation functions in the holographic symmetric orbifold CFT particularly relevant.

\subsection{On the problem of computing four-point functions in the D1-D5 CFT} 

One of the ongoing programs for computing four-point functions in $\AdS_3/\CFT_2$ 
uses `microstate geometries' as a tool
\cite{Galliani:2016cai,
Galliani:2017jlg,
Bombini:2017sge,
Giusto:2018ovt,
Tian:2019ash,
Bombini:2019vnc,
Giusto:2019pxc,
Giusto:2020neo,
Ceplak:2021wzz}.
Microstate geometries are horizonless solutions
of Type IIB supergravity 
that are asymptotically  $\AdS_3 \times \bbS^3 \times M$.
They are part of the conjectured `fuzzball' resolution of black holes formed by bound D1-D5 branes wrapping $\bbT \times M$, with $M$ being $\bbT^4$ or K3 \cite{Lunin:2001jy,Skenderis:2008qn}. The dual $\CFT_2$, called `the D1-D5 CFT', is a ${\cal N} = (4,4)$ superconformal theory in the moduli space of the 
symmetric orbifold $M^N / S_N$.
A vast collection of such geometries has now been found, largely due to the development of the fuzzball program. In particular, $\frac14$-BPS geometries have been completely classified, being dual to superpositions of Ramond ground states; classes of $\frac18$-BPS geometries are known as well \cite{Lunin:2001jy,Skenderis:2008qn,Skenderis:2006ah,Kanitscheider:2007wq,Kanitscheider:2006zf,Taylor:2007hs}.
In the semi-classical limit where $N \gg 1$, the central charge of the D1-D5 CFT, $c = 6N$, is very large.
As $c\to \infty$, an operator is said to be `heavy' if its conformal weight scales as $h_H \sim c$, or `light' if its weight $h_L$ is fixed and finite.
Operators dual to specific (microstate) geometries are heavy: for example, the Ramond ground states have $h_H = \frac1{24} c$.  On the other hand, probe-like fields in the bulk are light.  
If $O_H$ and $O_L$ are heavy and light, respectively, the four-point function
\be
\Big\langle \bar O_H(\infty) \, \bar O_L(1) \, O_L(z,\bar z) \, O_H(0) \Big\rangle \label{HHLLintro}
\ee
can be regarded as a two-point function of $O_L$ in the state $\ket{O_H}$.
In the bulk, this corresponds to the propagator of the light field dual to $O_L$ in the asymptotically $\AdS_3 \times \bbS^3$ geometry dual to $\ket{O_H}$.

Relating the bulk and CFT descriptions requires, first of all, that one knows the precise translation of $O_H$ into a microstate geometry, and of $O_L$ into a bulk field.
Pages of the holographic dictionary were written some time ago \cite{Skenderis:2008qn,Skenderis:2006ah,Kanitscheider:2007wq,Kanitscheider:2006zf,Taylor:2007hs,Giusto:2015dfa}, others more recently 
\cite{Rawash:2021pik,Giusto:2019qig,Tormo:2018fnt}.
From the bulk perspective, several examples of heavy-light correlators based on microstate geometries
\cite{Galliani:2016cai,
Galliani:2017jlg,
Bombini:2017sge,
Giusto:2018ovt,
Bombini:2019vnc,
Giusto:2019pxc,
Giusto:2020neo,
Ceplak:2021wzz}
have been studied, the computation of (\ref{HHLLintro}) amounting to solving a wave equation in the fixed space-time background.
One of the interesting uses of these correlators is to contrast them against known universal semi-classical properties of correlators in AdS.
 At large $c$, the form of conformal blocks of heavy-light four-point functions is constrained by AdS symmetries
\cite{Fitzpatrick:2014vua,Fitzpatrick:2015zha,Fitzpatrick:2015dlt,Fitzpatrick:2016ive,Hijano:2015rla,Hijano:2015qja,Alkalaev:2015wia,CarneirodaCunha:2016zmi}, resulting in phenomena associated with the thermality of AdS black holes, such as ``spurious singularities'' outside of OPE limits, that contradict the unitarity of the CFT.  
It has been argued that these paradoxical properties should be resolved by corrections appearing at order $1/c$, but in \cite{Galliani:2016cai,Bombini:2017sge} explicit computations with known microstate geometries revealed examples of heavy-light correlators that are unitary already at leading order in $c$. 

The holographic dictionary between SUGRA and the D1-D5 CFT in the free orbifold point
is based on the proper identification of protected states, BPS operators, their OPE algebra, and the corresponding three-point functions
\cite{Taylor:2007hs,Giusto:2015dfa,Rawash:2021pik,Giusto:2019qig,Tormo:2018fnt}.
Now, 
\emph{four}-point functions are not guaranteed to be protected when the free CFT is deformed in moduli space, even if they involve only protected operators. This is because four-point functions typically depend also on non-BPS (i.e.~non-protected) fields that might appear in channels of the OPEs between pairs of twisted operators. 
This is true, in particular, of the deformation modulus, which has twist 2.
Thus, while the holographic dictionary allows us to identify which heavy/light SUGRA fields correspond to which heavy/light CFT operators,
there is generally a mismatch between computing the heavy-light function (\ref{HHLLintro}) in terms of linear fluctuations around microstate geometries, or as the corresponding correlator in the free CFT.
Explicit examples of this can be found, for instance, in Refs.\cite{Galliani:2016cai,Galliani:2017jlg}; the correlators computed in \cite{Galliani:2016cai} in the SUGRA and in the free CFT descriptions match, but those computed in \cite{Galliani:2017jlg}, involving less symmetric heavy fields, do not. 
So, to completely fix the dictionary between four-point functions, it is necessary to determine the conditions that select contributions (at leading order) only from OPE channels containing BPS operators. The computation of (\ref{HHLLintro}) in the free orbifold point, as well as a study of the non-BPS content of the different OPE channels, appears to be an important step towards this goal.

Computing functions like (\ref{HHLLintro}), even at the free point, is not always easy. The twisted boundary conditions of the symmetric orbifold $M^N / S_N$ often causes correlation functions to be rather complicated.
The orbifold has twisted sectors corresponding to permutations $\prod_n (n)^{N_n} \in S_N$, where $(n)$ is a cycle of length $n$, and the multiplicities $[N_n]$ form a partition $\sum_n n N_n = N$. 
The fact that permutations are uniquely decomposable into disjoint cycles can be interpreted as saying that multi-cycle fields are ``composite'', while single-cycle fields are indecomposable.
In this sense, single- and multi-cycleness in the orbifold CFT$_2$ is analogous to single- and mutli-traceness in four-dimensional ${\cal N} = 4$ SYM. 
Heavy fields, such as the Ramond ground states, are multi-cycle, while light fields, and in particular ``elementary'' fields, are generically single-cycle.
Typically, Ramond ground states are in fact made of many cycles, also known as `strands', with different lengths and ``spins''.
Hence (\ref{HHLLintro}) is typically a complicated function, with all the $S_N$ monodromy properties and selection rules that ensue.
One way to still have a not-very-complicated function is to take the light fields in (\ref{HHLLintro}) to be \emph{un}twisted. This simplifies the permutations to such an extent that one does not even need to resort to covering surfaces. Examples of such computations have been considered in many places  \cite{Galliani:2016cai,Galliani:2017jlg,Giusto:2018ovt,Ceplak:2021wzz,Bombini:2017sge}, leading to very interesting results as mentioned above. 
It is well known, however, that the complete holographic bulk-boundary dictionary --- for, say,  light NS chiral fields with conformal dimension one or two --- must include both untwisted \emph{and} twisted (with twist 2) fields with equal conformal dimensions.

\subsection{A summary of our results}

In the present paper we consider examples of four-point functions with the simplest configuration of \emph{twisted} light NS fields.
That is, our goal is to study correlators where \emph{all fields are non-trivially twisted} and, besides, some of the fields (e.g.~the heavy Ramond ground states) can be \emph{multi}-cycle.

The paper can be divided in two parts. In Sects.\ref{SectMultiCyclCorr}-\ref{SectCompFunct}, we study  general properties of correlation functions with multi-cycle fields in $M^N/S_N$ orbifolds. 
In Sect.\ref{SectFuncandMast}, we apply our general results to the D1-D5 CFT, computing a collection of four-point functions with Ramond ground states, NS chiral fields and the deformation modulus.

More precisely, in Sect.\ref{SectMultiCyclCorr} we study generic $Q$-point functions of twisted fields, and extract their decomposition into components associated with equivalence classes of permutations in $S_N$. 
(To improve clarity, detailed derivations of the results of Sect.\ref{SectMultiCyclCorr} are presented App.\ref{SectCountinFact}.)
We follow the work of \cite{Pakman:2009zz}, but with some relevant differences. First, we keep the twists generic, not restricted to single cycles. 
Second, the derivation, in Ref.\cite{Pakman:2009zz}, of the $N$-dependence of twisted correlators relies heavily on a diagrammatic interpretation of connected functions, while ours does not. Instead, we use a construction of equivalence classes of twist permutations entering in a given $Q$-point function. 
This is also a technology developed in \cite{Pakman:2009zz}: the equivalence classes are in one-to-one correspondence with diagrams for connected correlators.
Hence, although we do not resort to the diagrammatic interpretation, our analysis is in fact an application of the methodology of \cite{Pakman:2009zz} to (often disconnected) correlation functions with generic, multi-cycle fields.
That is a powerful technique, relating an orbifold CFT to the geometry of coverings of the Riemann sphere, via the conjugacy classes of twists, being thus related to `Hurwitz theory' (see e.g.~\cite{cavalieri_miles_2016}). 
Here we try to outline how this language can be used explore symmetries of twisted correlation functions. 

Specifically, we want to compute four-point functions involving (light) fields $Z_{[\ell]}$ with single-cycle twists of length $\ell = 2$, and (heavy) multi-cycle fields
$[\prod_{\zeta,n} ( \bar X^\zeta_{[n]} )^{N^{\zeta}_n} ]$,
with \emph{arbitrary} twist given by a partition of $N$,
\be \label{XXZZXXintro}
\begin{aligned} 
\Big\langle
\Big[ \prod_{\zeta,n} ( \bar X^\zeta_{[n]} )^{N^{\zeta}_n} \Big] (\infty) 
\, \bar Z_{[2]} (1)
\, Z_{[2]} (v,\bar v)\Big[ \prod_{\zeta,n} ( X^\zeta_{[n]} )^{N^{\zeta}_n} \Big] (0)
\Big\rangle ,
\\
\text{where}
\quad
\sum_{n,\zeta} N^{\zeta}_n \, n = N .
\end{aligned}
\ee
Here we use an index $\zeta$ to possibly distinguish between distinct components of the multli-cycle field which have the same cycle length $n$. For example, in the case where the multi-cycle field is a Ramond ground state, $\zeta$ indicates the R-charges of the strands.
We focus on twists $\ell = 2$ for the single-cycle $Z$ fields for two reasons.
First, it is the simplest non-trivial twist.
Second, the interesting moduli that deform the free orbifold CFT into an interacting theory dual to SUGRA solutions lie in the twisted sector with $\ell = 2$ \cite{David:2002wn,Avery:2010er}.
We will specifically consider the marginal deformation operator $\Oint$ which is a scalar under all SU(2) symmetries of the ${\cal N} = (4,4)$ superconformal algebra. 
We also consider NS chirals with $\ell=2$ which include, in particular, another set of operators with dimension one, the ``middle-cohomology'' NS chirals.

The function (\ref{XXZZXXintro}) is typically \emph{disconnected}.
By this, we mean that it factorizes into products of functions involving only some of the operators that compose the multi-cycle field --- not only products of two-point functions, but also of three- and four-point functions with ``smaller'' composite fields. In other words, the \emph{`disconnected four-point functions'} addressed in this paper are not ``bubble diagrams''; in fact, they are still dynamical objects.
It is well-known that twisted correlation functions are associated with ramified covering surfaces of the Riemann sphere
\cite{Lunin:2000yv,Lunin:2001pw}.
The nomenclature `disconnected' also agrees with the fact that, since the correlator factorizes, its associated covering surface can be seen as a product of disconnected surfaces.

One important information to be extracted from correlation functions is how they depend on $N$ or, at least, how they scale with large $N$. 
For connected correlators, the exact dependence found in \cite{Pakman:2009zz} reproduces the result of \cite{Lunin:2000yv},
\be \label{scalg}
\text{Connected single-cycle $Q$-point function} 
\sim \sum_{\bf g} N^{- {\bf g} + 1 - \frac12 Q}
\ee
where ${\bf g}$ is the genus of the covering surface.
For multi-cycle fields, the disconnected functions are associated with disconnected covering surfaces, for which the ${\bf g}$ is not well-defined. Still, we find a natural generalization of (\ref{scalg}), featuring the Euler invariant $\chi$ instead of ${\bf g}$,
\be \label{scalchi}
\left[\begin{aligned}
&\text{Disconnected* multi-cycle $Q$-point function} 
\\
&\qquad \qquad\quad \text{with $R \geq Q$ cycles}
\end{aligned}
\right]
\sim \sum_\chi N^{ \frac12 ( \chi - R)} .
\ee
The Euler invariant $\chi$ \emph{is} a well-defined, additive property of disconnected surfaces. For connected surfaces/correlators, it reduces to $\chi = 2 - 2{\bf g}$, and, if the fields are single-cycle, i.e.~$R = Q$, Eq.(\ref{scalchi}) reduces to (\ref{scalg}).
The reason for the * in Eq.(\ref{scalchi}) is that the formula requires some assumptions about the twists:
the number of cycles and their lengths must both be kept \emph{fixed} when $N \to \infty$.
These assumptions are very natural for connected single-cycle functions, but they do not hold for some of the most important examples of multi-cycle fields.
In particular, they do not hold for functions like (\ref{XXZZXXintro}),  unless $N^\zeta_1 \to \infty$. 
To find how functions that do not fulfill the assumptions of (\ref{scalchi}) 
depend on $N$ can be a rather difficult problem in general, that is strictly dependent on the twists of all fields entering the correlator.
Let us note that many results that we derive in Sect.\ref{SectMultiCyclCorr} were previously found by Dei and Eberhardt in \cite{Dei:2019iym}.

In Sect.\ref{SectCompFunct}, we apply the language developed in Sect.\ref{SectMultiCyclCorr} for generic $Q$-point functions, to study (\ref{XXZZXXintro}) in full detail.
Generalizing our previous works \cite{Lima:2020nnx,Lima:2021wrz}, where similar functions were considered, here we work at the level of $M^N/S_N$ orbifolds, i.e.~focusing only on the twists, not on the specific form of the fields $X^\zeta_{[n]}$ and $Z_{[2]}$.
Because transpositions are the simplest non-trivial elements of $S_N$, we are able to derive in detail the structure of these four-point functions, including the explicit way it factorizes into connected parts 
\be \label{connfunintro}
\Big\langle
\Big[ \bar X^{\zeta_1}_{[n_1]} \bar X^{\zeta_2}_{[n_2]} \Big] (\infty) 
\, \bar Z_{[2]} (1)
\, Z_{[2]} (v,\bar v)\
\Big[ \bar X^{\zeta_1}_{[n_1]} \bar X^{\zeta_2}_{[n_2]} \Big] (0)
\Big\rangle ,
\ee
containing only double-cycle components of the original multi-cycle field, multiplied by ``symmetry factors''. The double-cycle functions (\ref{connfunintro}) always appear in the factorization of (\ref{XXZZXXintro}) in association with a covering surface of genus zero.
While, for connected correlators with the same number of twists, the genus-zero contributions always dominate over higher genera, in the factorization of (\ref{XXZZXXintro}) there are also genus-one contributions, but with only one single-cycle component $X^{\zeta}_{[n]}$, instead of the composite double-components seen in (\ref{connfunintro}). By themselves these single-cycle, genus-one functions contribute at order $N^{-2}$, which is the same as the double-cycle, genus-zero functions (\ref{connfunintro}). 
But we show that, when multiplied by their symmetry factors, the genus-zero contributions do dominate when $N$ is large.

Let us note that, apart from the argument that the genus-zero functions dominate at large $N$, we will not take $N$ to be large in our formulas, so most of our formulas are exact at \emph{finite} $N$.
This is why, through most of the paper, we avoid the nomenclature `heavy' and `light' fields, preferring `multi-cycle' and `single-cycle' fields instead. 
It should be kept in mind, nevertheless, that, in the D1-D5 CFT examples we consider, the multi-cycle fields are almost always heavy (Ramond) fields, the single-cycle fields are always light, and that heavy-light correlators in the semi-classical limit are an important part of our motivations.

Focusing on (\ref{connfunintro}), the appropraite genus-zero covering map was derived in \cite{Lima:2020nnx}. Here we present a  detailed analysis of the geometry of the covering surfaces, and the relation between coverings and permutation classes.
The goal is to understand how the geometry dictated by the twists controls the form of the correlation functions.
The connected functions can be 
decomposed into ${\bf H}$ `Hurwitz blocks', where ${\bf H}$ is the Hurwitz number of different coverings of the sphere.%
\footnote{%
Hurwitz blocks correspond to diagrams in the language of \cite{Pakman:2009zz}.}
These blocks are defined by the roots of an algebraic equation, which cannot be found in closed form when $n_1 \neq n_2$, but nevertheless fix many properties of the correlation functions.
In particular,  they determine the twists of the fields appearing in the OPE channels,
\begin{align} \label{OPEsintro}
\begin{aligned}
\bar Z_{[2]} \times  Z_{[2]} 
&=  C_{Z \bar Z U} \ \{ U_{[1]} \} + C_{Z \bar Z S} \ \{ S_{[3]} \}
\\
Z_{[2]} \times \big[ X^{\zeta_1}_{[n_1]}  X^{\zeta_2}_{[n_2]} \big]
&= 
C_{ [\bar X\bar X] \bar Z Y} \ \big\{  Y_{[n_1+n_2]} \big\}
\\
&\quad
+
C_{ \bar X  \bar Z [W X]} 
B_{\bar X X}
\ \Big\{  \big[ W_{[n_1-n_2]} X^{\zeta_2}_{[n_2]} X^{\zeta_2}_{[n_2]} \big] \Big\} 
\end{aligned}
\end{align}
where $C$s are structure constants, $B$ a two-point function normalization, and curly brackets indicate conformal families.
The appearance of the composite field containing the operator $W_{[n_1-n_2]}$, with twist length $n_1-n_2$, is the result of an interesting interaction between the twist permutations in the $v \to 0$ channel, which was previously overlooked in our papers \cite{Lima:2020nnx,Lima:2021wrz}, and is now analyzed in detail within this more general context of $M^N/S_N$ orbifolds.
When $n_1 = n_2$, we also find a special symmetry between covering surfaces (or equivalence classes, or Hurwitz blocks), that allow us to compute the correlators in closed form on the base sphere, while ``reducing the Hurwitz number by half''.

In Sect.\ref{SectFuncandMast} we turn our attention from $M^N/S_N$ orbifolds in general to the D1-D5 CFT (at the free-orbifold point) specifically. 
We derive a pair of ``master formulas'' that encompass many different choices for the operators in (\ref{XXZZXXintro})-(\ref{connfunintro}).
The multi-cycle fields can be Ramond ground states or composite NS chirals, and the single-cycle fields can be Ramond fields, NS chirals or the scalar deformation modulus taking the CFT away from the free-orbifold point.
With these functions, we can use the technology of Sect.\ref{SectCompFunct} to extract conformal data.
Besides the twists, we can find the dimensions of the operators and the structure constants in the OPEs (\ref{OPEsintro}).
In Refs.\cite{Lima:2020nnx,Lima:2021wrz} this analysis was performed when $Z_{[2]}$ is the interaction modulus $\Oint$, and the $X^\zeta_{[n]}$ are Ramond ground states of the $n$-twisted sector. Here, with our general functions, we can extend these results to find the OPEs between $\Oint$ and NS chirals, between NS chirals and Ramond ground states, and also between single-cycle and composite NS chirals themselves.
In the latter case, we note that the form of the correlation functions is restricted by the NS chiral ring, and show that we can recover some known structure constants \cite{Jevicki:1998bm,Dabholkar:2007ey,Pakman:2009ab}
by taking $n_2 =1$, reducing the composite field to a single-cycle one. 

The D1-D5 CFT's ${\cal N} = (4,4)$ superconformal algebra has a symmetry under `spectral flow' \cite{Schwimmer:1986mf}, that changes weights and R-charges of fields, and relates states in the NS and Ramond sectors. 
In \S\ref{SectSpecFlow}, we discuss the effect of spectral flow on four-point functions, and show how it connects specific pairs of functions derived with our master formulas.

We close with a brief discussion of our results and a few comments concerning the derivation of a new family of four-point functions
related to D1-D5-P superstrata. 
These four-point functions, which involve excitations of the left-moving twisted Ramond ground states, can be found in terms of the correlators calculated in the present paper, using standard $\mathcal {N}=4$  super-conformal Ward identities.

\section{Multi-cycle correlators on $S_N$ orbifolds}	\label{SectMultiCyclCorr}

The  $M^N / S_N$ orbifold is made by $N$ identical copies of a ``seed theory'' in $M$, each copy labeled by an index $I = 1,\cdots , N$, and all independent, so that the total central charge is $c = Nc_{\rm{seed}}$.
The Hilbert space decomposes into twisted sectors created by `bare twist fields' $\s_g(z)$. The permutation $g$ acts on copy indices of operators going around the twist, 
$O_I(e^{2\pi i} z ) \s_g(0) =  O_{g(I)}(z) \s_g(0)$.
Every $g \in S_N$ can be uniquely decomposed as a product of disjoint cycles of different lengths  $n_i$,
 $(n_i) = (I_1,\cdots,I_{n_i}) \in {\mathbb Z}_{n_i}$, 
\be	\label{gPartition}
{\textstyle g = \prod_{n} (n)^{N_n} ,	\qquad \sum_{n} n N_n = N }.
\ee
The conformal weight of $\s_g$, with $g$ given in (\ref{gPartition}), is the sum 
\be	\label{twistdim}
h^\s_{[N_n]} = \sum_n N_n h^\s_{n} ,
\quad 
h_n^\s = \frac{1}{4} \Big( n - \frac{1}{n} \Big) = \tilde h^\s_n 
\ee	
where $h^\s_n$ is the dimension of the single-cycle components 
\cite{Dixon:1986qv,Lunin:2000yv}.
The same is true for anti-holomorphic weight $\tilde h^\s_{[N_n]}$, and the total dimension is $\Delta^\s_{[N_n]} = h^\s_{[N_n]} + \tilde h^\s_{[N_n]}$.
The weight (\ref{twistdim}) is the same for any $g$ in the conjugacy class 
$[g] = \{ h g h^{-1} \, | \, h \in S_N\}$, associated with the partition
\be
[N_n] \equiv \{ N_n \in {\mathbb N} \ | \,  \textstyle\sum_{n} n N_n = N  \} .
\ee

Twists corresponding to individual permutations are not invariant under actions of $S_N$. An invariant field can be built by summing over the orbit of $g \in  S_N$ under the action of $S_N$ by conjugacy, 
\be	\label{Sninvg}
\s_{[g]} \equiv \frac{1}{\scr S_{[g]}} \sum_{h \in S_N} \s_{h g h^{-1}} .
\ee 
The factor  ${\scr S}_{[g]}$  ensures that the $S_N$-invariant two-point function normalization is the same as that of its (non-$S_N$-invariant) components.
In \S\ref{SectCountinFactTwp} we show that 
\be	\label{Sninvgb}
{\scr S}_{[g]} = \sqrt{ N! |\Cent [g] | } 
\ee
where the order of the centralizer of $g$ in $S_N$ is%
	\footnote{
	We recall the definition of $\Cent[g]$ in (\ref{CentDef}). See e.g.~\cite{sagan2001symmetric} for a derivation of formula (\ref{Sninvg}).
	}
\be	\label{OrdCentg}
\big| \Cent[g] \big| = \prod_n  N_n! \, n^{N_n} 
\quad
\text{for}
\quad
g =\prod_n (n)^{N_n} , \quad \sum_n N_n n = N ,
\ee
The result (\ref{Sninvgb}) was previously found in \cite{Dei:2019iym}.
For single cycles $g = (n)(1)^{N-n}$, it yields the usual normalization factor \cite{Arutyunov:1997gt,Lunin:2000yv,Pakman:2009zz} which we will denote by ${\scr S}_{n}$,
and for double cycles $g = (n_1) (n_2) (1)^{N-n_1-n_2}$ we obtain a normalization denoted by ${\scr S}_{n_1,n_2}$,
\be
{\scr S}_{n} = \sqrt{N! ( N-n)! n} ,
\qquad
{\scr S}_{n_1,n_2} = \sqrt{N! (N-n_1-n_2)! n_1n_2} .
\ee
Excited twisted fields can also be combined into $S_N$-invariant operators, in the same way as (\ref{Sninvg}), and with the same normalization (\ref{Sninvgb}).

\subsection{Twisted $Q$-point functions}	\label{SectQpointmain}

$Q$-point functions of twisted operators are subject to selection rules associated with the permutations carried by the fields.
A fundamental property of a twisted correlator, possibly with a collection $\scr X$ of excitations, is that the permutations must compose to the identity $\id \in S_N$,
\be
\Big\langle {\scr X} \prod_{i=1}^Q \s_{g_i}  (z_i ) \Big\rangle
\neq 0
\quad
\text{only if}
\quad
g_1 \cdots g_Q = \id  \label{compto1}
\ee
otherwise the function would have ill-defined monodromy.
If the $g_i$ can be separated into two disjoint sets $\{g_i\} = \{g_k\} \sqcup \{g_\ell\}$, such that one set commutes with the other, the function factorizes,
\be	\label{Qfact}
\Big\langle {\scr X} \prod_{i} \s_{g_i}  (z_i ) \Big\rangle
 =
\Big\langle {\scr Y} \prod_k \s_{g_k}  (z_k ) \Big\rangle
\
\Big\langle {\scr Y}' \prod_\ell \s_{g_\ell}  (z_\ell ) \Big\rangle 
\ee
where ${\scr Y}$ and ${\scr Y}'$ are excitations of the respective sets of bare twists. A function which cannot be factorized in such a way is called `connected'.
In this section, we will be interested only on the $S_N$-related properties of twisted correlators, so we now consider functions of bare twists $\s_{[g]}$ only.
The $Q$-point function of invariant operators is the sum
\be	\label{sumQs1}
\Big\langle \s_{[g_1]} (z_1) \cdots \s_{[g_Q]}(z_Q) \Big\rangle
	=
	\frac{1}{\prod_i {\scr S}_{[g_i]}}
	\sum_{\substack{h_1 \in S_N \\ \cdots \\ h_Q\in S_N}}
	\Big\langle \s_{h_1 g_1 h_1^{-1}} (z_1) \cdots \s_{h_Q g_Q h_Q^{-1}}(z_Q) \Big\rangle .
\ee
Many terms of the sum of the r.h.s.~vanish because they do not satisfy the condition (\ref{compto1}). It is convenient to replace the individual sums over the orbits of the $\{ g_i \}$ by sums of different equivalence classes of permutations that do satisfy (\ref{compto1}).
Let $p_i = h_i g_i h_i^{-1}$ be the permutations in the r.h.s.~of (\ref{sumQs1}), and consider an ordered sequence $\{p_1, \cdots, p_Q\}$ that satisfies (\ref{compto1}), 
\be	\label{pcompot1}
p_1 \; p_2 \; \cdots \; p_Q = \id .
\ee 
This
will also be satisfied by every other sequence in the \emph{equivalence class} $\a$ defined by 
\be	\label{equivaM}
\a : \quad
\{p_1, p_2 , \cdots, p_Q\} \sim \{ k p_1 k^{-1} , k p_2 k^{-1} , \cdots , k p_Q k^{-1} \} , 
\ee
i.e.~a \emph{global} conjugation of every $p_i$ by the \emph{same} $k \in S_N$.
Moreover, all correlators $\langle \s_{p_1} (z_1) \cdots \s_{p_Q}(z_Q)\rangle$ with the $\{p_i\}$ in a given class $\a$ will be equal by symmetry, because the global conjugation only relabels every copy in the twists, and all copies are identical.
If we denote the set of all such conjugacy classes by $\Class$, the sum over orbits in (\ref{sumQs1}) can be replaced by a sum over all $\a \in \Class$, where we take one representative correlator for each class $\a$, and multiply it by a ``symmetry factor'' ${\cal N}_\a$, counting the number of permutations in $\a$. It is convenient to separate the classes according to the number ${\bf c}$ of distinct copies that participate in \emph{non-trivial} cycles (i.e.~cycles of length $n > 1$).%
	\footnote{%
	For example, for $N = 9$, the permutation $(1,2,3)(5,6)$,
	has ${\bf c} = 5$ non-trivial copies, namely copies $I = 1,2,3,5,6$. 
	Copies $I =4,7,9$ are trivial, as they do not participate in any non-trivial cycle. See the discussion below (\ref{alclassecxapl}) for examples of \emph{classes} of permutations with different ${\bf c}$.
	}
This number is, of course a class property, so we can decompose $\Class = \cup_{\bf c} \Class_{\bf c}$. In the end, the r.h.s.~of Eq.(\ref{sumQs1}) becomes
\begin{align}	\label{pdsqMa2}
\begin{aligned}[b]
\Big\langle \s_{[g_1]} (z_1) \cdots \s_{[g_Q]}(z_Q) \Big\rangle
	=
	\frac{1}{\prod_i {\scr S}_{[g_i]}}
	\sum_{\bf c}
	\sum_{\a_{\bf c} \in \Class_{\bf c}}
	{\cal N}_{\a_{\bf c}}
	\Big\langle \s_{p_1^{\a_{\bf c}}} (z_1) \cdots \s_{p_Q^{\a_{\bf c}}}(z_Q) \Big\rangle
\end{aligned}
\end{align}

In Appendix \ref{SectCountinFact} we give several examples of classes $\a$ and discuss the set $\Class$ in detail.
Some of the classes are made of permutations that factorize, as in (\ref{Qfact}), one or more times. The type of factorization is, also, a class property. In App.\ref{SectCountinFact} we show that the symmetry factor ${\cal N}_{\a_{\bf c}}$ is basically the same for every class $\a_{\bf c} \in \Class_{\bf c}$, 
\be	\label{symmfaccalN}
{\cal N}_{\a_{\bf c}} = \frac{N!}{(N-{\bf c})!} \left( \prod_{i=1}^Q \big| \Cent[g_i] \big| \right) \nu_{\a_{\bf c}} .
\ee
The only class-dependent factor, $\nu_{\a_{\bf c}}$, is given by Eq.(\ref{formwas}). In classes $\a_{\bf c}$ where no two-point function factorizes, $\nu_{\a_{\bf c}} = 1$; in classes $\a_{\bf c}$ where there is a factorization of $d$ two-point functions with cycles $n_j$, $j = 1,\cdots, d$, we have $\nu_{\a_{\bf c}} = 1 / \prod_j n_j$.
Eqs.(\ref{pdsqMa2}) and (\ref{symmfaccalN}) contain the exact $N$-dependence of the twisted $Q$-point function,
\begin{align}	\label{pdsqMa2N}
\Big\langle \prod_{i=1}^Q \s_{[g_i]} (z_1)  \Big\rangle
	=
	\left[
	\prod_{i=1}^Q \sqrt{\frac{| \Cent[g_i] |}{N!}}
	\right]
	\sum_{\bf c} 
	\frac{N!}{(N - {\bf c})!}
	\sum_{\a_{\bf c} \in \Class_{\bf c}}
	\nu_{\a_{\bf c}}
	\Big\langle \prod_{i=1}^Q \s_{p_i^{\a_{\bf c}}} (z_i)  \Big\rangle
\end{align}
This formula generalizes a result of \cite{Pakman:2009zz}, which only considers connected functions.
	The \emph{connected} classes  $\a \in \Class_{\bf g}$ can be described in a diagrammatic language developed in \cite{Pakman:2009zz}; each class $\a$ corresponds to a different diagram, and the sum in Eq.(\ref{QptNDoubNchi}) is a ``sum over diagrams''.

\subsubsection{The large-$N$ limit}

The way Eq.(\ref{pdsqMa2N}) depends on $N$ seems to dwell solely in the coefficients multiplying the last sum over $\a_{\bf c} \in \Class_{\bf c}$.
If this was the case, it would suffice to expand these coefficients as functions of $N$ to find scaling of the function as $N \to \infty$. This works for single-cycle correlators with cycles of fixed length \cite{Pakman:2009zz}, but when we consider multi-cycle fields, there are subtleties.
A multi-cycle twist may be allowed to have a large number of cycles; an important example is $g_i = (n)^{N/n} \in S_N$.
So the centralizers of $g_i$ may depend on $N$, in Eq.(\ref{pdsqMa2N}).
Also,  the number of terms in the sum over $\a_{\bf c} \in \Class_{\bf c}$ may be very large, scaling with $N$.
In summary, determining the scaling of a multi-cycle $Q$-point function in the large-$N$ limit is a problem that depends intrinsically on the specific properties of the twists involved. The detailed analysis of a relatively simple case is the subject of Sect.\ref{SectQpointmain} below.

But, under certain assumptions, we can find an interesting generalization of the results of \cite{Pakman:2009zz}.
Let us isolate trivial cycles in the twist permutations,
\be	\label{giconjN}
g_i = (1)^{N- \sum_j N^{(i)}_j n^{(i)}_j}  \prod_j (n^{(i)}_j)^{N^{(i)}_j} ,
\quad
 \quad n^{(i)}_j > 1 
\ee
such that the order of the centralizers are
\be	\label{prNmninNnij}
\big| \Cent[g_i] \big| = \left( N- \textstyle\sum_j N^{(i)}_j n^{(i)}_j \right)! \prod_j N_j! \ n_j^{N_j}  .
\ee
If we now 
\emph{assume that the cycle lengths $n_j^{(i)}$ and their multiplicities $N_j^{(i)}$ are fixed as $N \to  \infty$}, we can use Stirling's formula to find
\be	\label{lcneNlsar}
\sqrt{\frac{| \Cent[g_i] |}{N!}} 
	\approx 
	\sqrt{\textstyle\prod_j N^{(i)}_j! \ (n^{(i)}_j)^{N^{(i)}_j} }
	\
		e^{-\frac12 \sum_j N^{(i)}_j n^{(i)}_j}
	\	
		N^{- \frac12 \sum_j N^{(i)}_j n^{(i)}_j}
\ee
expanding the factor $N! / (N-{\bf c})!$ as well, Eq.(\ref{pdsqMa2N}) becomes, up to order $1/N$,
\begin{align}	\label{pdsqMa3}
\Big\langle \prod_{i=1}^Q \s_{[g_i]} (z_1)  \Big\rangle
	\approx
	\sum_{\bf c} 
	N^{ {\bf c} - \frac12 \sum_i\sum_j N^{(i)}_j n^{(i)}_j }
	\varpi
	\sum_{\a_{\bf c} \in \Class_{\bf c}}
	\nu_{\a_{\bf c}}
	\Big\langle \s_{p_1^{\a_{\bf c}}} (z_1) \cdots \s_{p_Q^{\a_{\bf c}}}(z_Q) \Big\rangle
\end{align}
The factor
\be	\label{varpifom}
\varpi 
	= e^{{\bf c} -\frac12 \sum_i \sum_j N^{(i)}_j n^{(i)}_j} 
		\sqrt{\textstyle\prod_i\prod_j N^{(i)}_j! \ (n^{(i)}_j)^{N^{(i)}_j} }
\ee
is the same for all classes $\a_{\bf c}$, depends on the lengths $n_r$ and on ${\bf c}$ but not on $N$. 
Let $R$ be the total number of non-trivial cycles in \emph{all} the permutations $g_i$, 
and $n_r > 1$, $r = 1, \cdots, R$, be their lengths,
in such a way that we can write the sum in the exponent of $N$ in Eq.(\ref{pdsqMa3}) as
$
\sum_i\sum_j N^{(i)}_j n^{(i)}_j 
\equiv
 \sum_{r=1}^R n_r .
$
Inserting this into Eq.(\ref{pdsqMa3}) leads to Eq.(\ref{QptNDoubN}).
\begin{align}	\label{QptNDoubN}
\Big\langle \prod_{i=1}^Q \s_{[g_i]} (z_1)  \Big\rangle
	=
	\sum_{\bf c} 
	N^{ {\bf c} - \frac12 \sum_{r=1}^R n_r }
	\Big[1 + \rm{O}(\tfrac1N) \Big]
	\varpi
	\sum_{\a_{\bf c} \in \Class_{\bf c}}
	\nu_{\a_{\bf c}}
	\Big\langle \prod_{i=1}^Q \s_{p_i^{\a_{\bf c}}} (z_i)  \Big\rangle
\end{align}
Note that the total number of these cycles is $R \geq Q$, and if all $g_i$ are \emph{single}-cycle permutations, $R = Q$.
If we further assume that the sum over classes ${\a_{\bf c} \in \Class_{\bf c}}$ in Eq.(\ref{QptNDoubN}) also do not depend on $N$, then we have found the leading large-$N$ scaling of the function.
This assumption about the sum over classes is not unrelated to the assumption used to derive (\ref{lcneNlsar}). If there is a finite number of cycles with fixed (and finite) lengths $n_r$, then it is reasonable to expect a finite number of non-vanishing classes satisfying the condition (\ref{pcompot1}). (This is true, for example, in the case of single-cycle fields.)

Eq.(\ref{QptNDoubN}) can be written as
\begin{align}	\label{QptNDoubNchi}
\Big\langle \prod_{i=1}^Q \s_{[g_i]} (z_1)  \Big\rangle
	=
	\sum_{\chi} 
	N^{ \frac12 (\chi -  R) }
	\Big[1 + \rm{O}(\tfrac1N) \Big]
	\varpi( n_r)
	\sum_{\a_{\chi} \in \Class_{\chi}}
	\nu_{\a_\chi}
	\Big\langle \prod_{i=1}^Q \s_{p_i^{\a_\chi}} (z_i)  \Big\rangle
\end{align}
if we define the number
\be	\label{EulerCh}
\chi({\bf c}) \equiv 2 {\bf c} + R -  \sum_{r=1}^R n_r .
\ee
We can interpret $\chi$ as the \emph{Euler characteristic of covering surfaces}, as follows.
It is well-known that a connected twisted correlator is associated with  a ramified covering surface $\Sigma$ of the `base sphere' $\bbS^2_\rm{base}$ \cite{Lunin:2000yv}. 
In a \emph{connected} correlator, the number $R$ of non-trivial cycles is the number of ramification points of $\Sigma$, the number ${\bf c}$ of distinct copies entering these cycles is the number of sheets of $\Sigma$, and the order of the ramification point associated with the cycle $(n_r)$ is its length $n_r$. With this ramification data, the Riemann-Hurwitz formula gives the \emph{genus} of the (connected) covering surface $\Sigma$ to be
\be
  \label{RiemHurwForm}
 {\bf g} = 1 - {\bf c} - \frac12 R + \frac{1}{2} \sum_{r=1}^R n_r 
\ee
which is compatible with (\ref{EulerCh}), i.e.
$\chi = 2 - 2{\bf g}.$
But some of the classes $\a_{\bf c}$ may give \emph{disconnected} correlators, which are products of $\ell$ connected functions. The latter are each associated with a covering surface $\Sigma_i$, and we can assoaciate the factorized correlator with their disjoint union $\Sigma_1 \sqcup \cdots \sqcup \Sigma_\ell$.
The ${\bf c}$ non-trivial copies and the non-trivial cycles will be split among the factorized correlators in such a way that Eq.(\ref{EulerCh}) gives, schematically
\be
\textstyle \chi( \Sigma_1 \sqcup \cdots \sqcup \Sigma_\ell) = \chi(\Sigma_1) + \cdots + \chi( \Sigma_\ell) ,
\ee
which is the appropriate behavior of the Euler characteristic.
Note that the maximum value of the Euler invariant is $\chi = 2$ when the covering surfaces have ${\bf g} = 0$, followed by $\chi = 0$ for ${\bf g} = 1$,  and for higher genera $\chi < 0$.
So, in Eq.(\ref{QptNDoubNchi}), as in the standard case, the leading-$N$ contribution to the correlator comes from (possibly disconnected) the zero-genus covering surfaces.%

Eq.(\ref{QptNDoubNchi}) is a natural generalization, for disconnected functions, of the well-known scaling of connected $Q$-point functions as $\sim \sum_{\bf g} N^{- {\bf g} + 1 - \frac12 Q}$\cite{Lunin:2000yv}.
Of course, for single-cycle, connected functions, our derivation above can be reduced to that of \cite{Pakman:2009zz}.
See also the more general results of \cite{Dei:2019iym}.

Let us stress that formulas (\ref{QptNDoubN}) and (\ref{QptNDoubNchi}) for the large-$N$ scaling only hold under certain assumptions about the twists of the multi-cycle fields. Essentially, we are assuming that, when $N \to \infty$, the number of cycles in the correlation function does not proliferate --- hence, although the function may be disconnected, it disconnects into a product of a finite number of connected functions/covering surfaces.

\subsubsection{The monodromy of classes}

\begin{figure}
\centering
\includegraphics[scale=0.25]{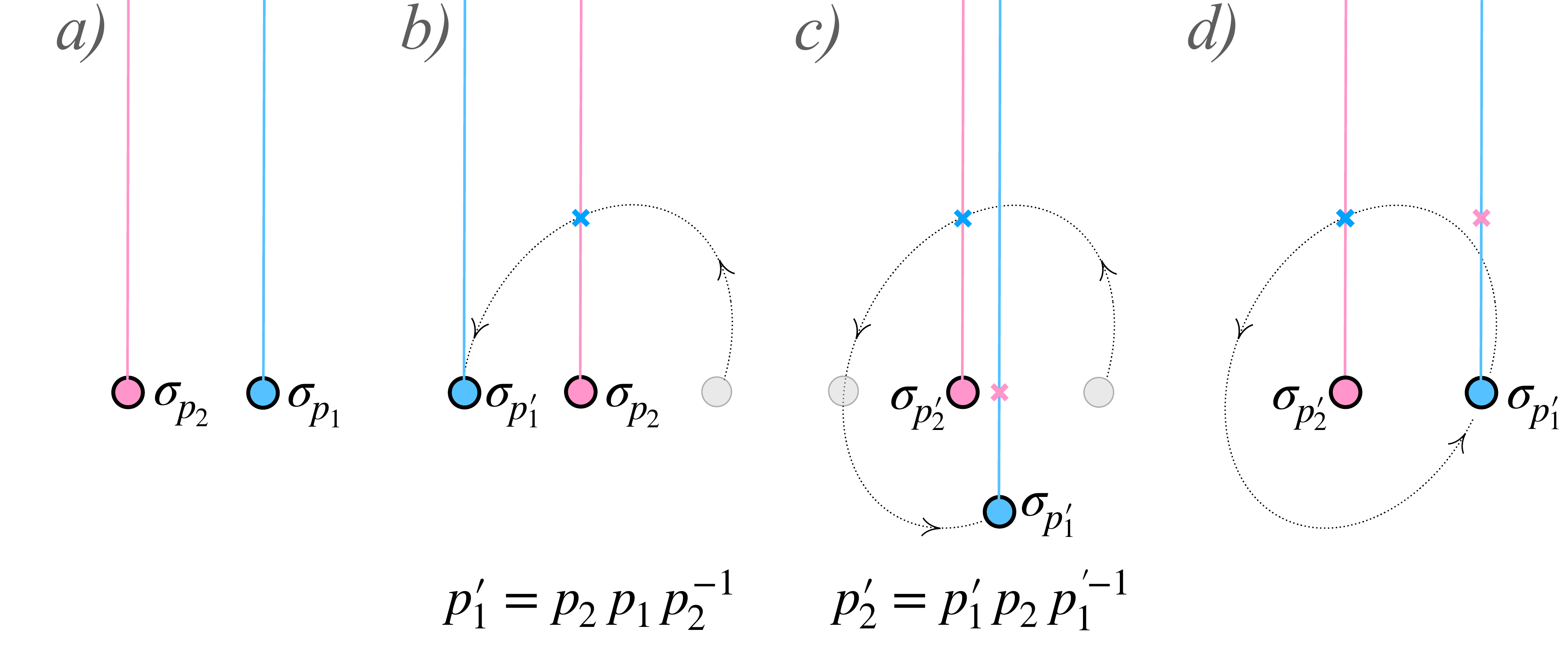}
\caption{%
Moving twist operators on $\bbS^2_\rm{base}$.
Each twist creates a branch cut (dotted lines). 
Whenever a twist crosses a branch cut, its permutation changes.
}
\label{twist_monodromy}
\end{figure}

The twist $\s_{p}(z)$ creates a branch cut  at $z \in \bbS_\rm{base}^2$. When an operator crosses it, the copy indices are permuted by the action of $p$.
If $\s_{p_1}$ crosses the branch cut of $\s_{p_2}$ counterclockwise
(Fig.\ref{twist_monodromy}b),
$p_2$ acts on $p_1$ by left conjugation, and $\s_{p_1}$ becomes $\s_{p_1'}$ where $p_1' = p_2 p_1 p_2^{-1}$.
(If the branch cut is crossed clockwise, $p_2$ acts by right conjugation on  $p_1 \mapsto p_2^{-1} p_1 p_2$.)
Completing the circular movement of the first twist, the second one crosses a branch cut and is also affected (Fig.\ref{twist_monodromy}c). The final configuration (Fig.\ref{twist_monodromy}d) has two different twists than the initial one (Fig.\ref{twist_monodromy}a),
but the product of the permutations is preserved:
\be	\label{p1p2linp1p2}
p_2' p_1' = p_2 p_1 .
\ee
Thus, when we rotate the $\s_{p_i}(z_i)$ around each other inside a $Q$-point function we obtain a function with different twists.
The condition (\ref{compto1}) is, however, preserved, as a result of (\ref{p1p2linp1p2}).

Suppose we start with a twisted $Q$-point function whose permutations belong to the equivalence class $\a$. After rotating the twists as in Fig.\ref{twist_monodromy}, the final permutations belong to a different class $\a' \neq \a$.
The fact that moving twist fields around each other moves between the different equivalence classes $\a \in \Class$ was called 
``channel crossing symmetry'' in \cite{Pakman:2009zz}.
	(Since each class is associated with a diagram, ``channel crossing'' is a symmetry of the set of all diagrams under the monodromies of a connected correlation function \cite{Pakman:2009zz}.)

In summary, correlation functions of individual permutations, such as (\ref{compto1}), do not have well-defined monodromies, because individual twists are altered when they go around another twist.
But the $S_N$-invariant functions (\ref{sumQs1}) do have well-defined monodromies, because they are a sum over all equivalence classes $\a$, hence  explicitly channel-crossing symmetric.

\subsection{Untwisted operators}

When the correlation function contains operators in the untwisted sector, the discussion above must be modified.
In this case, the sum over conjugacy orbits of (\ref{Sninvg}) is not a good definition for an $S_N$-invariant untwisted operator.
Instead, it should be replaced by a fully symmetrized tensor product 
\be	\label{untwscomp}
\big[ \prod_i (X^{\zeta_i}_{[1]})^{p_i} \big] 
	\equiv 
	\frac{1}{{\scr S}}
	 \Symm [ \otimes_i  \big( X^{\zeta_i}_{I_1^{(i)}} \otimes \cdots \otimes  X^{\zeta_i}_{I_{p_i}^{(i)}} \big) ] ,
\quad
{\scr S} = \sqrt{\frac{N!}{(N - \sum_i p_i)! \prod_i (p_i!)}} .
\ee
Here it should be understood that the copies $I$ entering the symmetrized tensor product are all different.
The normalization factor ${\scr S}$, whose structure is different from the one in (\ref{Sninvg}), 
counts the number of equivalent terms in the symmetrized product, and is derived in \S\ref{SectNormofUntw}.
When there is only one untwisted field, we have a simple sum over copies:
$X_{[1]} \equiv N^{-\frac12} \sum_{I=1}^N X_I$.
For the  $N$-fold product of two components $X^{\zeta_1}_I = X_I$ and $X^{\zeta_2}_I = Y_I$, we have
\be \label{scrSnpqq}
\big[ \prod_{i=1}^2 (X^{\zeta_i}_{[1]})^{p_i} \big] 
	=
	\frac1{\scr S}
	\Symm
	 \big[ X^{\otimes p} Y^{\otimes(N-p)}  \big] ,
\quad
{\scr S} 
	= \sqrt{{N \choose p} }.
\ee
Fields with this structure appear, for example, in \cite{Galliani:2017jlg}.

\bigskip

There is a way to extend the definition (\ref{untwscomp}) to products of composite \emph{twisted} fields, which is widely used in the literature concerned with fuzzballs and microstate geometries. A detailed derivation of the normalization factor analogous to the one in (\ref{untwscomp}) can be found e.g.~in \cite{Bombini:2019vnc}. 
 This type of construction of $S_N$-invariant twisted fields is different from the sum over orbits that we use here, and the normalization factor in \cite{Bombini:2019vnc} differs from the one in Eq.(\ref{Sninvg}).%
	\footnote{%
	The sum over the orbits in $S_N$ has many ``repeated'' terms because the centralizer of the permutation is non-empty, as discussed extensively in Appendix \ref{SectCountinFact}; these repeated terms are duly accounted for in the normalization factor. In contrast, following \cite{Bombini:2019vnc} we sum over less terms.
	}
We note that, although it seems perhaps less intuitive than the straightforward symmetrization of copies, the sum over orbits is particularly useful for correlators whose fields are all twisted, as it is amenable to the equivalence-class decomposition of \cite{Pakman:2009zz,Pakman:2009ab,Pakman:2009mi}
discussed in \S\ref{SectQpointmain}.

\section{Four-point functions with two fields of twist two}	\label{SectCompFunct}

From now on, we will be interested in four-point functions of the type
\be	\label{Ahhll}
A(v,\bar v) 
= 
\Big\langle 
	\bar {\scr X}_{[\Nzn]} (\infty) 
	\bar Z_{[2]} (1) Z_{[2]} (v,\bar v)
	{\scr X}_{[\Nzn]} (0) 
\Big\rangle
\ee
where $v$ is a anharmonic ratio.
 $Z_{[2]}$ is an $S_N$-invariant single-cycle field of length 2, and ${\scr X}_{[\Nzn]}$ is a generic multi-cycle field
with  $[\Nzn] = \{ \Nzn \in {\mathbb N} \ |  \sum_{\zeta,n}  n \Nzn = N \}$
\be \label{MultiscrO}
{\scr X}_{[\Nzn]}
	\equiv	\frac1{{\scr S}[\Nzn]}
		\sum_{h \in S_N}
		\Big[  \prod_{\zeta,n} ( X^{\zeta}_{h (n) h^{-1}} )^\Nzn \Big]
	\equiv
		\Big[ \prod_{\zeta,n} ( X^{\zeta}_{[n]} )^\Nzn \Big] 	
\ee
It is usual to interpret single-cycle fields as ``winding stands'', see e.g.~\cite{Bena:2015bea,Lunin:2001jy}.
In this language, 
$Z_{(2)}$ and $X^\zeta_{(n)}$ are excitations of a 2-wound and an $n$-wound strand, respectively. The index $\zeta$ labels possibly different excitations of the multiple strands that make up ${\scr X}_{[\Nzn]}$,
and the bar over $\bar X^\zeta_{(n)}$ indicates a field with opposite charges.
For example, in the D1-D5 CFT, $\zeta$ labels different SU(2) charges.%
	\footnote{%
	We use $X$ to denote arbitrary fields. They are not to be confused with the bosons $X^{\dot AA}_I$ of the seed CFT defined in App.\ref{SectN44CFT}, which do not appear in the main text.}
In the D1-D5 CFT, we will mostly be interested in the case where $Z_{[2]}$ is a NS chiral or the interaction operator, and $X^\zeta_{(n)}$ are Ramond ground states, but, at this point, we focus only on the twist structure, which rules the factorization properties of the full correlator.

\subsection{Factorization}	\label{SectFactoriz}
	
The correlation function (\ref{Ahhll}) factorizes because the cycles of  $Z_{[2]}$ and $\bar Z_{[2]}$ can overlap with at most four of the cycles of ${\scr X}_{[N^{(s)}_n]}$ and $\bar {\scr X}_{[N^{(s)}_n]}$.
To find the exact way the factorization occurs, recall that the r.h.s.~of (\ref{Ahhll}) is a multiple sum over orbits, and, apart from normalization factors, each term has the structure
\be	\label{TermXXZZ}
A_{\a_{\bf c}} =
\Big\langle
\big[ \prod_{\zeta,n} ( \bar X^\zeta_{(n)} )^\Nzn \big] (\infty)
\	\bar Z_{(2)} (1)
\	Z_{(2)} (v,\bar v)
\big[ \prod_{\zeta,n} ( X^\zeta_{(n)} )^\Nzn \big] (0)
\Big\rangle .
\ee
The term (\ref{TermXXZZ}) factorizes in two different ways, depending on how the cycles of the four operators overlap.
The first is a four-point function with only one component of each heavy field
(we omit position dependences for brevity)
\bsub	\label{TermXXZZB}
\be	\label{TermXXZZf1}
\Big\langle
 \bar X^{\zeta_1}_{(n_1)} 
\	\bar Z_{(2)}
\	 Z_{(2)}
\
  X^{\zeta_1}_{(n_1)} 
\Big\rangle ,
\ee
and the other possibility is a four-point function with double-cycle fields
\be	\label{TermXXZZf0}
\Big\langle
\big[ \bar X^{\zeta_1}_{(n_1)}  \bar X^{\zeta_2}_{(n_2)} \big] 
\	\bar Z_{(2)}
\	 Z_{(2)}
\
\big[  X^{\zeta_1}_{(n_1)} X^{\zeta_2}_{(n_2)} \big] 
\Big\rangle .
\ee	\esub
This restricted factorization follows because in both Eqs.(\ref{TermXXZZB}) there is, implicit, a product of factorized two-point functions
$\langle \bar X^{\zeta}_{(n)}(\infty)  X^{\zeta}_{(n)^{-1}}(0)  \rangle = 1,$
and the fields $X^{\zeta}_{(n)}$ and $\bar X^{\zeta}_{(n)}$ whose cycles do not overlap with $ Z_{(2)}$ nor $\bar  Z_{(2)}$ must all match in such a way that none of these two-point function vanishes, see \S\ref{SectQpointAppFact}.%
	\footnote{%
	See also \cite{Lima:2021wrz}. Here we derive the factorization in a slightly different way.
	}

Besides (\ref{TermXXZZB}) there is, sometimes, a third possible type of factorization of (\ref{TermXXZZ}), resulting in a product of three-point functions. This only happens for some special configurations of the cycles in the composite field $[ \prod_{\zeta,n} ( X^\zeta_{(n)} )^\Nzn ]$, and for special configurations of the R-charges, including that of $Z_{[2]}$.
Since this type of factorization is not generically present, we will ignore it in the remaining of this paper, and henceforth it should be always understood that the composite fields are not such that this factorization occurs. 
Still, we note that, in the cases where it does occur, the contribution of the factorized three-point functions can be determined in a similar way as we derive the contributions of the connected four-point functions below.
We give a more detailed discussion of this case in \S\ref{SectQpointAppFact}.

Applying Eq.(\ref{pdsqMa2N}) to the four-point function (\ref{Ahhll}), we get
\begin{align}
A(v,\bar v) 
=
	\frac{| \Cent[g] | | \Cent[(2)] |}{N!}
	\sum_{\bf c} 
	\frac{1}{(N - {\bf c})!}
	\sum_{\a_{\bf c} \in \Class_{\bf c}}
	\nu_{\a_{\bf c}}
	A_{\a_{\bf c}}(v,\bar v) .
\end{align}
Here $g = \prod_n (n)^{N_n}$, with $\sum_n n N_n = N$, is the permutation in the multi-cycle field. 
The number of active copies is constrained by the conjugacy class of $g$.
We now assume, for simplicity, that $N_1 = 0$, i.e.~that all cycles entering the correlation function are non-trivial.
Then all $N$ copies enter the correlation function non-trivially, and%
	\footnote{%
	If $N_1 > 0$, then (in a given class $\a_{\bf c}$) ${\bf c} = N - N_1 + \la$, where $\la$ is the number of copies which are trivial in $g$, but participate in  the cycles of $Z_{(2)}$ and/or of $\bar Z_{(2)}$.
	}
\be	\label{cfcN}
{\bf c} = N		\qquad (\text{if $g = \textstyle\prod_n (n)^{N_n}$, with all $n >1$}).
\ee
Assuming (\ref{cfcN}), and using (\ref{OrdCentg}),
\begin{align}	\label{facAhlltaa}
\begin{split}
A(v,\bar v) 
&=
	\frac{(\prod_{n>1} N_n! n^{N_n}) (N-2)! 2}{N!}
	\sum_{\a \in \Class}
	\nu_{\a}
	A_{\a}(v,\bar v) 
\\
&=
	\frac{2 (N-2)!  \prod N_n! }{N!}
	\Bigg[
	\sum_{\a_0}
	n_1 n_2
	\Big\langle
	\big[ \bar X^{\zeta_1}_{(n_1)}  \bar X^{\zeta_2}_{(n_2)} \big] 
		\bar Z_{(2)}
		 Z_{(2)}
	\big[  X^{\zeta_1}_{(n_1)} X^{\zeta_2}_{(n_2)} \big] 
	\Big\rangle_{\a_0}
\\
&\qquad\qquad\qquad\qquad\qquad
	+
	\sum_{\a_1}
	n
	\Big\langle
	 \bar X^{\zeta}_{(n)} 
	\bar Z_{(2)} 
	Z_{(2)} 
	X^{\zeta}_{(n)}  
	\Big\rangle_{\a_1}
	\Bigg] .
\end{split}
\end{align}
The classes $\a$ can be divided into two subgroups, denoted $\a_0$ and $\a_1$, according to whether they factorize as in (\ref{TermXXZZf1}) or (\ref{TermXXZZf0}), respectively.
In each case, many two-point functions factorize, and inserting the factors $\nu_{\a}$ given by (\ref{formwas}) we find the final result in (\ref{facAhlltaa}).
Among the terms in the classes $\a_0$, are all possible pairings of components $\bar X^{\zeta_i}_{(n_i)}$, and of components $X^{\zeta_i}_{(n_i)}$, from the multi-cycle fields.
The classes with the same pairing reconstruct the connected part of the $S_N$-invariant function with double-cycles and with the number of colors restricted to ${\bf c} = n_1+n_2$, that is
\begin{align}	\label{conng0withcl}
\begin{split}
&
\Big\langle
\big[
	\bar X^{\zeta_1}_{[n_1]} 
	\bar X^{\zeta_2}_{[n_2]}
\big] (\infty) 
\,	\bar Z_{[2]}(1) 
\,	Z_{[2]} (v,\bar v)
\big[
	X^{\zeta_1}_{[n_1]} 
	X^{\zeta_2}_{[n_2]} 
\big](0)
\Big\rangle_0
\\
&\qquad
	\equiv
\Bigg[
	\frac{2 (N-2)! (N-n_1-n_2)! n_1n_2}{N!}
\\
&\qquad\qquad
	\times
	\sum_{\bf c} \frac1{(N-{\bf c})!} \sum_{\a_{\bf c}} 
		\Big\langle
	\big[ \bar X^{\zeta_1}_{(n_1)}  \bar X^{\zeta_2}_{(n_2)} \big] 
	\	\bar Z_{(2)}
	\	 Z_{(2)}
	\
	\big[  X^{\zeta_1}_{(n_1)} X^{\zeta_2}_{(n_2)} \big] 
	\Big\rangle_{\a_{\bf c}}
\Bigg]_{{\bf c} = n_1+n_2}	
\\
&\qquad
	=
	\frac{2 (N-2)! n_1n_2}{N!}
	\sum_{\a_0} 
		\Big\langle
	\big[ \bar X^{\zeta_1}_{(n_1)}  \bar X^{\zeta_2}_{(n_2)} \big] 
	\	\bar Z_{(2)}
	\	 Z_{(2)}
	\
	\big[  X^{(\zeta_1)}_{(n_1)} X^{\zeta_2}_{(n_2)} \big] 
	\Big\rangle_{\a_0} .
\end{split}	
\end{align}
We assumed $n_1 \neq n_2$; for $n_1 = n_2 = n$, the overall coefficient in the r.h.s.~must be multiplied by $2!$.
The index $0$ in the function $\langle \cdots \rangle_0 \sim \sum_{\a_0} \langle \cdots \rangle_{\a_0}$
indicates that its associated covering-surfaces have genus ${\bf g} = 0$, as given by the Riemann-Hurwitz formula (\ref{RiemHurwForm}) with ${\bf c} = n_1 + n_2$. 
Similarly, the classes $\a_1$, which have ${\bf c} = n$, reconstruct the connected function
\begin{align}
\Big\langle
	\bar X^{\zeta}_{[n]} (\infty) 
\,	\bar Z_{[2]}(1) 
\,	Z_{[2]} (v,\bar v)
	X^{\zeta}_{[n]} (0)
\Big\rangle_1
	=
	\frac{2 (N-2)! n}{N!}
	\sum_{\a_1} 
	\Big\langle
	 \bar X^{\zeta}_{(n)}  \bar Z_{(2)} Z_{(2)} X^{\zeta}_{(n)} 
	\Big\rangle_{\a_1} 
\end{align}
whose index indicates that the associated covering surfaces have genus ${\bf g} =1$.

Combining everything, we gather that Eq.(\ref{facAhlltaa}) gives
\begin{align}	\label{Axxzz}
\begin{split}
A(v,\bar v) 
&= 
\Big\langle
\Big[ \prod_{{\zeta},n} ( \bar X^{\zeta}_{[n]} )^\Nzn \Big] (\infty) 
\,	\bar Z_{[2]} (1)
\,	Z_{[2]} (v,\bar v)
\Big[ \prod_{\zeta,n} ( X^{\zeta}_{[n]} )^\Nzn \Big] (0)
\Big\rangle 
\\
&=
	\Big( \prod_{n} N_n! \Big)
	\Bigg[
	\sum_{\substack{n_1 \neq n_2 \\ \zeta_1 \neq \zeta_2}}
	\left( N_{n_1}^{\zeta_1} N^{\zeta_2}_{n_2} \right)^2
	\Big\langle
	\Big[ \bar X^{\zeta_1}_{[n_1]} \bar X^{\zeta_2}_{[n_2]} \Big]  
	\,	\bar Z_{[2]}
	\,	Z_{[2]} 
	\Big[ X^{\zeta_1}_{[n_1]}  X^{\zeta_2}_{[n_2]} \Big]
	\Big\rangle_0
\\
& \phantom{\Big( \prod_{n} N_n! \Big)\Bigg[ }
	+
	\sum_{\substack{n_1 = n_2 = n  \\ \zeta_1 \neq \zeta_2}}
	\left( N_{n}^{\zeta_1} N^{\zeta_2}_{n} \right)^2
	\Big\langle
	\Big[ \bar X^{\zeta_1}_{[n]} \bar X^{\zeta_2}_{[n]} \Big]  
	\,	\bar Z_{[2]}
	\,	Z_{[2]} 
	\Big[ X^{\zeta_1}_{[n]}  X^{\zeta_2}_{[n]} \Big]
	\Big\rangle_0
\\
& \phantom{\Big( \prod_{n} N_n! \Big)\Bigg[ }
	+
	\sum_{n , \zeta}
	{\scr P}^2( \Nzn) 
	\Big\langle
	\Big[ \bar X^{\zeta}_{[n]} \bar X^{\zeta}_{[n]} \Big]  
	\,	\bar Z_{[2]}
	\,	Z_{[2]} 
	\Big[ X^{\zeta}_{[n]}  X^{\zeta}_{[n]} \Big]
	\Big\rangle_0
\\
& \phantom{\Big( \prod_{n} N_n! \Big)\Bigg[ }
	+
	\sum_{n , \zeta}
	(\Nzn)^2
	\Big\langle
	\bar X^{\zeta}_{[n]} 
	\,	\bar Z_{[2]}
	\,	Z_{[2]} 
	 X^{\zeta}_{[n]}
	\Big\rangle_1
\Bigg]	.
\end{split}
\end{align}
The coefficients in each sum are `symmetry factors', given by the number of equivalent ways of forming pairs of components from the original multi-cycle fields; 
see \cite{Lima:2021wrz}.
(They are squared because there are two multi-cycle fields.) The function ${\scr P}(q)$ is the number of ways to pair $q$ objects, 
\be
{\scr P}(2p) = \frac{(2p)!}{p! 2^p} ,
\qquad
{\scr P}(2p+1) = \frac{(2p)!}{p! 2^p} + 2p .
\ee

We have reduced the four-point function with two full multi-cycle fields to a sum of connected four-point functions with, at most, double-cycle fields. 
Note that to arrive at Eq.(\ref{Axxzz}) we have not used the large-$N$ approximation. 
Although the connected functions $\langle \cdots \rangle_0$ and $\langle \cdots \rangle_1$ have genera ${\bf g} =0$ and ${\bf g}=1$, respectively,
we see from Eq.(\ref{QptNDoubNchi}) that, 
for large $N$, \emph{both} scale as $\sim N^{-2}$. This is because $\langle \cdots \rangle_0$ has an extra pair of cycles giving an extra pair of ramification points to the covering surface.
The symmetry factors, which depend on the multiplicities $\Nzn$, also depend on $N$ because they are constrained by 
$\sum_{n,\zeta} \Nzn \, n = N$. Hence, depending on the configuration of this partition, and on how the large-$N$ limit is taken (e.g.~leaving the cycle's lengths fixed or not), the symmetry factors can also become large.
It is to be expected that if some of the $\Nzn$ grow parametrically with $N$, the terms with ${\scr P}(\Nzn)$ dominate the r.h.s.~of (\ref{Axxzz}), as they contain factorials. In this case, the genus-one functions end up being subleading.

As a concrete example, consider the multi-cycle field%
	\footnote{%
	We tacitly assume that $\max(n_1, n_2) \neq 2\min (n_1,n_2)$, so that the type of three-point function factorization discussed below (\ref{TermXXZZB}) does not exist.
	}
\be
{\scr X} = \Big[ (X^{\zeta_1}_{[n_1]})^{2p} (X^{\zeta_2}_{[n_2]})^{2q} \Big] ;
	\quad 2p \, n_1 + 2q \, n_2 = N ;
	\quad
	n_1 > n_2 > 1, 
	\quad
	\zeta_1 \neq \zeta_2.
\ee
In this case, the second sum in (\ref{Axxzz}) is void:
\begin{align}	\label{Apqex1}
\begin{split}
&
\Big\langle
\Big[ (\bar X^{\zeta_1}_{[n_1]})^{2p} (\bar X^{\zeta_2}_{[n_2]})^{2q} \Big] 	\bar Z_{[2]} 
	Z_{[2]} 
\Big[ (X^{\zeta_1}_{[n_1]})^{2p} (X^{\zeta_2}_{[n_2]})^{2q} \Big]
\Big\rangle 
\\
&\qquad
=
	(2p)! \, (2q)! 
	\Bigg[
	( 4pq )^2
	\Big\langle
	\Big[ \bar X^{\zeta_1}_{[n_1]} \bar X^{\zeta_2}_{[n_2]} \Big]  
	\,	\bar Z_{[2]}
	\,	Z_{[2]} 
	\Big[ X^{\zeta_1}_{[n_1]}  X^{\zeta_2}_{[n_2]} \Big]
	\Big\rangle_0
\\
&\qquad  \qquad\qquad
	+
	\Big( \frac{(2p)!}{p! 2^p} \Big)^2
	\Big\langle
	\Big[ \bar X^{\zeta_1}_{[n_1]} \bar X^{\zeta_1}_{[n_1]} \Big]  
	\,	\bar Z_{[2]}
	\,	Z_{[2]} 
	\Big[ X^{\zeta_1}_{[n_1]}  X^{\zeta_1}_{[n_1]} \Big]
	\Big\rangle_0
\\
&\qquad  \qquad\qquad
	+
	\Big( \frac{(2q)!}{q! 2^q} \Big)^2
	\Big\langle
	\Big[ \bar X^{\zeta_2}_{[n_2]} \bar X^{(\zeta_2)}_{[n_2]} \Big]  
	\,	\bar Z_{[2]}
	\,	Z_{[2]} 
	\Big[ X^{\zeta_2}_{[n_2]}  X^{\zeta_2}_{[n_2]} \Big]
	\Big\rangle_0
\\
&\qquad  \qquad\qquad
	+
	(2p)^2
	\Big\langle
	\bar X^{\zeta_1}_{[n_1]} 
	\,	\bar Z_{[2]}
	\,	Z_{[2]} 
	 X^{\zeta_1}_{[n_1]}
	\Big\rangle_1
	+
	(2q)^2
	\Big\langle
	\bar X^{\zeta_2}_{[n_2]} 
	\,	\bar Z_{[2]}
	\,	Z_{[2]} 
	 X^{\zeta_2}_{[n_2]}
	\Big\rangle_1
\Bigg]	.
\end{split}
\end{align}
Since
$n_1 (p / N) + n_2 (q / N) = \frac12$,
if we keep the cycles' lengths fixed in the large-$N$ limit, there must be a large number of both components, i.e.~$p, q \gg 1$.
Using Stirling's formula, we see that the ${\bf g}=1$ terms in the last line are subleading to all terms with double-cycle fields.

An even simpler example is a field with only one type of component,
\be	\label{XtoNn}
{\scr X} = \Big[ (X^\zeta_{[n]})^{2p} \Big] , \quad 2p = N / n.
\ee
The four-point function simplifies further,
\begin{align}	\label{Apqex2}
\begin{split}
\Big\langle
\Big[ (\bar X^\zeta_{[n]})^{\frac{N}{n}} \Big] 	
\bar Z_{[2]} 	Z_{[2]} 
\Big[ ( X^\zeta_{[n]})^{\frac{N}{n}} \Big]
\Big\rangle 
=
	\Big(\frac{N}{n}\Big)! 
	\Bigg[
&	\Bigg( \frac{(\frac{N}{n})!}{(\frac{N}{2n})! 2^{\frac{N}{2n}}} \Bigg)^2
	\Big\langle
	\Big[ \bar X^\zeta_{[n]} \bar X^\zeta_{[n]} \Big]  
	\,	\bar Z_{[2]}
	\,	Z_{[2]} 
	\Big[ X^\zeta_{[n]}  X^\zeta_{[n]} \Big]
	\Big\rangle_0
\\
&
	+
\left( \frac{N}{n} \right)^2
	\Big\langle
	\bar X^\zeta_{[n]} 
	\,	\bar Z_{[2]}
	\,	Z_{[2]} 
	 X^\zeta_{[n]}
	\Big\rangle_1
\Bigg]	.
\end{split}
\end{align}
For $N/n \gg 1$, we find again that the genus-one term is strongly suppressed.

\subsection{Genus-zero covering surfaces and Hurwitz blocks} 	\label{SectHurwBlocks}

We have seen that the main ingredient of the four-point functions (\ref{Axxzz}) are the connected functions
\be	\label{An1n2zz}
A_{n_1,n_2}^{\zeta_1,\zeta_2}(v,\bar v) 
\equiv
\Big\langle
\big[
	\bar X^{\zeta_1}_{[n_1]} 
	\bar X^{\zeta_2}_{[n_2]}
\big] (\infty) 
\,	\bar Z_{[2]}(1) 
\,	Z_{[2]} (v,\bar v)
\big[
	X^{\zeta_1}_{[n_1]} 
	X^{\zeta_2}_{[n_2]} 
\big](0)
\Big\rangle_0 .
\ee
From now on we omit the label $0$, and always assume that we are dealing with the connected function with a genus-zero covering surface, which is obtained with the covering map 
\cite{Lima:2020nnx}
\be		\label{coverm}
z(t) = \left( \frac{t}{t_1}\right)^{n_1} 
	\left( \frac{t-t_0}{t - t_\infty} \right)^{n_2} 
	\left( \frac{t_1-t_\infty }{t_1-t_0} \right)^{n_2}.
\ee
The ethos of a covering map \cite{Lunin:2001pw} is to cover the ``base Riemann sphere'' $\bbS^2_\rm{base} \ni z$, where (\ref{An1n2zz}) is evaluated, with a ramified surface $\Sigma_{\bf g} \ni t$ of genus $\bf g$, whose ramification points have the property of trivializing the twists in (\ref{An1n2zz}). 
The map (\ref{coverm}) defines such a covering surface with ${\bf g} = 0$, i.e.~a covering of the sphere by the sphere.
The pair of (disjoint) twist insertions at $z = 0$ lift to the pair of ramification points $t = 0$ and $t = t_0$ with ramifications $n_1$ and $n_2$. The same happens to the pair of twists at $z = \infty$. The single-cycle twists at $z =1$ and $z = v$ must, each, be lifted to one ramification point, which we call $t = t_1$ and $t = x$, respectively. 
At these points, the map must have the correct monodromy,
 i.e.~the derivative must be factorizable as $z'(t) \sim (t-t_1)(t-x)$. 
This imposes relations among the parameters $t_0,t_1,t_\infty$ and $x$, that can be satisfied by choosing
\begin{align}		\label{tim}
t_0 = x-1 ,
\qquad
t_1 = \frac{(x-1) (x - 1 + \frac{n_1}{n_2})}{x + \frac{n_1}{n_2}} ,
\qquad
t_\infty = \frac{x ( x - 1 + \frac{n_1}{n_2})}{x + \frac{n_1}{n_2}} .
\end{align}
The asymmetry between $n_1$ and $n_2$ in Eqs.(\ref{coverm})-(\ref{tim}) is ``fictitious'': it stems from a freedom in parametrizing the covering map \cite{Lima:2020nnx}. Without loss of generality, we will consider $n_1 \geq n_2$.

The covering surfaces with the same ramification data, i.e.~the same number of ramification points with fixed orders and positions on $\bbS^2_\rm{base}$, are not unique. The number ${\bf H}$ of such surfaces it is a Hurwitz number \cite{Pakman:2009ab,Pakman:2009zz}.
It is the number of inverses of 
\be	\label{zxeqv}
 v = z(x) = \Bigg( \frac{x+ \frac{n_1}{n_2}}{x-1} \Bigg)^{n_1+n_2} 
	\Bigg( \frac{x}{x  + \frac{n_1 - n_2}{n_2} } \Bigg)^{n_1-n_2}  
\ee
found by inserting (\ref{tim}) into (\ref{coverm}). This is equivalent to the algebraic equation
\be
v \, ( x-1 )^{n_1 +n_2} \, ( x \tfrac{n_1 - n_2}{n_2} )^{n_1-n_2}
-
( x + \tfrac{n_1}{n_2} )^{n_1+n_2} \, x^{n_1 - n_2} = 0 ,
	\label{uxm}
\ee
with 
\be
{\bf H} = 2\max(n_1,n_2)
\ee
 roots $x_{\frak a}(v)$, ${\frak a} = 1, \cdots, {\bf H}$.
Also, as explained in \cite{Pakman:2009zz} (see also \cite{Pakman:2009ab}), 
there is also a correspondence between covering surfaces and the equivalence classes $\a$ that compose the $S_N$-invariant function (\ref{An1n2zz}), cf.~(\ref{conng0withcl}),
\be \label{A12sol}
\begin{split}
A^{\zeta_1, \zeta_2}_{n_1,n_2} (v,\bar v)
	&= \frac{2 n_1n_2 (N-2)!}{N!}
	\sum_{ \substack{ \a \in \Class \\ {\bf g} = 0} }
	\Big\langle
	\big[ \bar X^{\zeta_1}_{(n_1)_\a}  \bar X^{\zeta_2}_{(n_2)_\a} \big] 
	\bar Z_{(2)_\a}
	 Z_{(2)_\a}
	\big[  X^{\zeta_1}_{(n_1)_\a} X^{\zeta_2}_{(n_2)_\a} \big] 
	\Big\rangle
\\	
	&=
	\frac{2 n_1n_2 (N-2)!}{N!}
	\sum_{\frak a = \frak1}^{{\bf H}} \big| A^{\zeta_1,\zeta_2}_{n_1,n_2}( x_{\frak a}(v)) \big|^2 
\end{split}
\ee
In other words, in the first line there are ${\bf H}$ different equivalence classes $\a$ with ${\bf g} = 0$, 
each class associated to one of the solutions $x_{\frak a}(u)$ in the second line.
So the sum in (\ref{A12sol}) is over the ${\bf H} = 2\max (n_1,n_2)$ topologically distinct covering surfaces with $R = 6$ ramification points of orders given by the twists, and ${\bf c} = n_1 + n_2$ sheets, as per the Riemann-Hurwitz equation (\ref{RiemHurwForm}).

As $v$ sweeps  $\bbS^2_{\rm{base}}$,
each function $x_{\frak a}(v)$ fills one out of ${\bf H}$ disjoint regions, which compose again the entire Riemann sphere.
We will call this regions \emph{`Hurwitz regions'}.
 Eq.(\ref{A12sol}) shows that the $S_N$-invariant function 
$A^{\zeta_1, \zeta_2}_{n_1,n_2} (v,\bar v)$,
with domain on $\bbS^2_{\rm{base}}$, 
is decomposable as a sum of ${\bf H}$ 
\emph{`Hurwitz blocks'},
derived from the function 
$A^{\zeta_1, \zeta_2}_{n_1,n_2}(x)$,
with domain in the $x$-plane.
It is crucial that all Hurwitz blocks are summed for the total function to be $S_N$-invariant, because each block correspond to one of the equivalence classes $\a$. As discussed by the end of \S\ref{SectQpointmain}, when one twisted operator revolves around another (c.f.~Fig.\ref{twist_monodromy})
the equivalence classes are shuffled.
Hence  a missing Hurwitz block makes the monodromies of the correlation function not well defined.

It is often possible to compute the ``Hurwitz block function'' $A^{\zeta_1, \zeta_2}_{n_1,n_2}(x)$ in closed form. We will do this in Sect.\ref{SectFuncandMast} for a varied collection of operators.
But even when this is the case, it is, in general, not possible to write the $S_N$-invariant function itself in closed form, 
because
the $x_{\frak a}(v)$ are the roots of Eq.(\ref{uxm}), which has order higher than 5 for almost all twists.

\subsubsection{Exact Hurwitz bloks for composite fields with equal cycles}

\begin{figure}
  \begin{minipage}[c]{0.55\textwidth}
    \includegraphics[width=\textwidth]{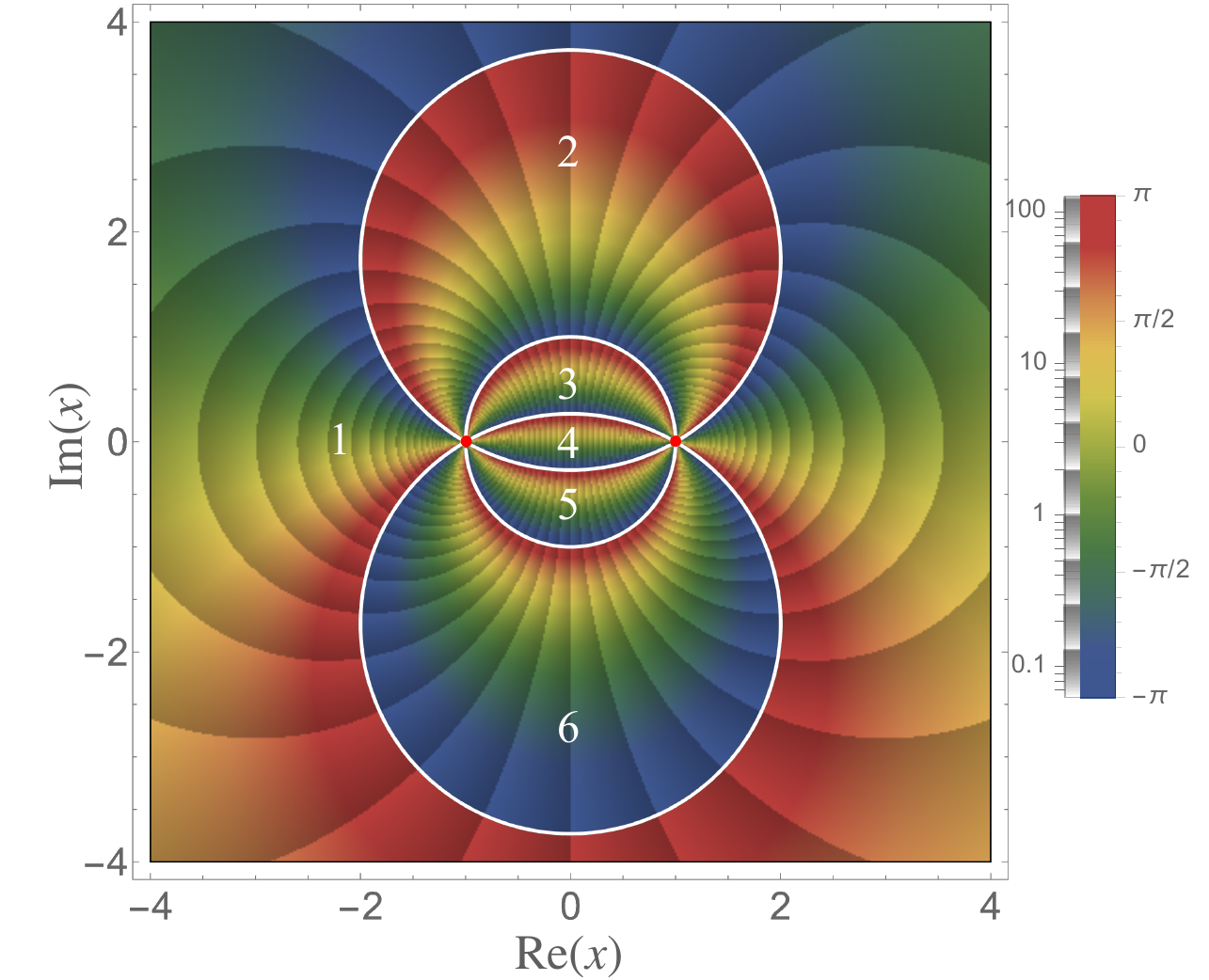}
  \end{minipage}\hfill
  \begin{minipage}[c]{0.4\textwidth}
    \caption{
	Plot of $v(x)$ for $n_1 = n_2 = 3$. 
	Colors depicts the phase of $v(x)$, 
	and the shading follows the curves of constant $|v(x)|$, as indicated by the scales.
	The $x$-plane is divided into ${\bf H} = 6$ regions, $v(x)$ taking all values in ${\mathbb C}$ inside of each.
    } \label{Fig_u_of_x_n1_eq_n2}
  \end{minipage}
\end{figure}

The exception is when $n_1 = n_2$. The polynomial (\ref{uxm}) simplifies to
\be	\label{zxeqvnn}
 v = z(x) = \left( \frac{x+ 1}{x-1} \right)^{2n} 
\ee
and we can find its ${\bf H} = 2n$ solutions exactly:
\be	\label{xfraka2n}
x_{\frak a}(v) = - \frac{1 + v^{\frac1{2n}} e^{\frac{{\frak a} \pi i}{n} } }{1 - v^{\frac1{2n}} e^{\frac{{\frak a} \pi i}{n} } } ,
\qquad
 {\frak a} = 0, 1, 2, \cdots, 2n -1 ,
\ee
where 
$v^{\frac1{2n}}$ is a (single) $2n$th root of $v$.
The division of the $x$-plane into ${\bf H} = 2n$ disjoint Hurwitz regions can be clearly seen in the plot of $v(x)$, shown in Fig.\ref{Fig_u_of_x_n1_eq_n2} for $n = 3$. 
The shading of the plot follows the contours where $|v(x)| = \rm{constant}$, distinguishing the curves traced on the $x$-plane when $v$ goes in circles around the origin of the base sphere.
All regions meet at the two critical points $x = \pm 1$.

%
%

\begin{figure}
\centering
\includegraphics[scale=0.35]{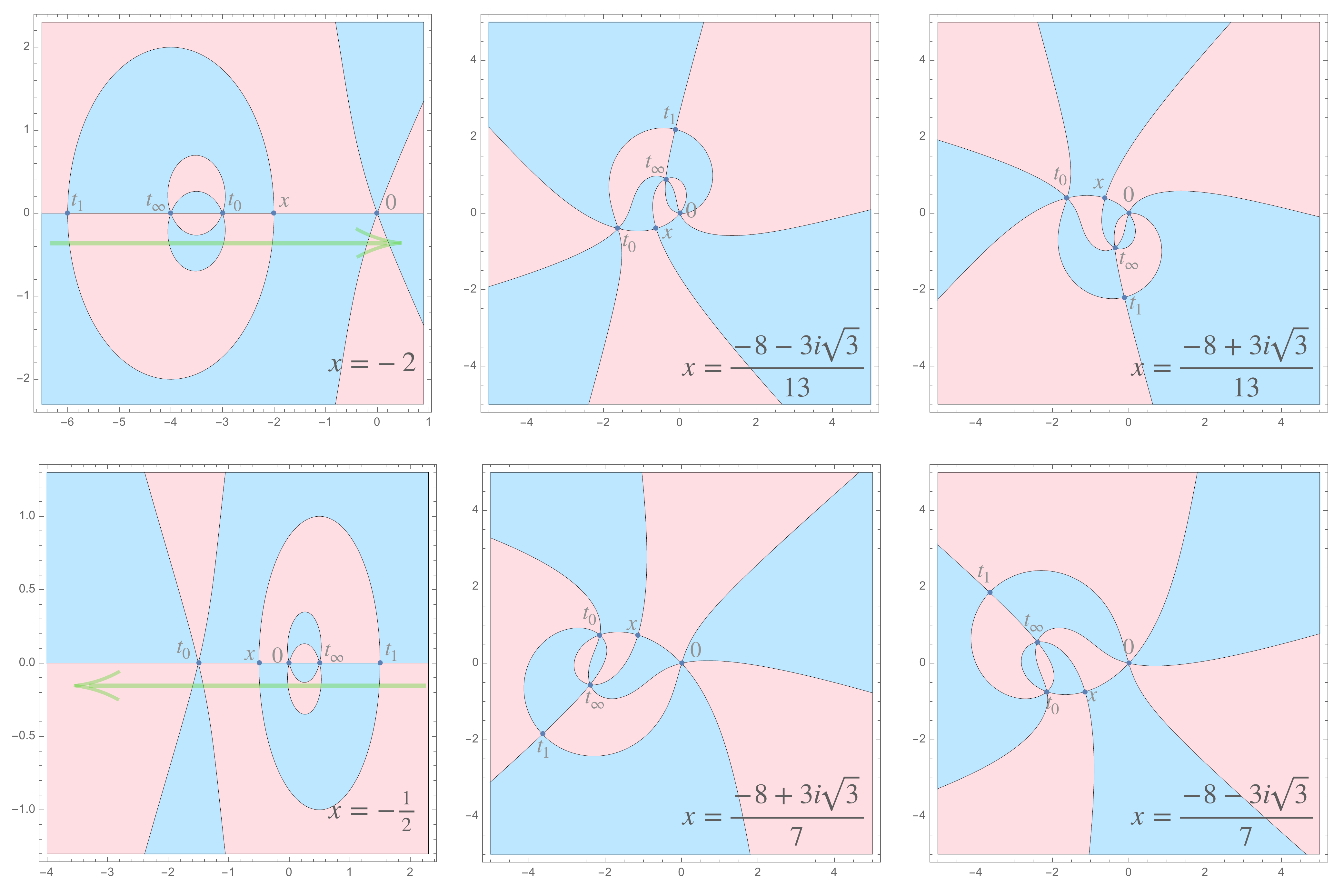}
\caption{%
Covering surfaces for $n_1=n_2=3$. 
In each panel, the horizontal axis is $\rm{Re}(t)$ and the vertical axis $\rm{Im}(t)$. 
Blue patches are the preimages of the upper-half plane $\rm{Im}(z) > 0$, and pink patches the preimages of the lower-half plane $\rm{Im}(z) < 0$, under the covering map (\ref{coverm}).
The positions of the ramification points $t_0,t_1,t_\infty$ depend on the position of the ramification point $x$ according to (\ref{tim}). The 6 panels correspond to the 6 solutions $x_{\frak a}$ of Eq.(\ref{uxm}) for $z(x) = v = \frac1{729}$.
}
\label{Fig_Covers_n1_eq_n2}
\end{figure}

As stated above, each of the ${\bf H}$ regions of the $x$-plane are associated, on the one hand, to a topologically distinct ramified covering of $\bbS^2_\rm{base}$ and, on the other hand, to a distinct class $\a$ of permutations satisfying (\ref{pcompot1}).
But in the case of $n_1 = n_2$, there is a subtlety.
The functions (\ref{xfraka2n}) can be grouped in $n$ pairs related by inversion:
\be	\label{xfrakaandiv}
x_{\frak a + n} = \frac{1}{x_{\frak a}} ,
\ee
for ${\frak a} = 0, \cdots n-1$.
This is a global conformal transformation of the $x$-plane, which suggests that the two solutions $x_{\frak a}$ and $x_{\frak a+n}$ describe covering surfaces with the same topology. 
This is indeed the case, as illustrated in Fig.\ref{Fig_Covers_n1_eq_n2}, again with the example of $n=3$.
There are $2n = 6$ solutions of $v = z(x)$ for fixed $v$. 
The values of $x_{\frak a}$ for the panels in the bottom and upper rows are related by inversion (\ref{xfrakaandiv}).
Surfaces in the same row are all topologically distinct.
But comparing the pairs of surfaces in the same columns, we find that they are equivalent. Let us focus on the first column, with the pair $x_{\frak 0} = -2$ and $x_{\frak 3} = 1 / x_{\frak 0} = -1/2$. 
Rotating the upper panel 180$^\circ$, we do get the same topology of the bottom panel, but with the positions of the points $t = t_0$ and $t = 0$ swapped in relation to the other ramification points. This is highlighted by the green arrows in Fig.\ref{Fig_Covers_n1_eq_n2}; following the arrow in the upper panel we find the sequence 
$t_1 \to t_\infty \to t_0 \to x \to 0$. 
Rotating the plane we get an arrow in the opposite direction, to be contrasted with that indicated in the bottom panel: 
$t_1 \to t_\infty \to 0 \to x \to t_0$.
The relative positions of every point are the same, except for $t = 0$ and $t = t_0$, which are swapped.

As ramification points, $t = 0$ and $t = t_0$ are equivalent: they are both the preimages of $z = 0$, and with equal ramification because they correspond to cycles of equal length. In this sense, swapping these two points does not matter, and the number of distinct ramified coverings is reduced to 
${\bf H} = n$.
This ``reduction by half of the Hurwitz number'' when $n_1 = n_2 = n$ can also be seen from the perspective of $\bf H$ as counting the number of different equivalence classes $\a$.
An explicit construction of the $2 \max(n_1,n_2)$ different classes in the function (\ref{A12sol}) can be found in Appendix B of Ref.\cite{Lima:2020nnx}. It is clear from the construction given there that when $n_1 = n_2$ the otherwise distinct $2\max(n_1,n_2)$ inequivalent classes are grouped in pairs, and only $n$ distinct classes remain. Eq.(\ref{xfrakaandiv}) is the manifestation of this pairing in terms of the geometry of the covering surfaces.

But there is still one further subtlety. 
In (\ref{A12sol}) we may have different excitations of the strands $n_1$ and $n_2$, even if the strands have the same length.
Then the ramification points $t = 0$ and $t = t_0$ are ``decorated'' with different operators, and are distinct, even though they are equivalent with regard to the twist structure.

In summary, for functions with double-cycles of the same length, the $S_N$-invariant four-point function (\ref{A12sol}) is
\be \label{A12soln}
A^{\zeta_1, \zeta_2}_{n_1,n_2} (v,\bar v)
	= \frac{4 n^2 (N-2)!}{N!}
	\sum_{\frak a = \frak0}^{{\frak2n-\frak1} } \big| A^{\zeta_1,\zeta_2}_{n,n}( x_{\frak a}(v)) \big|^2 
\ee
where $x_{\frak a}(v)$ are given in \emph{closed form} by Eq.(\ref{xfraka2n}).
Furthermore, when, in addition to the twists having cycles of the same length, the excitations are also equal, i.e.~$\zeta_1 = \zeta_2$, then the Hurwitz blocks have the symmetry
\be	\label{nnSymmSFD}
A^{\zeta, \zeta}_{n,n}( x_{\frak a}(v)) = A^{\zeta,\zeta}_{n,n}\left( \tfrac{1}{x_{\frak a}(v)} \right)
\ee
and only half the terms in the sum (\ref{A12soln}) are independent.

\subsubsection{Composite fields with unequal cycles}

\begin{figure}
\centering
\includegraphics[scale=0.6]{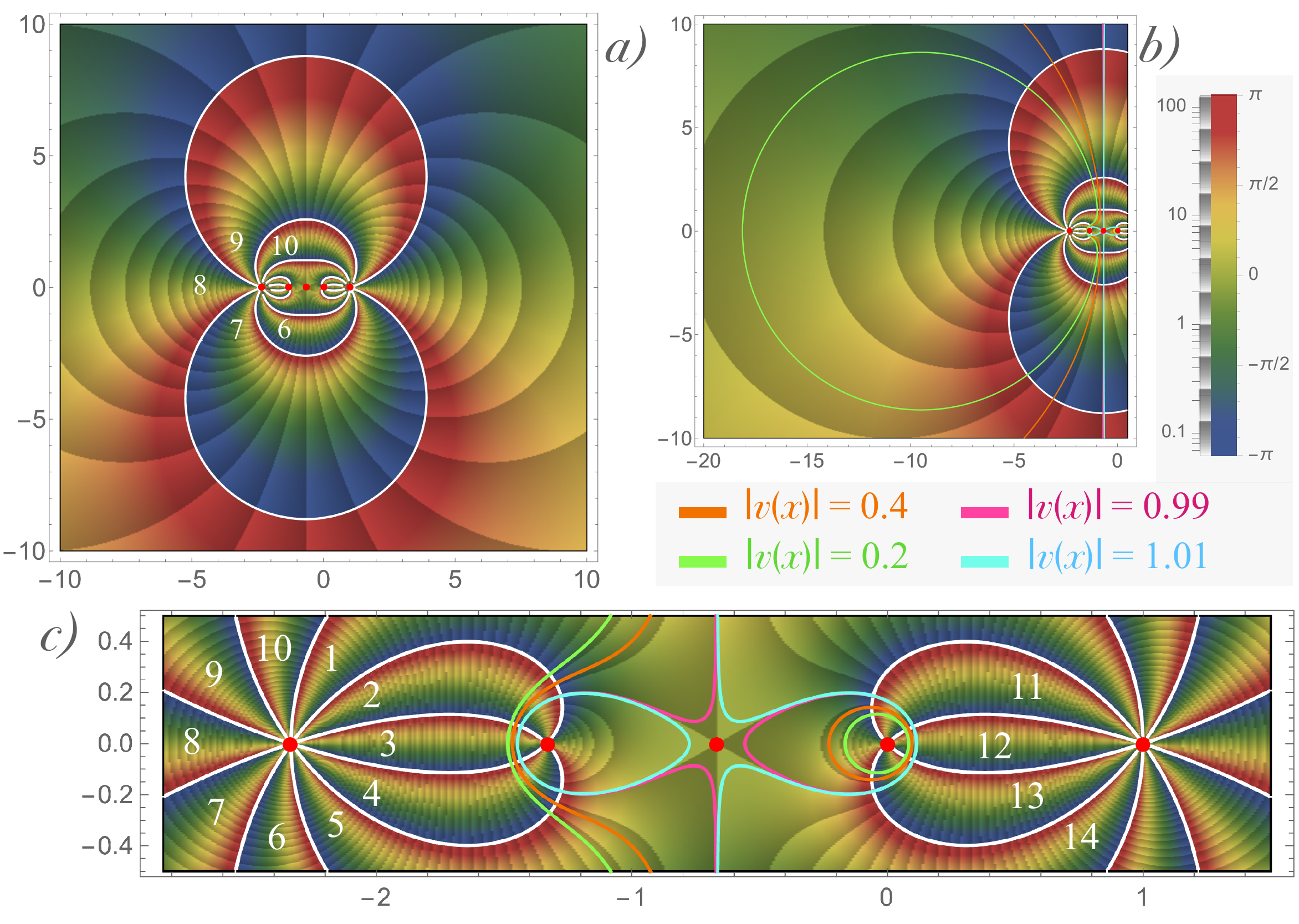}
\caption{%
	Plot of $v(x)$
	for $n_1 =7$, $n_2 = 3$. 
	The frames' horizontal and vertical axes are $\rm{Re}(x)$, and $\rm{Im}(x)$, respectively.
	Colors depict the phase of $v(x)$, 
	and shading follows the curves of constant $|v(x)|$, as indicated by the scales.
	\emph{a)} Complete borders of the Hurwitz regions, with ``outer'' regions labeled. 
	\emph{b-c)} Four contours with $|v(x)| = \rm{constant}$. 
	\emph{c)} Close up of the inner region near the critical points $x = - \frac{n_1}{n_2}$, $x = - \frac{n_1 - n_2}{n_2}$, $x = - \frac{n_1 - n_2}{2n_2}$, $x = 0$, $x = 1$ (red dots from left to right, respectively),
	with all 14 different regions labeled.
}
\label{Fig_u_of_x_n1_neq_n2}
\end{figure}

The geometry of the $x$-plane is more complicated when $n_1 \neq n_2$. In Fig.\ref{Fig_u_of_x_n1_neq_n2} we show it for $n_1 = 7$, $n_3 = 3$.
There are ${\bf H} = 2 \max(n_1,n_2) = 14$ regions, $v(x)$ taking all values in ${\mathbb C}$ inside of each.
(The number of regions can be found by counting, say, the different red streaks in the plot, where $\rm{Arg}(v) \lesssim \pi$.)
The main difference from Fig.\ref{Fig_u_of_x_n1_eq_n2} is that in Fig.\ref{Fig_u_of_x_n1_neq_n2} there is an inner region with three critical points,
\be	\label{critxpon}
 x = - \tfrac{n_1 - n_2}{n_2}, \quad x = - \tfrac{n_1 - n_2}{2n_2}, \quad x = 0 ,
 \ee
which collapse into a trivial one when $n_1 = n_2$.
The trefoil structure of the innermost regions
(labeled 1, 5 and 14 in Fig.\ref{Fig_u_of_x_n1_eq_n2}c)
around the middle-point 
$x = - \frac{n_1-n_2}{2n_2}$
is the same for any values of $n_1 \neq n_2$.
Increasing the difference $n_1 - n_2$ increases just the number of  ``petals'' (labeled 2, 3 and 4) between $x = - \frac{n_1}{n_2}$ and $x = - \frac{n_1-n_2}{n_2}$, 
as well as the symmetric ones (labeled 11, 12 and 13) between $x = 0$ and $x = 1$.
There are always $n_1 - n_2$ such petals at each side.
The petals and the trefoil are associated with the twist structure of the four-point function's OPE channels.

\subsection{OPEs and Hurwitz blocks}

The four-point function (\ref{An1n2zz}), there are two inequivalent OPE limits:
\bsub\label{OPchanss}
\begin{align}
v &\to 1 
\qquad
&\text{with OPE}
\qquad
&\bar Z_{[2]} \times Z_{[2]}
	\label{OPEchanv1}
\\
v &\to 0
\qquad
&\text{with OPE}
\qquad
&Z_{[2]} \times \big[ X^{\zeta_1}_{[n_1]}  X^{\zeta_2}_{[n_2]} \big]
	\label{OPEchanv0}
\end{align}
\esub
In $v \to \infty$, the OPE is equivalent to (\ref{OPEchanv0}) in what concerns the twists, but with the double-cycle fields having opposite charges. (That is, the operators appearing in the fusion rules are different, but with the same twists discussed below.)

The critical points of $v(x)$ correspond to OPE limits.
Although it is not possible to find $x_{\frak a}(v)$ in closed form for $n_1 \neq n_2$, we can find them asymptotically for $v \approx 0,1$.
At $v =1$, there are two critical points 
\be	\label{xp4nsCompv1}
x_\aleph(1) = \infty	\quad  \text{and} 	\quad x_\gimel(1) =  -\frac{n_1-n_2 }{2n_2} .
\ee
The root $x_\aleph$ has multiplicity one, and $x_\gimel$ has multiplicity three.
Explicitly, expanding $v(x)$ in the vicinity of (\ref{xp4nsCompv1}) and inverting the series,
\bsub	\label{xaxp4nsComp}
\begin{align}
\begin{split}		 \label{channelid}
x_\aleph (v) 
	&= 
	- \frac{4n_1}{1-v} + \frac{n_2 - n_1 + 4n_1n_2}{2n_2}
	+ \Bigg[ \frac{n_1}{3} - \frac{1}{24 \, n_1} - \frac{n_1}{24\, n_2^2} \Bigg] (1-v) 
\\
	&\qquad\qquad\qquad\qquad\quad
	+ \Bigg[ \frac{n_1}{6} - \frac{1}{48 \, n_1} - \frac{n_1}{48 \, n_2^2} \Bigg] (1-v)^2
	+ \rm{O}(1-v)^3
\end{split}
\\
\begin{split}	\label{channels3}
x_\gimel(v) 
	&= 
	 - \frac{n_1-n_2 }{2n_2}
		+ 
		\frac{3^{\frac13}}{4} 
		\Big( \frac{(n_1^2 - n_2^2)^2}{n_1 n_2^4} \Big)^{\frac13}
		(1 - v)^{\frac{1}{3}} 
\\
	&	- \frac3{40} \frac{n_1^2 + n_2^2}{n_1 n_2^2} (1-v)
		+ 
		\frac1{8 \cdot 3^{\frac23}} 	\Big( \frac{(n_1^2 - n_2^2)^2}{n_1 n_2^4} \Big)^{\frac13}
		(1-v)^{\frac43}
		+
		\rm{O}(1-v)^{\frac53}
\end{split}
\end{align}
\esub
These functions are plotted as magenta and cyan contours with $|v(x)| \approx 1$ in Fig.\ref{Fig_u_of_x_n1_neq_n2}b-c.
The contours extend to $x = \infty$ in a single direction, but move towards $x_\gimel$ from three different directions in the inner trefoil region.

When $v = 0$, all ${\bf H} = 2n_1$ roots are easily found:%
	\footnote{As stated before, we are assuming, without loss of generality, that $n_1 > n_2$; see \cite{Lima:2020nnx}.}
\be	\label{xp4nsCompv0}
x_\beth(0) = - \frac{n_1}{n_2}	
\quad  \text{and} \quad 
x_\daleth (0) =  0 .
\ee
with multiplicities $n_1+n_2$ and $n_1 - n_2$, respectively. 
Expanding $v(x)$ near these points and inverting the series expansion, we find
\bsub	\label{vto0xs}
\begin{align}
\begin{split}	\label{xcaneeplsnn}
x_\beth(v) 
	&=
	-\frac{n_1}{n_2}
	-
	\left(1+\frac{n_1}{n_2} \right) 
	\left( \frac{n_1}{n_2} \right)^{\frac{n_2-n_1}{n_1+n_2}} 
	v^{\frac{1}{n_1+n_2}} 
\\
	&\qquad
	\times
	\Bigg[
	1 
	+
	\left( \frac{n_1}{n_2} \right)^{- \frac{n_1-n_2}{n_1+n_2}} 
	\left( \frac{n_1}{n_2} + \frac{n_2}{n_1} -1 \right)
	v^{\frac{1}{n_1+n_2}} 
	+
	\rm{O}( v^{\frac{2}{n_1+n_2}} )
	\Bigg]
\end{split}
\\
\begin{split}	\label{xcaneeminnn}
x_\daleth(v) 
	&=
	\left( 1 - \frac{n_1}{n_2} \right) 
	\left( \frac{n_2}{n_1} \right)^{\frac{n_1+n_2}{n_1-n_2}} 
	v^{\frac{1}{n_1-n_2}} 
\\
	&\qquad
	\times
	\Bigg[
	1 
	+ 
	\left( \frac{n_2}{n_1} \right)^{\frac{n_1+n_2}{n_1-n_2}} 
	\left( \frac{1}{n_2} + \frac1{n_2} + \frac{n_1}{n_2^2} \right)
	v^{\frac{1}{n_1-n_2}} 
	+
	\rm{O}( v^{\frac{2}{n_1-n_2}} )
	\Bigg]
\end{split}
\end{align}\esub
The two functions can be visualized in Fig.\ref{Fig_u_of_x_n1_neq_n2}b-c as the orange and green contours with $|v(x)| \approx 0$.
The contours split in two closed parts. One part,
given by $x_\daleth(v)$,
 circles around $x = - \frac{n_1}{n_2}$, as shown in Fig.\ref{Fig_u_of_x_n1_neq_n2}b. They avoid $n_1-n_2$ regions in the petals-trefoil patch in Fig.\ref{Fig_u_of_x_n1_neq_n2}c,
thus crossing a total of ${\bf H} - (n_1-n_2) = n_1 + n_2$ regions.
The other closed contour,
given by $x_\gimel(v)$, encircles $x = 0$, passing over the $n_1-n_2$ remaining regions.
As $v \to 0$, the contour $x_\beth(v)$ tightens around $x = - \frac{n_1}{n_2}$, and $x_\daleth(v)$ tightens (much faster) around $x = 0$.

\subsubsection{Fusion rules}

The twists of operators resulting from the OPEs (\ref{OPchanss}) must appear as branches of the correlation function
$A_{n_1,n_2}^{\zeta_1,\zeta_2}(v,\bar v)$.
The branches correspond to the multiplicity of roots $x_{\frak a}$ in the coincidence limits. 
The multiplicities can be read both from the leading powers in the expansions (\ref{xaxp4nsComp}) and (\ref{xp4nsCompv0}), and also from the number of regions around the critical points discussed above in Figs.\ref{Fig_u_of_x_n1_eq_n2} and \ref{Fig_u_of_x_n1_neq_n2}.

For $v \to 1$, since $x_\aleph(v)$ has no branch cuts, the OPE has an \emph{untwisted} field $U_{[1]}$, while the third-order branch of $x_\gimel(v)$ indicates an operator $S_{[3]}$ of twist $3$. 
This agrees with the composition of permutations:
a product of two transpositions is either the identity or a cycle of length three,
\be
[2] \times [2] = \id + [3] .
\ee 
Hence the OPE (\ref{OPEchanv1}) reads
\be	\label{fusZZ}
\bar Z_{[2]} \times  Z_{[2]} =  C_{Z \bar Z U} \ \{ U_{[1]} \} + C_{Z \bar Z S} \ \{ S_{[3]} \}
\ee
where $C$s are structure constants and $\{ \cdots \}$ indicates conformal families.

Similarly, in the OPE (\ref{OPEchanv0}), two types of resulting permutations contribute to the four-point function,
\be	\label{2nnnn1}
[2] \times [(n_1)(n_2)] 
	= [ n_1 + n_2] + [(n_1 - n_2)(n_2)  (n_2)] .
\ee
In the first term in the r.h.s., a transposition $(2)$ joins two cycles $(n_1)(n_2)$ into a single cycle of length $n_1+n_2$; this is what we find from the branch cut of $x_\beth(v)$ in (\ref{xcaneeplsnn}). 
In the other type of contribution, the transposition \emph{splits} the longer cycle in two: $(2) \times (n_1) = (n_1 - n_2) (n_2)$.
The resulting cycle of length $n_1 - n_2$ is seen in the branch cut of $x_\daleth(v)$ in (\ref{xcaneeminnn}).
The total fusion rule extracted from the four-point function is
\be	\label{OPEW}
\begin{split}
Z_{[2]} \times \big[ X^{\zeta_1}_{[n_1]}  X^{\zeta_2}_{[n_2]} \big]
	&= 
	C_{ [\bar X\bar X] \bar Z Y} \ \big\{  Y_{[n_1+n_2]} \big\}
\\
&\quad
	+
	C_{ \bar X  \bar Z [W X]} 
	B_{\bar X X}
	\ \Big\{  \big[ W_{[n_1-n_2]} X^{\zeta_2}_{[n_2]} X^{\zeta_2}_{[n_2]} \big] \Big\}	
\end{split}\ee
where $B_{\bar X X}$ is the normalization constant of a two-point function.

The appearance of the operator $W_{[n_1-n_2]}$ is an example of how there are non-trivial interactions in the fusion rules of composite, multi-cycle twisted fields, and it deserves a more detailed discussion.
The terms in the r.h.s.~of Eq.(\ref{OPEW}) come from different types of equivalence classes $\a$ in the sum (\ref{A12sol}).
Consider the following examples of representatives of classes contributing to each term, for $n_1 = 5$, $n_2 = 3$ (we label cycles by the corresponding twist position):
\bsub
\begin{align}
\id 
&= 
(1,2,3,4,5)_\infty (6,7,8)_\infty \times (1,6)_1 \times (1,4)_v \times (8,7,6,5,4)_0 (3,2,1)_0	\label{classJoin}
\\
\id 
&= 
(1,2,3,4,5)_\infty (6,7,8)_\infty \times (1,4)_1 \times (4,6)_v \times (8,7,6,5,4)_0 (3,2,1)_0	\label{classSplit}
\end{align}
\esub
Taking the limit $v \to 0$ in each case,
\bsub
\begin{align}
(1,4)_v \times (8,7,6,5,4)_0 (3,2,1)_0
	&= (8,7,6,5,4, 3,2,1)_*	\label{limclassJoin}
\\
(4,6)_v \times (8,7,6,5,4)_0 (3,2,1)_0
	&= (4,5)_* (8,7,6)_* (3,2,1)_0	\label{limclassSplit}
\end{align}
\esub
where new cycles resulting from the composition are marked by a star.
So, keeping track of the cycles,  the composite operator in the r.h.s.~of (\ref{OPEW}) is, schematically, 
\be
Z_{(2)_v} \times \big[ X^{\zeta_1}_{(n_1)_0}  X^{\zeta_2}_{(n_2)_0} \big]
	= 
	C_{ \bar X  \bar Z [W X]} 
	B_{\bar X X}
	\ \Big\{  \big[ W_{(n_1-n_2)_*} X^{\zeta_2}_{(n_2)_*} X^{\zeta_2}_{(n_2)_0} \big] \Big\}	.
\ee
The branch cut of (\ref{xcaneeminnn}) only ``sees'' the cycle of length $n_1-n_2$  from $W_{(n_1-n_2)}$ because, in this OPE, $Z_{(2)_v}$ does not interact with $X^{\zeta_2}_{(n_2)_0}$, and $X^{\zeta_2}_{(n_2)_*}$ factorizes from the correlation function.
The factorization can be seen comparing (\ref{classSplit}) and (\ref{limclassSplit}). The cycle $(n_2)_\infty$ in the composite operator at $\infty$, and the cycle $(n_2)_*$, resulting from the OPE, are inverses of each other,
$(n_2)_\infty (n_2)_* = \id$.
They also commute with the remaining cycles, so
\begin{align} \label{classSplitafter}
\begin{split}
\id 
&= 
(1,2,3,4,5)_\infty (6,7,8)_\infty \times (1,4)_1 \times 
		 (4,5)_* (8,7,6)_* (3,2,1)_0
\\
&= 
\Big[ (6,7,8)_\infty \times (8,7,6)_* \Big]
	\times
\Big[ (1,2,3,4,5)_\infty \times (1,4)_1 \times  (4,5)_*  (3,2,1)_0 \Big]
\end{split}
\end{align}
(Note that there is no factorization of $(6,7,8)_\infty$ in (\ref{classSplit}), before the OPE.)
The effect of the OPE inside the four-point function is
\begin{align}
\begin{split}
&
	\Big\langle
	\big[ \bar X^{\zeta_1}_{(n_1)_\infty}  \bar X^{\zeta_2}_{(n_2)_\infty} \big]
	\bar Z_{(2)_1}
	\big[  W_{(n_1-n_2)_*}  X^{\zeta_2}_{(n_2)_*} X^{\zeta_2}_{(n_2)_0} \big] 
	\Big\rangle
\\
&\qquad\qquad\qquad\qquad\qquad
=	
	\Big\langle
	 \bar X^{\zeta_2}_{(n_2)_\infty} 
	 X^{\zeta_2}_{(n_2)_*} 
	\Big\rangle
\,
	\Big\langle
	\bar X^{\zeta_1}_{(n_1)_\infty}  
	\bar Z_{(2)_1}
	\big[  W_{(n_1-n_2)}  X^{\zeta_2}_{(n_2)_0} \big] 
	\Big\rangle
\end{split}
\end{align}
The factorized two-point function cannot vanish (unless the branch cut (\ref{xcaneeminnn}) is absent from the four-point function expansion), which explains why the new operator with twist $(n_2)_*$ must be  $X^{\zeta_2}_{(n_2)_*}$.
A similar reasoning explains the product of constants in (\ref{OPEW})
\be
B_{\bar X X} 
	\equiv
	\Big\langle 
	\bar X^{\zeta_2}_{[n_2]} (\infty)  X^{\zeta_2}_{[n_2]} (0)
	\Big\rangle
	,
\quad
C_{ \bar X  \bar Z [W X]} 
	\equiv
	\Big\langle 
	\bar X^{\zeta_2}_{[n_1]} (\infty)  \bar Z_{[2]} (1) 
	\big[ W_{[n_1-n_2]} X^{\zeta_2}_{[n_2]} \big] (0)
	\Big\rangle .
\ee
We assume the strands $X^\zeta_{[n]}$ are individually normalized, hence $B_{\bar X X} = 1$.

Note that classes (\ref{classJoin}) and (\ref{classSplit}), which differ only by the cycles $(2)_1$ and $(2)_v$, are related by channel crossing symmetry.
Starting from (\ref{classJoin}) and moving $(1,4)_v$ counterclockwise around $(1,6)_1$, as in Fig.\ref{twist_monodromy}, we obtain (\ref{classSplit}).
Hence the operators in the r.h.s.~of the OPE (\ref{OPEW}) are intimately related.

\subsubsection{Composite fields with equal cycles}

When $n_1 = n_2$, the solutions in the coincidence limits are
\begin{align}
\begin{aligned}
\infty &= x_\aleph(1) = x_{\frak 0}(1) ,
	&& \text{(multiplicity 1)}
\\
1 &= x_\beth(0) = x_{\frak a}(0) \quad \forall \ {\frak a} 
	&& \text{(multiplicity $2n$)}
\end{aligned}
\end{align}
as can be seen directly from (\ref{xfraka2n}).
Solutions $x_\gimel(v)$ and $x_\daleth(v)$ are missing.
It follows that, for $n_1 = n_2$, there is no operator $S_{[3]}$ in the $v \to 1$ channel, and no $W_{[n_1-n_2]}$ in the $v \to 0$ channel.
This can also be understood from the perspective of $S_N$ selection rules.
For example, $\s_{[3]}$ disappears because there is no three-point function satisfying (\ref{compto1}) with a cycle of length 3 and two double-cycle twists $[(n)(n)]$.

\section{Four-point functions of twisted fields in the D1-D5 CFT}	\label{SectFuncandMast}

We now turn to the D1-D5 CFT at the free orbifold point. Conventions for the notation of fields are given in App.\ref{SectN44CFT}.
The holomorphic Ramond ground states of the $n$-twisted strands can be written in bosonized language as
\bsub	\label{RamondFields}
\begin{align}
R^{\pm}_{(n)}(z) 
	&=
	\exp \left( \pm \frac{i}{2n} 
	\sum_{I = 1}^n \big[ \phi_{1,I}(z) - \phi_{2,I}(z)  \big] 
	\right) \s_{(n)}(z)	
\label{Rampmnnoninv}
\\
R^{\dot1}_{(n)}(z) 
	&= 
	\exp \left( - \frac{i}{2n} 
	\sum_{I = 1}^n \big[ \phi_{1,I}(z) + \phi_{2,I}(z)  \big] 
	\right) \s_{(n)}(z)	
\label{Rampmnnoninv}
\\
R^{\dot2}_{(n)}(z) 
	&= 
	\exp \left( + \frac{i}{2n} 
	\sum_{I = 1}^n \big[ \phi_{1,I}(z) + \phi_{2,I}(z)  \big] 
	\right) \s_{(n)}(z)	
\label{Rampmnnoninv}
\end{align}\esub
All have conformal weight
$h^\R_n = \frac{n}{4} = \frac{n c_\rm{seed}}{24}$,
 the correct weight of a spin filed in a CFT with central charge $n c_\rm{seed}$. are distinguished by their SU(2) charges $(j , {\frak j})$.%
	\footnote{We denote by $j$ and ${\frak j}$ the eigenvalues of the components $J^3$ and ${\frak J}^3$, not of the Casimirs.}
 For $R^\pm_{(n)}$ and $R^{\dot A}_{(n)}$, respectively,
$(j = \pm \tfrac12 , \ \frak j = 0 )$
and
$( j = 0 , \ \frak j = \pm \tfrac12 )$.
NS chiral primaries can be expressed in bosonized form as
\bsub	\label{NSchirals}
\begin{align}
O^{(0)}_{(n)} (z) &=  
	\exp \Bigg(
	 \frac{i(n-1)}{2n} \sum_{I = 1}^n
	 \Big[
	 \phi_{1,I}(z)  - \phi_{2,I}(z)
	 \Big] 
	 \Bigg) \s_{( n)}(z)
\\
O^{(2)}_{(n)} (z) &=  
	\exp \Bigg(
	 \frac{i(n+1)}{2n} \sum_{I = 1}^n
	 \Big[ 
	 \phi_{1,I}(z)  - \phi_{2,I}(z)
	 \Big] 
	 \Bigg) \s_{(n)}(z)
\\
O^{(1\pm)}_{(n)}  (z) &=  
	\exp \Bigg(
	\sum_{I = 1}^n
	\Bigg[
		\frac{i(n\pm1)}{2n}  \phi_{1,I}(z) 	
		- \frac{i(n\mp1)}{2n}  \phi_{2,I}(z) 
	\Bigg]
	\Bigg)
		\s_{(n)}(z)
\end{align}\esub
(see e.g.~\cite{Jevicki:1998bm,Dabholkar:2007ey,Pakman:2009ab}). They have conformal weights and R-charges
\begin{align}	\label{NShj}
h^{(0)}_n = \tfrac12 (n-1) = j^{(0)}_n ;
\qquad
h^{(2)}_n =  \tfrac12 (n+1) = j^{(2)}_n ;
\qquad
h^{(1)}_n =  \tfrac12 n = j^{(1)}_n .
\end{align}
Anti-chiral operators
$\bar O^{(p)}_{(n)}(z)$
have opposite R-charges and the same dimensions.
The labels $(p)$ in refer to the associated cohomology of $\bbT^4$. The two middle-cohomology fields $O^{(1\pm)}_{(n)}$ have degenerate dimension and R-charge, but are distinguished by the global SU(2) charges ${\frak j}^{(p)}_n$, with respect to which the other fields are neutral.
There are analogous fields in the anti-holomorphic sector.

The operator that drives the theory away from the free orbifold point is a specific excitation of the 2-twisted lowest weight NS chiral with super-current modes, 
\begin{align}	\label{DeformwithMo}
\begin{split}
O^{(\inter)}_{[2]}(z , \bar z ) 
	&\equiv \e_{A  B} G^{-  A}_{-\frac{1}{2}} \tilde G^{ \dot -  B}_{-\frac{1}{2}} 
	O^{(0,0)}_{[2]}(z, \bar z) 	
\\
	&= \e_{A  B} 
	\oint \frac{\dd w}{2\pi i}
	\oint \frac{\dd\bar w}{2\pi i}
	G^{-  A}(w) \tilde G^{ \dot -  B}(\bar w) 
	O^{(0,0)}_{[2]}(z, \bar z) 	.
\end{split}
\end{align}
This is an exactly marginal deformation, with dimensions 
$h = 1 = \tilde h$, 
and it is a singlet of all the SU(2)s, with $j = \tilde\jmath = 0$, ${\frak j} = \tilde{\frak j} = 0$.

From the single-cycle fields above, we can build multi-cycle, $S_N$-invariant fields such as the Ramond ground states of the full orbifold,
\be	\label{RamondComp}
R_{[N^{\zeta}_n]}
\equiv
\Big[ \prod_{\zeta,n} ( R^{\zeta}_{[n]} )^{N^{\zeta}_n} \Big] 
\quad
\text{with} \quad
h^\R = \frac{N}{4} , 
\quad
j = \sum_{\zeta,n} {N^{\zeta}_n} \, j_{\zeta} .
\quad
{\frak j} = \sum_{\zeta,n} N^{\zeta}_n {\frak j}_\zeta ,
\ee
and also composite NS chirals
\begin{align}	\label{NScomp}
O_{[N^p_n]}
\equiv
\Big[ \prod_{p,n} ( O^{(p)}_{[n]} )^{N^p_n} \Big] ,
\quad
\text{with}
\quad
h &= j 
	=
\frac{N}{2} + \frac{\text{\# $O^{(2)}$s} - \text{\# $O^{(0)}$s}}{2}
\end{align}
where $\#O^{(p)}$ denotes the number of strands of the type $p$ entering the composite fields.
In both Eqs.(\ref{RamondComp})-(\ref{NScomp}), the multiplicities form a partition of $N = \sum_{\zeta,n} n N^{\zeta}_n$.
In the large-$N$ limit, the Ramond ground states (\ref{RamondComp}) are heavy, $h^\R \sim N$. The multi-cycle NS chirals (\ref{NScomp}) can also be heavy, if the number of lowest-cohomology components is parametrically small.

\subsection{Formulas for two classes of connected functions}	\label{SectMasterForm}

\begin{table}
\begin{center}
\begin{tabular}{| r || c c c c |}
\hline 
\hline
\hline
&$O^{(0)}_{[2]}$
&$O^{(2)}_{[2]}$
&$O^{(1+)}_{[2]}$
&$O^{(1-)}_{[2]}$
\\
\hline
$\{\alpha , \beta \}$
& {\footnotesize $\{1,-1 \}$}
& {\footnotesize $\{3 , -3 \}$}
& {\footnotesize $\{3 , -1 \}$}
& {\footnotesize $\{1 , -3 \}$}
\\
\hline 
\hline
\hline
&$O^{(1+)}_{[n_1]} O^{(2)}_{[n_2]}$
&$O^{(1+)}_{[n_1]} O^{(0)}_{[n_2]}$
&$O^{(1+)}_{[n_1]} O^{(1+)}_{[n_2]}$
&$O^{(1+)}_{[n_1]} O^{(1-)}_{[n_2]}$
\\
\hline
$\{\hat \sigma , \hat \varrho \}$
& {\footnotesize $\{n_1+1 , 1-n_1 \}$}
& {\footnotesize $\{n_1+1 , 1-n_1 \}$}
& {\footnotesize $\{n_1+1 , 1-n_1 \}$}
& {\footnotesize $\{n_1+1 , 1-n_1 \}$}
\\
$\{ \check \sigma , \check \varrho \}$ 
& {\footnotesize $\{ n_2+1 , -1-n_2 \}$}
& {\footnotesize $\{ n_2-1 , 1-n_2 \}$}
& {\footnotesize $\{ n_2+1 , 1-n_2 \}$}
& {\footnotesize $\{ n_2-1 , -1-n_2 \}$}
\\
\hline 
\hline
\hline
&$O^{(1-)}_{[n_1]} O^{(2)}_{[n_2]}$
&$O^{(1-)}_{[n_1]} O^{(0)}_{[n_2]}$
&$O^{(1-)}_{[n_1]} O^{(1-)}_{[n_2]}$
&$R^{\pm}_{[n_1]} R^{\mp}_{[n_2]}$
\\
\hline
$\{\hat \sigma , \hat \varrho \}$
& {\footnotesize $\{n_1-1 , -1-n_1 \}$}
& {\footnotesize $\{n_1-1 , -1-n_1 \}$}
& {\footnotesize $\{n_1-1 , -1-n_1 \}$}
& {\footnotesize $\{\pm1 , \mp1 \}$} 
\\
$\{ \check \sigma , \check \varrho \}$ 
& {\footnotesize $\{ n_2+1 , -1-n_2 \}$}
& {\footnotesize $\{ n_2-1 , 1-n_2 \}$}
& {\footnotesize $\{ n_2-1 , -1-n_2 \}$}
& {\footnotesize $\{ \mp 1 , \pm1 \}$}
\\
\hline 
\hline
\hline
&$O^{(2)}_{[n_1]} O^{(2)}_{[n_2]}$
&$O^{(0)}_{[n_1]} O^{(2)}_{[n_2]}$
&$O^{(0)}_{[n_1]} O^{(0)}_{[n_2]}$
&$R^{\pm}_{[n_1]} R^{\pm}_{[n_2]}$
\\
\hline
$\{\hat \sigma , \hat \varrho \}$
& {\footnotesize $\{ n_1+1 , -1-n_1 \}$}
& {\footnotesize $\{n_1-1 , 1-n_1 \}$}
& {\footnotesize $\{n_1-1 , 1-n_1 \}$}
& {\footnotesize $\{\pm1 , \mp1 \}$}
\\
$\{ \check \sigma , \check \varrho \}$ 
& {\footnotesize $\{ n_2+1 , -1-n_2 \}$}
& {\footnotesize $\{ n_2+1 , -1-n_2 \}$}
& {\footnotesize $\{ n_2-1 , 1-n_2 \}$}
& {\footnotesize $\{\pm1 , \mp1 \}$}
\\
\hline 
\hline
\hline
&$R^{\dot1}_{[n_1]} R^{\pm}_{[n_2]}$
&$R^{\dot2}_{[n_1]} R^{\dot2}_{[n_2]}$
&$R^{\dot1}_{[n_1]} R^{\dot2}_{[n_2]}$
&$R^{\dot1}_{[n_1]} R^{\dot1}_{[n_2]}$
\\
\hline
$\{\hat \sigma , \hat \varrho \}$
& {\footnotesize $\{-1 , -1 \}$}
& {\footnotesize $\{+1 , +1 \}$}
& {\footnotesize $\{-1 , -1 \}$}
& {\footnotesize $\{-1 , -1 \}$}
\\
$\{ \check \sigma , \check \varrho \}$ 
& {\footnotesize $\{ \pm1 , \mp1 \}$}
& {\footnotesize $\{ +1 , +1 \}$}
& {\footnotesize $\{ +1 , +1 \}$} 
& {\footnotesize $\{ - 1 , -1 \}$}
\\
\hline
\end{tabular}
\caption{Parameters
 for making 
 $Z^{\{ \a,\b \}}_{[2]}$
and 
 $[X^{\{\hat\s,\hat\varrho\}}_{[n_1]} X^{\{\check\s , \check\varrho\}}_{[n_2]} ]$
 into different NS chirals or Ramond fields.
}
\label{ZNSR}
\end{center}
\end{table}

We want to compute explicitly correlation functions of the type (\ref{A12sol}), using the fields above. 
We can do a rather generic computation, if we define the following ``adjustable'' operators
(from which we build $S_N$-invariant combinations)
\begin{align}		\label{ZGen}
Z^{\{\a,\b\}}_{(2)}
	&\equiv
	\exp \Bigg[
	 \frac{i}{4} 
	 \sum_{I = 1}^2
	\Big(
	\a
	\phi_{1,I}
	+
	\b
	 \phi_{2,I}
	 \Big) 
	 \Bigg] \s_{(1,2)}
\\	 
\begin{split}\label{Rn1n2Gen}
\Big[ 
	X^{\{\hat \sigma , \hat \varrho \}}_{(n_1)} 
	X^{\{ \check \sigma , \check \varrho \}}_{(n_2)} 
	\Big] 
	&\equiv 
	\exp \Bigg[ 
	\frac{i}{2n_1}
	\sum_{I=1}^{n_1} 
	\big( \hat \sigma \phi_{1,I} + \hat \varrho \phi_{2,I} \big) 
\\	&\qquad\quad
	+
	{i\over 2n_2}
	\sum_{I=n_1+1}^{n_1+n_2}
	 \big( 
	 \check \sigma \phi_{1,I} + \check \varrho \phi_{2,I} 
	\big)
	\Bigg]
	\sigma_{(1,\cdots, n_1)} 
	\sigma_{(n_1+1,\cdots, n_1+n_2)} 
\end{split}
\end{align} 
with holomorphic conformal weights 
\begin{align} \label{habZ} 
\hZ
	&= \frac{\a^2 + \b^2}{16} + \frac14 \left( 2 - \frac12 \right)
	= \frac{\a^2 + \b^2 + 6}{16} 
\\
	\label{hrsXX}
\hX 
	&= \frac{ \h\s^2 + \h\vro^2}{8n_1} + \frac{\c\s^2 + \c\vro^2}{8n_2}
		+
		\frac{n_1^2 - 1}{4n_1}
		+
		\frac{n_2^2 - 1}{4n_2}
\end{align}
The anti-holomorphic counterparts of (\ref{ZGen})-(\ref{Rn1n2Gen}) are completely analogous.
Adjusting the parameters $\a$, $\b$, $\hat\sigma$, $\hat\varrho$, $\check\sigma$, $\check\varrho$ we obtain all the Ramond or NS chiral double-cycle fields, following Table \ref{ZNSR}.
We can then compute 
\be	\label{AZv}
\AmZ{v} \equiv
	\Big\langle
	\Big[
	\bar X_{[n_1]}^{\{\hat \sigma , \hat \varrho \}} 
	\bar X_{[n_2]}^{\{ \check \sigma , \check \varrho \}} 
	\Big] (\infty) 
	\
	\bar Z_{[2]}^{\{\a,\b\}} (1) 
	\
	Z_{[2]}^{\{\a,\b\}} (v)
	\
	\Big[
	X_{[n_1]}^{\{\hat \sigma , \hat \varrho \}}
	X_{[n_2]}^{\{ \check \sigma , \check \varrho \}} 
	\Big](0) 
	\Big\rangle .
\ee
and find the desired cases fixing the parameters afterwards.
The twisted correlator (\ref{AZv}) can be computed in the way of Lunin and Mathur \cite{Lunin:2000yv,Lunin:2001pw}, or using conformal Ward identities to find a first-order differential equation, in what is known as the `stress-tensor method' \cite{Dixon:1986qv,Arutyunov:1997gi,Arutyunov:1997gt,Pakman:2009ab,Pakman:2009zz}.
The computation for generic $X$s, using both techniques, was done in detail in Appendix B of Ref.\cite{Lima:2021wrz} for the case where $Z_{[2]} = \Oint$. The case of $Z_{[2]} = O^{(p,p)}_{[2]}$ is much simpler, and will be described now, 
using the Lunin-Mathur technique. Details are left to App.\ref{General4ptNS}.

The fermionic exponentials in (\ref{AZv}) are lifted to ramification points on the covering surface, with an appropriate factor depending on the local behavior of the map (\ref{coverm}); see Eqs.(\ref{rim})-(\ref{bofx}). The resulting covering-surface correlator is a \emph{six}-point function, 
\be	\label{Gcoverx}
\AmZ{x}_{\rm{cover}} =
	\Big\langle
	\bar X^{\{\hat \sigma , \hat \varrho \}}(\infty) 
	\bar X^{\{ \check \sigma , \check \varrho \}} (t_\infty) 
	\bar Z^{\{\a,\b\}} (t_1) 
	Z^{\{\a,\b\}} (x)
	X^{\{ \check \sigma , \check \varrho \}} (t_0) 
	X^{\{\hat \sigma , \hat \varrho \}}(0) 
	\Big\rangle .
\ee
The relation between $\AmZ{x}_\rm{cover}$ and the base-sphere correlator $\AmZ{x}$ is
\be	\label{GeSLcover}
\AmZ{x} = e^{S_L} \AmZ{x}_{\rm{cover}}
\ee
where $S_L$ is a Liouville action induced by the covering map \cite{Lunin:2000yv}. 
In fact, $e^{S_L}$ is the correlation function of the bare twists within (\ref{AZv}), and is universal, independent of the specific excitations that define $X^{\zeta}$ and $Z$.
The algorithm by Lunin and Mathur \cite{Lunin:2000yv,Lunin:2001pw} to derive $S_L$ involves a careful regularization of the path integral around the ramification points. (See also \cite{Avery:2010qw} for a very detailed account.) An alternative, described in \cite{Lima:2020kek,Lima:2021wrz}, is to use the stress-tensor method to compute the bare-twist correlation function, bypassing the regularization procedure.%
	\footnote{%
	Computing the functions independently in both ways also gives an important cross-check of the final results, which we have performed.}
 The results for $\AmZ{x}_\rm{cover}$ and $S_L$ are given in Eqs.(\ref{GbosnAp}) and (\ref{SLofx}), respectively, yielding our desired master formula:
\bsub\label{GenricAapp}
\begin{align}
\AmZ{x} =
	C_Z
	\
	x^{K_1}
	(x-1)^{K_2}
	(x + \tfrac{n_1}{n_2})^{K_3}
	(x + \tfrac{n_1}{n_2} -1)^{K_4}
	(x + \tfrac{n_1 - n_2}{2n_2})^{K_5}
\end{align}
with the exponents
\begin{align}	\label{expontK}
\begin{split}
K_1	&= 
	\frac{(-n_1 + n_2) (\alpha ^2+\beta ^2+6)}{16} 
	- \frac{n_1 (2 - \c\vro^2 - \c\s^2)}{8 n_2}
	- \frac{n_2 (2 - \h\vro^2 - \h\s^2 )}{8 n_1}
	\\
&\qquad\qquad\qquad\quad
	+\frac{
	\a^2 + \b^2 
	+ 2 \big[1 - \c\vro \h\vro - \c\s \h\s  
	- \a (\c\s - \h\s) - \b (\c\vro - \h\vro) 
	\big]
	}{8}
\\
K_2 &=
	\frac{(+n_1 + n_2) (\alpha ^2+\beta ^2+6)}{16} 
	+ \frac{n_1 (2 - \c\vro^2 - \c\s^2)}{8 n_2}
	+ \frac{n_2 (2 - \h\vro^2 - \h\s^2 )}{8 n_1}
	\\
&\qquad\qquad\qquad\quad
	+\frac{
	\a^2 + \b^2 
	+ 2 \big[1 + \c\vro \h\vro +  \c\s \h\s  
	- \a (\c\s+\h\s) - \b (\c\vro+\h\vro) 
	\big]
	}{8}
\\
K_3 &=
	\frac{(-n_1 - n_2) (\alpha ^2+\beta ^2+6)}{16} 
	+ \frac{n_1 (2 - \c\vro^2 - \c\s^2)}{8 n_2}
	+ \frac{n_2 (2 - \h\vro^2 - \h\s^2 )}{8 n_1}
	\\
&\qquad\qquad\qquad\quad
	+\frac{
	\a^2 + \b^2 
	+ 2 \big[ 1 + \c\vro \h\vro +  \c\s \h\s 
	+ \a (\c\s+\h\s) +  \b (\c\vro+\h\vro)
	\big]
	}{8}
\\
K_4 &= 
	\frac{(+n_1 - n_2) (\a^2+ \b^2+6)}{16} 
	- \frac{n_1 (2 - \c\vro^2 - \c\s^2)}{8 n_2}
	- \frac{n_2 (2 - \h\vro^2 - \h\s^2 )}{8 n_1}
	\\
&\qquad\qquad\qquad\quad
	+\frac{
	\a^2 + \b^2 
	+ 2 \big[ 1 - \c\vro \h\vro -  \c\s \h\s 
	+ \a (\c\s-\h\s) +  \b (\c\vro-\h\vro)
	\big]
	}{8}
\\
K_5 &= 
	- \frac{2 + 3 (\a^2 + \b^2) }{8}
\end{split}
\end{align}\esub
Some constant factors involving $n_1$ and $n_2$ have been absorbed into the constant $C_Z$, which also takes into account the arbitrariness of normalization of the bare twists, and is fixed by the correct normalization of the correlation function in the identity OPE channel.

The function 
\be	\label{Aintv}
\Amint{v, \bar v}
\equiv
	\Big\langle
	\Big[
	\bar X_{[n_1]}^{\{\hat \sigma , \hat \varrho \}} 
	\bar X_{[n_2]}^{\{ \check \sigma , \check \varrho \}} 
	\Big] (\infty) 
	\
	\Oint (1) 
	\
	\Oint (v,\bar v)
	\
	\Big[
	X_{[n_1]}^{\{\hat \sigma , \hat \varrho \}}
	X_{[n_2]}^{\{ \check \sigma , \check \varrho \}} 
	\Big](0) 
	\Big\rangle ,
\ee
is more complicated than (\ref{AZv}) because the deformation operator $\Oint$ is not simply a fermionic exponential, but a linear combination of terms with bosonic factors and contributions from the integral defining the modes of the super-current excitation, cf.~(\ref{DeformwithMo}). 
Its computation, carried on in Appendix B of Ref.\cite{Lima:2021wrz}, is, nevertheless, completely analogous to the one above, including the same Liouville factor, because the twist structure is identical. 
In the end, we obtain 
\bsub	\label{Gcofx}
\begin{align}
\Amint{x, \bar x}
 =
 	\Big|
	{\cal A}(x) \big( 1 + {\cal B}(x) \big)
 	\Big|^2
\end{align}
where
\begin{align}
\begin{split}
{\cal A}(x) &= \frac{C_\rm{int}}{2} 
		\
		\frac{
		x^{P_1 - Q_1}
		(x-1)^{P_2 - Q_2}
		(x + \frac{n_1}{n_2})^{P_3 - Q_3}
		(x + \frac{n_1}{n_2} -1)^{P_4 - Q_4}
		}{
		(x + \frac{n_1 - n_2}{2n_2})^4
		}		
\\
{\cal B}(x) &=
		x^{2Q_1}
		(x-1)^{2Q_2}
		(x + \tfrac{n_1}{n_2})^{2Q_3}
		(x + \tfrac{n_1}{n_2} -1)^{2Q_4}
\end{split}
\end{align}
with the exponents
\begin{align}	\label{expontP}
\begin{split}
P_1	&= 
	+ n_2 - n_1
	- \frac{\c\vro \h\vro+\c\s \h\s - 6}{4}
	+ \frac{n_2 \left(\h\vro^2+\h\s^2-2\right)}{8n_1}
	+ \frac{n_1 \left(\c\vro^2+\c\s^2 -2 \right)}{8n_2}
\\
P_2 &=
	+ n_2 + n_1
	+ \frac{\c\vro \h\vro+\c\s \h\s + 6}{4}
	- \frac{n_2 \left(\h\vro^2+\h\s^2-2\right)}{8n_1}
	-\frac{n_1 \left(\c\vro^2+\c\s^2 -2\right)}{8n_2}
\\
P_3 &=
	-n_2 - n_1
	+ \frac{\c\vro \h\vro+\c\s \h\s +6}{4}
	-\frac{n_2 \left(\h\vro^2+\h\s^2-2\right)}{8n_1}
	-\frac{n_1 \left(\c\vro^2+\c\s^2-2\right)}{8n_2}
\\
P_4 &= 
	-n_2 + n_1
	- \frac{\c\vro \h\vro+\c\s \h\s - 6}{4}
	+ \frac{n_2 \left(\h\vro^2+\h\s^2-2\right)}{8 n_1 }
	+
	\frac{n_1 \left(\c\vro^2+\c\s^2 - 2\right)}{8n_2}
\end{split}
\end{align}
and
\begin{align}	\label{expontQ}
Q_1 = 
	\frac{-\c\s - \c\vro + \h\s + \h\vro}{4}
	= - Q_4 ,
\qquad
Q_2 =
	\frac{- \c\s - \c\vro - \h\s - \h\vro}{4}
	= - Q_3
\end{align}
\esub
This result is, in fact, more general than the one derived in Ref.\cite{Lima:2021wrz}, because in the latter case we had restricted our attention to double-cycle Ramond fields, for which 
$\c\s^2 = \c\vro^2 = \h\s^2 = \h\vro^2 = 1$, 
hence the last two terms in each exponent $P_i$ vanishes. The present result allow us to give also the correlators for 
$[X^{\{\hat\s,\hat\varrho\}}_{[n_1]} X^{\{\check\s , \check\varrho\}}_{[n_2]} ]$ 
made by NS chirals, using the dictionary in Table \ref{ZNSR}.%

\subsubsection*{Composite fields with equal cycles}

In \S\ref{SectHurwBlocks} we showed that when $n_1 = n_2 = n$, the covering maps develop a symmetry that reflects upon the Hurwitz blocks.
We can check this property, using our master formulas.
The correlators (\ref{Gcofx}) and (\ref{GenricAapp}) simplify considerably in this case,
\bsub\label{GenricAappnn}
\begin{align}	\label{GenricAappnna}
\AmZnn{x} =
	\frac1{(- 4 n)^{ 2 \hZ}} 	\
	x^{ K_0}
	(x-1)^{ K_-}
	(x + 1)^{ K_+}
\end{align}
where
\begin{align}
\begin{split}
K_0	&= 
	\frac{
	(\c\s  - \h\s )^2 + (\c\vro -  \h\vro)^2  
	}{4}
	- 
	\frac{
	\a^2 + \b^2 + 6
	}{8}
\\
K_\pm &=
	\frac{
	 \c\vro \h\vro +  \c\s \h\s \pm \a (\c\s+\h\s) \pm \b (\c\vro+\h\vro) 
	 }{4}
	-
	\frac{
	  \c\vro^2 + \c\s^2 + \h\vro^2 + \h\s^2 
	}{8}
	+ 
	\frac{
	(1\mp n)(\alpha ^2+\beta ^2 + 6) 
	}{8}
\end{split}
\end{align}\esub
and
\bsub	\label{Gcofxnn}
\begin{align}
\Amintnn{x}
 =
		\frac1{16 n^2}
		\
		x^{P_0}
		(x-1)^{P_- + Q}
		(x + 1)^{P_+ - Q} 
	\left[ 1 + \left( \frac{x+1}{x-1} \right)^{2Q} \right]
\end{align}
where
\begin{align}
\begin{split}
P_\pm &=
	2(1 \mp n)
	- \frac{ (\c\s-\h\s)^2 + (\c\vro - \h\vro )^2}{8} ,
\qquad
P_0	= 
	\frac{(\c\s - \h\s)^2 + (\c\vro - \h\vro)^2 - 8}{4} ,
\\
Q &= \frac{\c\s + \c\vro + \h\s + \h\vro}{4}
\end{split}
\end{align}
\esub

The symmetry (\ref{nnSymmSFD}) of the Hurwitz blocks can be checked explicitly. 
We see that 
$\AmZnn{x} = \AmZnn{1/x}$ and $\Amintnn{x} = \Amintnn{1/x}$
iff 
\be
0=
2 K_0 + K_+ + K_- 
= \frac{(\c\vro - \h\vro)^2 + ( \c\s - \h\s)^2}{4} 
=
2 P_0 + P_+ + P_- .
\ee
This only holds if 
$\c\vro = \h\vro$ and $\c\s = \h\s$, i.e.~if the strands
$	X_{[n]}^{\{\hat \sigma , \hat \varrho \}}
	=
	X_{[n]}^{\{ \check \sigma , \check \varrho \}} 
$
are identical, as expected from the discussion leading to Eq.(\ref{nnSymmSFD}).

Since the $x_{\frak a}(v)$ are expressible in closed form (\ref{xfraka2n}), we can write a closed formula for the correlation functions directly on the base sphere, 
\begin{align}
\label{AZvnn}
\AmZnn{v}
	= 
	\frac{4n^2 (N-2)!}{N!}
	\sum_{ {\frak a}=0}^{n-1}
	\left|
	\frac{	2^{K_+ + K_-}}{(4n)^{2\hZ}} 
	\frac{
	\left( v^{\frac1{2n}} e^{\frac{{\frak a} \pi i}{n}} \right)^{K_+}
	\left( 1 + v^{\frac1{2n}} e^{\frac{{\frak a} \pi i}{n}} 	\right)^{K_0}
	}{
	\left( 1 - v^{\frac1{2n}} e^{\frac{{\frak a} \pi i}{n} }	\right)^{2\hZ}
	}	
	\right|^2
\end{align}
where we have used the fact that $K_0 + K_- + K_+ = 2\hZ$.
When $n=1$, there are only two inverse functions, 
\be	\label{AZvnn1}
\begin{split}
&	\Big\langle
	\Big[
	X_{[1]}^{\{\hat \sigma , \hat \varrho \}} 
	X_{[1]}^{\{ \check \sigma , \check \varrho \}} 
	\Big]^{\dagger} (\infty) 
	\
	Z_{[2]}^{\{\a,\b\}\dagger} (1) 
	\
	Z_{[2]}^{\{\a,\b\}} (v)
	\
	\Big[
	X_{[1]}^{\{\hat \sigma , \hat \varrho \}}
	X_{[1]}^{\{ \check \sigma , \check \varrho \}} 
	\Big](0) 
	\Big\rangle
\\
&\qquad\qquad 
	= 
	2^{2 - 2 K_0 - K_+ - K_-} 
	\Bigg[
	\Bigg|
	\frac{
	( v^{\frac1{2}}  )^{K_+}
	( 1 + v^{\frac1{2}} 	)^{K_0}
	}{
	( 1 - v^{\frac1{2}}  )^{2\hZ}
	}	
	\Bigg|^2
	+	
	\Bigg|
	\frac{
	( v^{\frac1{2}} )^{K_+}
	( 1 - v^{\frac1{2}}  )^{K_0}
	}{
	( 1 + v^{\frac1{2}}  )^{2\hZ}
	}	
	\Bigg|^2
	\Bigg]
\end{split}
\ee
using the appropriate expression for the $N$-dependent factor.
Note that functions with $n =1$ scale as $N^0$.
There are only two non-trivial twists, hence  two ramification points in the covering surface of the connected correlators, so $R = 2$ in Eq.(\ref{QptNDoubNchi}).

\subsection{OPE limits}	\label{SectBlocksinHHLLchann}

We can now derive not only the twists but the conformal dimensions and structure constants of operators appearing in the OPE limits $v \to 1$ and $v \to 0$. 
In the channel $v \to 1$, the Huwitz blocks where $x \to x_\aleph(1) = \infty$, give
\be	\label{scrAxinf} 
{\scr A}(x_\aleph(v) ) 
	= C_Z x_\aleph^{2h_Z}(v)  \big[1 + \rm{O}(1 /x_\aleph(v)) \big] 
	= \frac{(-4n_1)^{2 h_Z} C_Z}{ (1-v)^{ 2 h_Z } }
	\Big[ 1 + \rm{O}(1-v) \Big]		
\ee
for both $\AmZ{x}$, where $h_Z$ is given by Eq.(\ref{habZ}),  and for $\Amint{x}$, where $h_Z =1$.
Looking at the power of the leading singularity, we see that the untwisted operator $U_{[1]}$ in the OPE (\ref{fusZZ}) is the identity. 
Since we have assumed that the individual cycle fields are normalized, the arbitrary constant in the correlator is now fixed to
\be	\label{normCs}
C_Z = \frac1{(- 4 n_1)^{ 2 \hZ}} .
\ee
The Hurwitz blocks where
 $x \to x_\gimel(1)$
 again have the same form for $\AmZ{x}$ and $\Amint{x}$,
\be
\begin{split}
{\scr A}(x_\beth(v)) 
	&=
	\rm{constant} \times \Big( x_\beth(v) + \frac{n_1 - n_2}{2n_2} \Big)^{-6h_L + 2} \Big[1 + \rm{O}\big( x_\beth(v) + \tfrac{n_1 - n_2}{2n_2} \big) \Big] 
\\
	&=
		\frac{	\rm{constant}}{(1-v)^{2 h_Z - \frac23} }
		\Big[
		1 
		+ \rm{O}(1-v)^{\frac13}
		\Big]
\end{split}		
\ee
with a constant that is readily computable but given by a cumbersome expression in general.
The power of the leading singularity shows that twist-three operator $S_{[3]}$ in the OPE (\ref{fusZZ}) is a primary with dimension $h_p = \frac23$, that is the bare twist $\s_{[3]}$.

In channel $v \to 0$, the function (\ref{GenricAapp}) expands as
\bsub \label{Amzu0chan} \begin{align}
\AmZ{x_\beth(v)} 
	&= \frac{ e^{i \psi} }{2^{K_5} (4n_1)^{2h_Z}}
		\left( \frac{n_1}{n_2} \right)^{K_1 - \frac{n_1-n_2}{n_1+n_2} K_3}
		\left( 1+ \frac{n_2}{n_2} \right)^{K_2 + K_3 + K_5}
		v^{ \frac{K_3}{n_1+n_2}}
		+\cdots
\\
\AmZ{x_\daleth(v)} 
	&= \frac{ e^{i \psi} }{2^{K_5} (4n_1)^{2h_Z}}
		\left( \frac{n_1}{n_2} \right)^{\frac{n_1+n_2}{n_1-n_2} K_1 +K_3}
		\left( 1-\frac{n_2}{n_2} \right)^{K_1+K_4 + K_5}
		v^{ \frac{K_1}{n_1-n_2}}
		+\cdots
\end{align}\esub
Here $e^{i\psi}$ simply denotes an unimportant phase that is not necesseraly the same in all functions.
The leading order coefficients give the structure constants in the OPE 
\be	\label{OPEWns}
\begin{split}
Z_{[2]} \times \big[ X^{\zeta_1}_{[n_1]}  X^{\zeta_2}_{[n_2]} \big]
	&= 
	C_{ [\bar X\bar X] \bar Z Y} \ \big\{  Y_{[n_1+n_2]} \big\}
\\
&\quad
	+
	C_{ \bar X  \bar Z [W X]} 
	B_{\bar X X}
	\ \Big\{  \big[ W_{[n_1-n_2]} X^{\zeta_2}_{[n_2]} X^{\zeta_2}_{[n_2]} \big] \Big\}	
\end{split}\ee
for the fields in Table \ref{ZNSR}. 
We can read the conformal weights from the leading powers (\ref{Amzu0chan}).
For the single-cycle field $Y_{[n_1+n_2]}$, 
\be \label{dimhY}
h_Y = \frac{K_3}{n_1+n_2} + h_Z + h_{XX} ,
\ee
where $h_Z$ and $h_{XX}$ are given in (\ref{habZ}).
Similarly, the dimension of the composite operator 
$[ W_{[n_1-n_2]} X^{\zeta_2}_{[n_2]} X^{\zeta_2}_{[n_2]} ]$
is
$\frac{K_1}{n_1-n_2} + h_Z + h_{XX}$,
but since we know the dimensions of the components $X^{\zeta_2}_{[n_2]}$, we can extract the dimension of $W_{[n_1-n_2]}$ alone,
\be	\label{dimhW}
h_W = \frac{K_1}{n_1-n_2} + h_Z + h_{XX} 
	- 2 \left( \frac{\c\s^2 + \c\vro^2}{8n_2} + \frac{n_2^2 - 1}{4n_2} \right)
\ee

The same analysis holds for the functions (\ref{Gcofx}) with the deformation operator $\Oint$, with weight $h =1$.
The leading-order expansions are
\bsub\begin{align}
\Amint{x_\beth(v)} 
	&= 
		e^{i \psi} 
		(n_1 + n_2)^{M_{2} - 4}
		n_1^{M_{1} - 2}
		n_2^{4 - M_1 - M_2}
		v^{ \frac{M_3}{n_1+n_2}}
		+\cdots
\\
\Amint{x_\daleth(v)} 
	&=
		e^{i \psi} 
		(n_1 - n_2)^{M_{4} - 4}
		n_1^{M_{3} - 2}
		n_2^{4 - M_3 - M_4}
		v^{ \frac{M_1}{n_1-n_2}}
		+\cdots
\end{align}\esub
where $M_i \equiv \max( P_i - Q_i , P_i + Q_i)$.
We can read the conformal data of the OPE
\be	\label{OPEWint}
\begin{split}
\Oint \times \big[ X^{\zeta_1}_{[n_1]}  X^{\zeta_2}_{[n_2]} \big]
	&= 
	C_{ [\bar X\bar X] \bar Z Y} \ \big\{  Y_{[n_1+n_2]} \big\}
\\
&\quad
	+
	C_{ \bar X  \bar Z [W X]} 
	B_{\bar X X}
	\ \Big\{  \big[ W_{[n_1-n_2]} X^{\zeta_2}_{[n_2]} X^{\zeta_2}_{[n_2]} \big] \Big\}	
\end{split}\ee
and find the weights
\be	
h_Y = \frac{M_3}{n_1+n_2} + 1 + h_{XX} ,
\quad
h_W = \frac{M_1}{n_1-n_2} + 1 + h_{XX} 
	- 2 \left( \frac{\c\s^2 + \c\vro^2}{8n_2} + \frac{n_2^2 - 1}{4n_2} \right)
\ee
where $h_{XX}$ is given in (\ref{habZ}).

\subsection{Functions with NS chiral fields and other examples}

Although the exponents (\ref{expontK}), (\ref{expontP}), (\ref{expontQ}) may look complicated functions, they are, in fact, usually very simple after the parameters of Table \ref{ZNSR} are inserted.
We now discuss some examples of functions and their conformal data.

\subsubsection{Single-cycle NS chirals, composite Ramond}  

Take $Z_{[2]}$ to be a middle-cohomology NS chiral, hence
$(h_Z, \tilde h_Z) = (1,1)$,
and the composite fields be made of be R-charged Ramond fields $R^{+\dot +}_{[n_i]}$. 
The function (\ref{GenricAapp}) is the same for both $O^{(1\pm,1\pm)}_{[2]}$,
\be	\label{FuncEx1pp}
A^{1\pm|+|+}_{n_1,n_2}(x)
	= 
	\frac1{16n_1^2}
	\frac{
	x^{1 - n_1 + n_2}
	(x-1)^{n_1+n_2}
	(x + \frac{n_1}{n_2})^{4-n_1-n_2}
	(x + \frac{n_1 - n_2}{n_2})^{1 + n_1 - n_2}
	}{ (x + \frac{n_1-n_2}{2n_2})^4}
\ee
The expansion of the $S_N$-invariant function (\ref{A12sol}) in the channel $v \to 1$ is 
\begin{align}	\label{s3exprro1}
\begin{split}
&\Big\langle 
	\Big[ \bar R^{+\dot+}_{[n_1]}  \bar R^{+\dot+}_{[n_2]} \Big] (\infty)
	\bar O^{(1\pm,1\pm)}_{[2]} (1)
	O^{(1\pm,1\pm)}_{[2]} (v,\bar v)
	\Big[ R^{+\dot+}_{[n_1]} 
	R^{+\dot+}_{[n_2]} \Big] (0)
\Big\rangle
=
\\
&	
	\frac{2n_1n_2 (N-2)!}{N!}
	\Bigg\{
	\Bigg|
	\frac{1}{(1-v)^2}
	\Bigg[
		 1 
		 - \frac12 \left( \frac1{n_1} + \frac1{n_2}  \right) (1-v)
		 + \rm{O}(1-v)^2 
	\Bigg]	
	\Bigg|^2
\\
	&\quad\quad
	+
	3
	\Bigg|
	\frac{1}{(1-v)^{\frac43}}
	\Bigg[
		 \frac{ (n_1 + n_2)^2}
		 	{12 
				\left[ 
				3  n_1^2n_2^2 (n_1^2 - n_2^2)^2   
				\right]^{\frac13}
			} 
		 + \frac{n_1 + n_2}{6 n_1 n_2} (1-v)^{\frac13}
		+ \rm{O}(1-v)^{\frac23}
	\Bigg]	
	\Bigg|^2 \Bigg\}
\\
	&\quad\quad
	+ \text{Non-singular Hurwitz blocks}
\end{split}
\end{align}
The factor of 3 in front of the terms $\sim (1-v)^{-\frac43}$ comes from the multiplicity of the function (\ref{channels3}).
The leading coefficients give products of structure constants in
$
	\bar O^{(1\pm,1\pm)}_{[2]} 
	\times
	O^{(1\pm,1\pm)}_{[2]}
\sim \{\id\} +  \{ \s_{[3]} \}
$
Note that, although the NS chirals' OPEs form a ring \cite{Jevicki:1998bm,Dabholkar:2007ey,Pakman:2009ab}, here the OPE is not between two chirals, but between a chiral and an anti-chiral field, which explains why the $\s_3$ block is not forbidden.

For the OPE  
$
O^{(1\pm,1\pm)}_{[2]} \times  [ R^{+\dot+}_{[n_1]}  R^{+\dot+}_{[n_2]} ]
$
in channel $v \to 0$, 
we find the following conformal weights for the operators $Y_{[n_1+n_2]}$ 
and $W_{[n_1 - n_2]}$
\begin{align}
h_Y = \frac{n_1+n_2}{4} + \frac{4}{n_1+n_2} ,
\qquad
h_W = \frac{n_1- n_2}{4} + \frac{1}{n_1-n_2} 
\end{align}
suggesting that these are fractional-mode excitations of Ramond ground states in  $(n_1\pm n_2)$-twisted strands. 
This should be expected, since the OPE of a NS field with a Ramond field is always in the Ramond sector.

\subsubsection{Composite NS chiral and interaction modulus}

In \cite{Lima:2021wrz} we have discussed four-point functions with $\Oint$ and composite Ramond fields. 
Here our generalized formula (\ref{Gcofx}) allow us to take the composite fields to be be made of NS chirals.
For example, for highest-weight chirals we find
\be	\label{O22OintO22}
A^{\rm{int}|2|2}_{n_1,n_2}(x)
	= 
	\frac1{16n_1^2}
	\frac{
	x^{1 - n_1 + n_2}
	(x-1)^{2+n_1+n_2}
	(x + \frac{n_1}{n_2})^{2-n_1-n_2}
	(x + \frac{n_1 - n_2}{n_2})^{1 + n_1 - n_2}
	}{ (x + \frac{n_1-n_2}{2n_2})^4}
\ee
and expanding the vacuum and $\s_3$ blocks, 
\begin{align}
\begin{split}
&
\Big\langle 
	\Big[ \bar O^{(2,2)}_{[n_1]} \bar O^{(2,2)}_{[n_2]} \Big] (\infty)
	\Oint (1)
	\Oint (v,\bar v)
	\Big[ O^{(2,2)}_{[n_1]} O^{(2,2)}_{[n_2]} \Big] (0)
\Big\rangle =
\\
&	
	\frac{2n_1n_2 (N-2)!}{N!}
	\Bigg\{
	\Bigg|
	\frac{1}{(1-v)^2}
	\Bigg[
		 1 
		 - \frac{1}{192} 
		 \left( \frac{5}{n_1^2} + \frac{5}{n_2^2}  
		 	+ \frac{6}{n_1n_2} 
			- 16 
		\right) (1-v)^2
		+ \rm{O}(1-v)^3
	\Bigg]	
	\Bigg|^2
\\
&	+
	\Bigg|
	\frac{1}{(1-v)^{\frac43}}
	\Bigg[
		 \frac{ (n_1 + n_2)^2}
		 	{12 
				\left[ 
				3  n_1^2 n_2^2 (n_1^2 - n_2^2)^2   
				\right]^{\frac13}
			}
\\
&\qquad\qquad\qquad\quad
	-
 	\left(  \frac{(n_1^2 - n_2^2)^2}{3^2 n_1^4 n_2^4} \right)^{\frac13}
	\frac{7 (n_1^2 + n_2^2) - 10 n_1 n_2}{80 (n_1 - n_2)^2}
		(1-v)^{\frac23}	
		+ \rm{O}(1-v)^{\frac33}
	\Bigg]	
	\Bigg|^2
	\Bigg\}
\\
&	
	+ \text{Non-singular Hurwitz blocks}
\end{split}
\end{align}
An important difference between the expansion above and (\ref{s3exprro1}) is the absence of the term of order $(1-v)$ in the identity block, and of the term of order $(1-v)^{\frac13}$ in the $\s_3$ block.
Hence there are no operators with $h=1$ in the OPE, a confirmation that $\Oint$ is, indeed, exactly marginal --- it does not couple to other operators of weight 1.
This is also found in functions with Ramond ground states \cite{Lima:2021wrz}.

\subsubsection{Functions with only NS chiral fields}

If we take every field in the correlator to be an NS chiral, the resulting function is constrained by the NS chiral ring.
Only a restricted number of three-point functions involving (single-cycle) NS chirals is non-vanishing \cite{Jevicki:1998bm,Dabholkar:2007ey,Pakman:2009ab}, and the OPEs of fields in the ring are non-singular. This reflects on the structure of the functions (\ref{GenricAapp}) at $x = 0$ and $x = - \frac{n_1}{n_2}$, i.e.~at the $v \to 0$ channel. 
Namely, powers of $x$ and $(x+ \frac{n_1}{n_2})$ are positive, so that there are no singularities, or zero, when the corresponding field is absent from the OPE.
These features can be seen in the list of formulas (\ref{NSfunctionappD}).

For example, using a schematic notation, we have
\be	\label{O22O22O1pm}
\Big\langle 
	\Big[ \bar O^{(2,2)}_{[n_1]} \bar O^{(2,2)}_{[n_2]} \Big] 
	\bar O^{(1+,1+)}_{[2]}
	O^{(1+,1+)}_{[2]}
	\Big[ O^{(2,2)}_{[n_1]} O^{(2,2)}_{[n_2]} \Big] 
\Big\rangle
	=
	\frac1{16n_1^2}
	\frac{
	x
	(x + \frac{n_1}{n_2})^{4}
	(x + \frac{n_1 - n_2}{n_2})
	}{ (x + \frac{n_1-n_2}{2n_2})^4}
\ee
which vanishes both at $x \to 0$ and $x \to - \frac{n_1}{n_2}$. Hence the OPE
$	O^{(1+,1+)}_{[2]}
	\times
	[ O^{(2,2)}_{[n_1]} O^{(2,2)}_{[n_2]} ] 
$
is void. This is not surprising, as there is no OPE 
$O^{(1+,1+)}_{[n]} \times O^{(2,2)}_{[m]}$
in the (single-cycle) NS chiral ring.

By contrast, the function
\be	\label{4ptO00s}
\Big\langle 
	\Big[ \bar O^{(0,0)}_{[n_1]} \bar O^{(0,0)}_{[n_2]} \Big] 
	\bar O^{(0,0)}_{[2]}
	O^{(0,0)}_{[2]}
	\Big[ O^{(0,0)}_{[n_1]} O^{(0,0)}_{[n_2]} \Big] 
\Big\rangle
	=
		- \frac{1}{4 n_1}
	\frac{ ( 1 - x)^2}{x + \frac{n_1 - n_2}{2n_2}}
\ee
is finite at both limits.  Eqs.(\ref{dimhY})-(\ref{dimhW}) give the dimensions
$h_Y = \frac12(n_1 + n_2 - 1)$
and
$h_W = \frac12 (n_1 - n_2 + 1)$.
The former is the correct dimension of a lowest-weight NS chiral of twist $n_1+n_2$, and the latter of a highest-weight chiral of twist $n_1 - n_2$.
Hence the OPE (\ref{OPEWns}) reads
\be
O^{(0,0)}_{[2]}
	\times
\Big[ O^{(0,0)}_{[n_1]} O^{(0,0)}_{[n_2]} \Big] 
	= 
	{\scr C}_1
	\big\{ O^{(0,0)}_{[n_1 + n_2]} \big\}
	+ 
	{\scr C}_2
	\Big\{
	\Big[ O^{(2,2)}_{[n_1-n_2]} O^{(0,0)}_{[n_2]}O^{(0,0)}_{[n_2]} \Big] 
	\Big\}
\ee
The appearance of $O^{(0,0)}_{[m]}$ and $O^{(2,2)}_{[m]}$ in the OPE with the composite field agrees with what one should expect from the single-cycle OPE of the chiral ring.
The structure constants squared, $|{\scr C}_1|^2$ and $|{\scr C}_2|^2$, can be read from value of (\ref{4ptO00s}) at $x = 0$ and $x = - \frac{n_1}{n_2}$, combined with the multiplicities and the ``dressing'' factor for $N$-dependence:
\begin{align}
\Big|
\Big\langle 
\bar O^{(0,0)}_{[n_1 + n_2]} 
O^{(0,0)}_{[2]}
\Big[  O^{(0,0)}_{[n_1]}  O^{(0,0)}_{[n_2]} \Big] 
\Big\rangle \Big|^2
&= 
\frac{2n_1n_2 (N-2)!}{N!}  \frac{ (n_1 + n_2)^3 }{(2 n_1 n_2)^2} 
\\
\Big|
\Big\langle 
\bar O^{(0,0)}_{[n_1]} 
O^{(0,0)}_{[2]}
\Big[  O^{(2,2)}_{[n_1-n_2]}  O^{(0,0)}_{[n_2]} \Big] 
\Big\rangle \Big|^2
&= 
\frac{2n_1n_2 (N-2)!}{N!} \frac{n_2^2}{(2 n_1)^2 (n_1- n_2)} 
\end{align}

As a third and final example, we consider 
\be	\label{func0022}
\Big\langle 
	\Big[ \bar O^{(2,2)}_{[n_1]} \bar O^{(0,0)}_{[n_2]} \Big]
	\bar O^{(0,0)}_{[2]}
	O^{(0,0)}_{[2]}
	\Big[ O^{(2,2)}_{[n_1]} O^{(0,0)}_{[n_2]} \Big]
\Big\rangle
	=
		- \frac{1}{4 n_1}
	\frac{ x^2}{x + \frac{n_1 - n_2}{2n_2}}
\ee
The function vanishes at $x = 0$, so there is no composite operator with $W_{[n_1-n_2]}$ in the OPE. But it is finite at $x = - \frac{n_1}{n_2}$, with an operator of dimension
$h_Y = \frac12 ( n_1 + n_2 +1)$,
i.e.~the highest-weight NS chiral:
\be
	O^{(0,0)}_{[2]}
	\times
	\Big[ O^{(2,2)}_{[n_1]} O^{(0,0)}_{[n_2]} \Big]
	= 
	{\scr C}_1
	\big\{ O^{(2,2)}_{[n_1 + n_2]} \big\}
\ee
The (square of the) structure constant can be read from by evaluating (\ref{func0022}) at $x = - \frac{n_1}{n_2}$ and using the multiplicity and dressing factor:
\be	\label{O22O00O22O00}
\Big|
\Big\langle 
\bar O^{(2,2)}_{[n_1 + n_2]} 
O^{(0,0)}_{[2]}
\Big[  O^{(2,2)}_{[n_1]}  O^{(0,0)}_{[n_2]} \Big] 
\Big\rangle \Big|^2
= 
\frac{2n_1n_2 (N-2)!}{N!}  \frac{ n_1^2 }{(2n_2)^2 (n_1+n_2)} 
\ee
If we take $n_2 = 1$ and $n_1=n >1$, the lowest-weight chiral in the composite field becomes the vacuum.
The $N$-dependent dressing factor, which is proportional to 
$| \Cent[(n_1)(n_2) (1)^{N-n_1-n_2}| = n_1n_2(N-n_1 - n_2)!$, 
becomes proportional to
$| \Cent[(n) (1)^{N-n}| = n (N-n)!$, 
so we must multiply (\ref{O22O00O22O00}) by a factor of $(N-n-1)$ to obtain the result
\be
\Big|
\Big\langle 
\bar O^{(2,2)}_{[n + 1]} 
O^{(0,0)}_{[2]}
  O^{(2,2)}_{[n]}  
\Big\rangle \Big|^2
= 
\frac{(N-2)! (N-n-1)}{N!}  \frac{ n^3 }{2 (n+1)} 
\ee
This matches precisely with a known structure constant computed, e.g.~in 
\cite{Dabholkar:2007ey}, providing a very non-trivial check of our results.

\subsection{The effect of spectral flow} \label{SectSpecFlow}

The ${\cal N} = 4$ superconformal algebra has an automorphism called `spectral flow' \cite{Schwimmer:1986mf}. The currents are transformed, and fermionic modes (and boundary conditions) are changed by a continuous parameter usually called spectral flow `units'.
Flow by $\xi$ units affects the R-charge and the Virasoro currents in such a way that the weight and R-charge of a field changes as
\be	\label{hjSFmap}
h \mapsto h_\xi = h + \xi j + \tfrac{c}{24} \xi^2 ,
\qquad
j \mapsto j_\xi = j + \tfrac{c}{12} \xi ,
\ee
while the super-currents $G^{\a A}(z)$ have their modes shifted by $\pm \frac12 \xi$. 
Since every NS chiral has $h = j$, their spectral flow by $\xi = -1$ gives a field with $h_{-1} = \frac1{24} c$, that is a Ramond ground state. Which NS chiral flows to which Ramond ground state is seen from the R-charges. For example, in the $n$-twisted sector, with $c = 6n$, the lowest weight NS field 
$O^{(0)}_{(n)}$ 
has R-charge $j = \frac{n-1}{2}$, so it flows to the Ramond ground state $R^-_{(n)}$, with R-charge
$j = \frac{n-1}{2} + \frac{6n}{12}\xi = - \frac12$. 
Overall, 
\be	\label{NStoRam}
\begin{aligned}
\ket{O^{(0)}_{[n]}} \xrightarrow[\ \xi = -1 \ ]{} \ket{R^-_{[n]}} ,
\qquad
\ket{O^{(2)}_{[n]}} \xrightarrow[\ \xi = -1 \ ]{} \ket{R^+_{[n]}} ,
\qquad
\ket{O^{(1\pm)}_{[n]}} \xrightarrow[\ \xi = -1 \ ]{} \ket{R^{\dot A}_{[n]}} 
\\
\ket{\bar O^{(0)}_{[n]}} \xrightarrow[\ \xi = -1 \ ]{} \ket{R^+_{[n]}} ,
\qquad
\ket{\bar O^{(2)}_{[n]}} \xrightarrow[\ \xi = -1 \ ]{} \ket{ R^-_{[n]} },
\qquad
\ket{\bar O^{(1\mp)}_{[n]}} \xrightarrow[\ \xi = -1 \ ]{} \ket{R^{\dot A}_{[n]}} 
\end{aligned}
\ee
Naturally, spectral flow relates pairs of functions involving these fields.
In fact, it is usual in the literature on the D1-D5 CFT to compute three-point functions with fields on the NS sector, and then relate these to functions on the Ramond sector (where SUGRA states live) via spectral flow; see for example \cite{Avery:2010qw,Avery:2009tu}.

Given a state $\ket{\Psi}$, the automorphism of the Hilbert space will map it to another state $\ket{\Psi}_\xi$, while an operator ${\scr O}(z)$ will be mapped to  ${\scr O}_\xi(z)$, with a linear operator $U_\xi$ such that
\be	\label{PsiscrOSF}
\ket{\Psi}_\xi = U_\xi \ket{\Psi} ,
\qquad
{\scr O}_\xi(z) = U_\xi {\scr O}(z) U_\xi^{-1},
\ee
preserving amplitudes $\bra{\Psi} {\scr O} \ket{\Psi}$. 
In the free orbifold CFT, the linear operator has a natural implementation in terms of the bosonized fermions, 
inserted at the origin (i.e.~at past infinity), 
\be	\label{UxiSF}
U_\xi(z) = \exp \left(  \frac{i\xi}{2}  \sum_{I=1}^N \Big[ \phi_{1,I}(z) - \phi_{2,I}(z) \Big] \right) ,
\qquad
U_\xi = U_\xi(0) .
\ee
This is an $S_N$-invariant operator, including all copies $I =1,\cdots, N$ of the free bosons that bosonize the fermions.
Moving $U_\xi$ past a bare twist $\s_{g}$, for any $g \in S_N$, only has the effect of shuffling the copies, which leaves $U_\xi$ invariant, hence 
$U_\xi \s_{g} = \s_{g} U_\xi$. Bosons also commute with $U_\xi$. 
Let ${\scr O}(z)$ be a primary fermionic field which can be written in bosonized language as an exponential of a linear combination of the $\phi_{r,I}$, 
multiplied (or not) by a bare twist $\s_g$. The most important examples of such fields are the composite NS chirals (\ref{NScomp}) and Ramond ground states (\ref{RamondComp}). 
Commutation with $U_\xi$ is then 
\be	\label{CommOU}
{\scr O} (z) U_\xi(0) 
	= z^{- j \xi} \ U_\xi(0) {\scr O}(z) ,
\quad \text{hence} \quad
{\scr O}_\xi (z) = U_\xi {\scr O}(z) U_\xi^{-1} = z^{j \xi} {\scr O}(z) .	
\ee
Here $j$ is the R-charge of ${\scr O}$.
The first equation follows from  the commutation of $U_\xi$ and $\s_g$, along with the well-known formula (see e.g.~\cite{francesco2012conformal}) for commuting a pair of exponentials: 
$
e^{k_a\phi_a(z)} e^{k'_b \phi_b(z')}
=  
e^{k_a k'_b \langle \phi_a(z) \phi_b(z') \rangle}
e^{k'_b \phi_b(z')} e^{k_a\phi_a(z)}
$,
where  $k_a, k_b' \in {\mathbb C}$, there are sums over $a,b$,
and the c-number exponential in the r.h.s.~includes the two-point function $\langle \phi_a(z) \phi_b(z') \rangle = \delta_{ab} \log(z-z')$, valid for our bosons; cf.~Eq.(\ref{2ptFuncpsipsi}).
The second equation in (\ref{CommOU}) also uses that
$U_\xi^{-1} = U_{-\xi} = U^\dagger_\xi$,
as readily seen from the explicit realization (\ref{UxiSF}).
Since $U_\xi$ commutes with bare twists and bosons, which are R-neutral,  formulas (\ref{CommOU}) actually hold for these fields as well.

To confirm that $U_\xi$ in (\ref{UxiSF}) is indeed the correct  operator leading to (\ref{hjSFmap}), we can look at how it affects the weight and the charge of a state $\ket{\scr O} = {\scr O}(0) \ket{\varnothing}$ generated by an operator that transforms as (\ref{CommOU}).
According to (\ref{PsiscrOSF}), we have
\begin{align}	\label{ketOxi}
\ket{\scr O}_\xi 
	= U_\xi \ket{\scr O}
	= \lim_{z \to 0} \left( U_\xi {\scr O}(z) U_\xi^{-1} \right) U_\xi \ket{\varnothing}
	= \lim_{z \to 0} z^{j \xi} {\scr O}(z) U_\xi \ket{\varnothing} .
\end{align}
The dimension of the state in the r.h.s. is a sum of the dimensions of ${\scr O}$ and $U_\xi$, plus a factor of $j\xi$ coming from $z^{j\xi}$. Since the exponential (\ref{UxiSF}) has weight $h = \frac{c}{24} \xi^2$ and R-charge $j = \frac{c}{12} \xi$, $c = 6N$, the result matches (\ref{hjSFmap}).
Alternatively, we can explicitly write the most general possible exponential
and take the OPE with (\ref{UxiSF}), 
\begin{align*}
{\scr O}(z) &= \exp \left( \tfrac{i}2 \textstyle\sum_{I = 1}^N \big[ \a_I \phi_{1,I}(z) + \b_I \phi_{2,I}(z) \big] \right) \s_g(z) ,
\ \ \text{with} \ \
j =  \textstyle\sum \tfrac14 (\a_I - \b_I) ,
\\
{\scr O}(z) U_\xi(0)
	&= z^{- j \xi}  \
	\exp \left( \tfrac{i}2 \textstyle\sum_{I = 1}^N \big[ (\a_I + \xi) \phi_{1,I} + (\b_I - \xi) \phi_{2,I} \big] \right) \s_g(0) .
\end{align*}
So the factor $z^{j\xi}$ cancels in Eq.(\ref{ketOxi}), giving
\be	\label{ketOxiExp}
\ket{\scr O}_\xi 
	= \exp \left( \frac{i}2 \sum_{I = 1}^N \Big[ (\a_I + \xi) \phi_{1,I}(0) + (\b_I - \xi) \phi_{2,I}(0) \Big] \right) \s_g(0) \ket{\varnothing} .
\ee
The weight and the charge of this last exponential again agree with (\ref{hjSFmap}). Further, by looking at (\ref{RamondFields})-(\ref{NSchirals}), Eq.(\ref{ketOxiExp}) explicitly reproduces the map (\ref{NStoRam}) between Ramond ground states and NS chiral states.

If we are considering just a specific $n$-twisted sector of Hilbert space generated by a bare twist $\s_{(n)}$, the sums over copies $I = 1, \cdots , N$ in all exponentials above can be replaced by sums over only the $n$ copies in the corresponding cycle $(n)$, say $I = 1, \cdots, n$.
This is possible because fields in different copies commute, so the normal-ordered exponential in (\ref{UxiSF}) can be readily factorized.%
	\footnote{%
	The factors of $U_\xi$ made by the copies $I'$ that do \emph{not} enter the operator ${\scr O} \sim \s_{(n)}$, they act on $\ket{\varnothing}$ to create a tensor product of untwisted Ramond fields $\ket{R^-_{I'}}$.
	}
We can in fact repeat the argument above, using these restricted sum over copies, to derive the transformation of the single-cycle fields (\ref{RamondFields})-(\ref{NSchirals}) more directly.
This restricted version of the $U_\xi$ operator also defines a notion of spectral flow on the individual $n$-twisted sectors (or $n$-twisted ``strands''), where the transformations (\ref{hjSFmap}) hold with $c = 6n < 6N$.
Although quite useful for some computations on the free orbifold, these individual flows are broken when the theory is deformed by $\Oint$, because its twist mixes different sectors, as discussed in \cite{Lima:2020kek}. Only the full spectral flow of the $c = 6N$ theory, involving all $N$ copies simultaneously, is preserved.

In order to relate our four-point functions by spectral flow, it is convenient to regard them as two-point functions on non-trivial states.
We can be rather general: consider a state $\ket{{\scr X}}$, created by an operator ${\scr X}(z)$ which transforms as in (\ref{CommOU}). 
Now consider the expectation value of a pair of conjugate operators $Z$ and $\bar Z$ on the flowed state $\ket{\scr X}_\xi = U_\xi \ket{\scr X}$.
Using the transposition property (\ref{CommOU}) twice, 
\begin{align}	\label{XZZXxi}
\begin{aligned}
\prescript{}{\xi}{\bra{\scr X}} \bar Z(1) Z(v) \ket{\scr X}_\xi
	&= \bra{\scr X} U_\xi^\dagger \  \bar Z(1) Z(v) \ U_\xi \ket{\scr X}
\\
	&= 
	v^{\xi j_Z} \bra{\scr X} U_\xi^\dagger \  \bar Z(1) \ U_\xi \ Z(v) \ket{\scr X}
\\
	&= 
	v^{\xi j_Z} \bra{\scr X} U_\xi^\dagger U_\xi \  \bar Z(1)   Z(v) \ket{\scr X}
\\
	&= 
	v^{\xi j_Z} \bra{\scr X} \bar Z(1)   Z(v) \ket{\scr X}
\end{aligned}
\end{align}
where $j_Z$ is the R-charge of $Z$, and passing $U_\xi$ over $\bar Z$ at $z =1$ gives a trivial factor. In the last line, we used that 
$U^\dagger_\xi = U_{-\xi} = U_\xi^{-1}$.
This computation relates correlators of the fields $Z$ and $\bar Z$ on different states $\ket{\scr X}$ and $\ket{\scr X}_\xi$. But looking at the r.h.s.~of the first line, we see that if we insert $\id = U_\xi U_\xi^{-1}$ between fields, to get
$
\bra{\scr X} ( U_\xi^{-1}  \bar Z(1) U_\xi ) ( U_\xi^{-1} Z(v) \ U_\xi ) \ket{\scr X} 
$, 
we can also find a relation between functions with flowed \emph{operators} on the (fixed) state $\ket{\scr X}$. In summary,
\begin{align}	\label{XZZXxiB}
\begin{aligned}
\bra{\scr X} \bar Z_\xi(1)   Z_\xi(v) \ket{\scr X}	
	=
	\prescript{}{\xi}{\bra{\scr X}} \bar Z(1) Z(v) \ket{\scr X}_\xi
	= v^{\xi j_Z} \bra{\scr X} \bar Z(1)   Z(v) \ket{\scr X} .
\end{aligned}
\end{align}

We can now apply these results to four-point functions of the type (\ref{Ahhll}), where the $Z$ fields carry a twist $\s_{[2]}$, and the states $\ket{{\scr X}}$ are created by the multi-cycle fields (\ref{MultiscrO}). 
Factorization lets us consider only the functions with double-cycle states in Eq.(\ref{Axxzz}), so we focus on the four-point functions (\ref{AZv}), which are given by the master formulas computed in \S\ref{SectMasterForm}.
We will omit the various indices of $\AmZ{v}$ for economy: 
\be	\label{AZvBe}
A(v) =
	\Big\langle
	\big[
	\bar X_{[n_1]}^{\{\hat \sigma , \hat \varrho \}} 
	\bar X_{[n_2]}^{\{ \check \sigma , \check \varrho \}} 
	\big] \Big|
	\
	\bar Z_{[2]}^{\{\a,\b\}} (1) 
	\
	Z_{[2]}^{\{\a,\b\}} (v)
	\
	\Big| \big[
	X_{[n_1]}^{\{\hat \sigma , \hat \varrho \}}
	X_{[n_2]}^{\{ \check \sigma , \check \varrho \}} 
	\big] 
	\Big\rangle 
\ee
and 
\be
A_\xi(v) =
	\prescript{}{\xi}{\Big\langle}
	\big[
	\bar X_{[n_1]}^{\{\hat \sigma , \hat \varrho \}} 
	\bar X_{[n_2]}^{\{ \check \sigma , \check \varrho \}} 
	\big] \Big|
	\
	\bar Z_{[2]}^{\{\a,\b\}} (1) 
	\
	Z_{[2]}^{\{\a,\b\}} (v)
	\
	\Big| \big[
	X_{[n_1]}^{\{\hat \sigma , \hat \varrho \}}
	X_{[n_2]}^{\{ \check \sigma , \check \varrho \}} 
	\big] 
	\Big\rangle_\xi 
\ee
are related as in (\ref{XZZXxiB}).
The R-charge of $Z_{[2]}^{\{\a,\b\}}$ is $j_Z = \tfrac14 (\a-\b)$, hence
\be
A_\xi(v) = 	v^{\frac{(\a - \b)\xi}{4}} A(v) .
\ee
The functions (\ref{GeSLcover}) are written in terms of $x$, that should be related to $v$ by the inverse covering maps $x_{\frak a}(v)$ and Eq.(\ref{A12sol}). Using Eq.(\ref{zxeqv}), we then have
\be	\label{AAxix}
\begin{split}
A_\xi(x) 
&= 
	x^{\frac14 (\a - \b)(n_1-n_2)\xi}
	(x-1)^{- \frac14 (\a - \b)(n_1+n_2)\xi}
\\	
&\qquad\qquad
	\times
	(x + \tfrac{n_1}{n_2})^{\frac14 (\a - \b)(n_1+n_2)\xi}
	(x + \tfrac{n_1}{n_2} -1)^{-\frac14(\a - \b)(n_1-n_2)\xi}
	\
	A(x) .
\end{split}
\ee	
Written this way, the shift in the exponents $K_i$ in (\ref{expontK}) is explicit.
	Let us emphasize that, although Eq.(\ref{AAxix}) is parameterized by $x$, we are performing a standard spectral flow on the base sphere.%
	\footnote{%
	Variants of the original \cite{Schwimmer:1986mf} automorphism of the superconformal algebra are known, e.g.~the `fractional spectral flow' related to fractional modes in twisted sectors \cite{deBeer:2019ioe}, and the recently introduced ``partial spectral flow'' \cite{Guo:2021uiu} that changes only two of the four fermions.}

For example, consider the function in Eq.(\ref{4ptO00s}), with only lowest weight NS chirals 
\be	\label{4ptO00sRel}
\Big\langle 
	\big[ \bar O^{(0)}_{[n_1]} \bar O^{(0)}_{[n_2]} \big] 
	\Big|
	\ \bar O^{(0)}_{[2]}  \ O^{(0)}_{[2]}  \
	\Big|
	\big[ O^{(0)}_{[n_1]} O^{(0)}_{[n_2]} \big] 
\Big\rangle
	=
		- \frac{1}{4 n_1}
	\frac{ ( 1 - x)^2}{x + \frac{n_1 - n_2}{2n_2}} .
\ee
Here 
$\a = -\b = 1$, see Table \ref{ZNSR}.
(We are using a schematic notation omitting the arguments of operators.) 
If we flow the double-cycle states by $\xi = -1$, we get the double-cycle Ramond state 
$
\ket{[O^{(0)}_{[n_1]} O^{(0)}_{[n_2]}]}_{\xi = -1} = 
\ket{[ R^-_{[n_1]} R^-_{[n_2]} ] }$, 
while flowing the anti-chiral state gives the conjugate Ramond state.
Hence, by Eq.(\ref{AAxix}),
\be
\begin{aligned}
&
\Big\langle 
	\big[ \bar O^{(0)}_{[n_1]} \bar O^{(0)}_{[n_2]} \big] 
	\Big|
	\ \bar O^{(0)}_{[2]}  \ O^{(0)}_{[2]} \
	\Big|
	\big[ O^{(0)}_{[n_1]} O^{(0)}_{[n_2]} \big] 
\Big\rangle \Big|_{\text{states flowed by $\xi = -1$}}
\\
&\qquad\quad
=
	- \frac{1}{4 n_1}
	\frac{ 
	x^{- \frac{n_1-n_2}{2} }
	(x -1)^{2 + \frac{n_1 + n_2}{2}}
	(x + \tfrac{n_1}{n_2})^{- \frac{n_1+n_2}{2} }
	(x + \tfrac{n_1}{n_2} -1)^{\frac{n_1-n_2}{2} }	
	}{x + \frac{n_1 - n_2}{2n_2}}
\\
&\qquad\quad
=
\Big\langle 
	\big[ \bar R^-_{[n_1]} \bar R^-_{[n_2]} \big] 
	\Big|
	\ \bar O^{(0)}_{[2]} \ O^{(0)}_{[2]} \
	\Big|
	\big[ R^-_{[n_1]} R^-_{[n_2]} \big] 
\Big\rangle	
\end{aligned}
\ee
the same result that we find if we apply the master formula (\ref{GenricAapp}) directly to the function in the last line.
As another example, take the function (\ref{O22O22O1pm}),
\be
\Big\langle 
	\big[ \bar O^{(2)}_{[n_1]} \bar O^{(2)}_{[n_2]} \big] \Big| 
	\ \bar O^{(1+)}_{[2]} \ 	O^{(1+)}_{[2]} \
	\Big| \big[ O^{(2)}_{[n_1]} O^{(2)}_{[n_2]} \big] 
\Big\rangle
	=
	\frac1{16n_1^2}
	\frac{
	x
	(x + \frac{n_1}{n_2})^{4}
	(x + \frac{n_1}{n_2} - 1)
	}{ (x + \frac{n_1-n_2}{2n_2})^4} .
\ee
Now $\a = 3$, $\b = -1$.
The flowed state is 
$\ket{[ R^+_{[n_1]} R^+_{[n_2]} ]}$, 
and formula (\ref{AAxix}) gives
\be
\begin{aligned}
&
\Big\langle 
	\big[ \bar O^{(2)}_{[n_1]} \bar O^{(2)}_{[n_2]} \big] \Big| 
	\ \bar O^{(1+)}_{[2]} \ O^{(1+)}_{[2]} \
	\Big| \big[ O^{(2)}_{[n_1]} O^{(2)}_{[n_2]} \big] 
\Big\rangle
	\Big|_{\text{states flowed by $\xi = -1$}}
\\
&\qquad\quad
=
	\frac1{16n_1^2}
	\frac{ 
	x^{1 -n_1+n_2}
	(x -1)^{n_1 + n_2}
	(x + \tfrac{n_1}{n_2})^{4 - n_1- n_2}
	(x + \tfrac{n_1}{n_2} -1)^{1 + n_1-n_2}	
	}{(x + \frac{n_1-n_2}{2n_2})^4}
\\
&\qquad\quad
=
\Big\langle 
	\big[ \bar R^+_{[n_1]} \bar R^+_{[n_2]} \big] 
	\Big|
	\ \bar O^{(1+)}_{[2]} \ O^{(1+)}_{[2]} \
	\Big|
	\big[ R^+_{[n_1]} R^+_{[n_2]} \big] 
\Big\rangle	
\end{aligned}
\ee
which is, again, what we find using the master formula (\ref{GenricAapp}) directly.

The interaction operator is more complicated than the exponential operator for which we have derived the transformation (\ref{CommOU}) above, but it has been shown \cite{Burrington:2012yq,Guo:2021uiu} that $\Oint$ is, in fact, invariant under spectral flow,%
	\footnote{%
	Here we mean the usual, ``original'' spectral flow; in \cite{Guo:2021uiu} the authors also discuss a ``partial'' spectral flow, under which $\Oint$ is (crucially) \emph{not} invariant.}
hence it actually does obey (\ref{CommOU}), being R-neutral.
Now, the first equation in the chain of equalities (\ref{XZZXxiB}) tells us that four-point functions including $\Oint$ and states related by spectral flow must be equal. For example, based solely  on spectral flow applied to the function (\ref{O22OintO22}), we conclude that
\begin{align}
\begin{aligned}
&
\Big\langle 
	\big[ \bar O^{(2)}_{[n_1]} \bar O^{(2)}_{[n_2]} \big] \Big|
	\Oint \ \Oint 
	\Big| \big[ O^{(2)}_{[n_1]} O^{(2)}_{[n_2]} \big]
\Big\rangle
\\
&\qquad
=
	\frac1{16n_1^2}
	\frac{
	x^{1 - n_1 + n_2}
	(x-1)^{2+n_1+n_2}
	(x + \frac{n_1}{n_2})^{2-n_1-n_2}
	(x + \frac{n_1 - n_2}{n_2})^{1 + n_1 - n_2}
	}{ (x + \frac{n_1-n_2}{2n_2})^4}
\\
&\qquad
=
\Big\langle 
	\big[ \bar O^{(2)}_{[n_1]} \bar O^{(2)}_{[n_2]} \big] \Big|
	\Oint \ \Oint 
	\Big| \big[ O^{(2)}_{[n_1]} O^{(2)}_{[n_2]} \big]
\Big\rangle	
	\Big|_{\text{states flowed by $\xi = -1$}}
\\
&\qquad
=
\Big\langle 
	\big[ \bar R^{+}_{[n_1]} \bar R^+_{[n_2]} \big] \Big|
	\Oint \ \Oint 
	\Big| \big[ R^+_{[n_1]} R^+_{[n_2]} \big]
\Big\rangle
\end{aligned}
\end{align}
This is, indeed, the correct function for Ramond fields found by the master formula, and previously known from \cite{Lima:2020nnx} (see Eq.(61) ibid.).

\section{Discussion and further developments}

The present paper tries to contribute to a problem that is particularly important for the fuzzball conjecture:
the complete description of the D1-D5 CFT at the free orbifold point and away from it. This requires the derivation of all three- and four-point functions involving the symmetric orbifold's Ramond and NS fields (and some of their excitations), the complete list of their OPEs and the full spectrum of the non-BPS fields that might appear at the OPE channels.

We have given here a detailed description of twisted $Q$-point functions in $M^N / S_N$ orbifolds, applying a technology of \cite{Pakman:2009zz} to correlators with multi-cycle twisted fields.
We have thoroughly analyzed a special class of relatively simple four-point functions where all operators are twisted: two being composite, multi-cycle fields and two being single-cycle fields with twists of length 2. We showed how to decompose these functions into connected parts where the multi-cycle fields are reduced to double-cycle fields, then studied these connected functions, with a detailed discussion of the geometry of the genus-zero covering surfaces. %

$Q$-point functions with multi-cycle fields are disconnected, and can become rather complicated. 
Even extracting the large-$N$ dependence is a task that strongly depends on the types of twist in the composite fields.
We have shown that if the fields are composite but have a finite number of cycles, i.e.~if the number of cycles does not grow with $N \to \infty$, then the function scales as $\sim \sum_\chi N^{ \frac12 ( \chi - R)}$, which is a natural generalization of the well-known formula $\sim \sum_{\bf g} N^{- {\bf g} + 1 - \frac12 Q}$ for connected functions, the genus ${\bf g}$ replaced by the Euler characteristic $\chi$.
But if the number of cycles in the composite field grows with $N$, this generalized formula does not apply. This happens for important types of composite fields, like Ramond ground states $[ (R_{[n]} )^{N  / n}]$, with $n$ fixed, that source well-known Lunin-Mathur geometries \cite{Lunin:2001fv}.
In our examples of functions involving these types of field, the total $N$-dependence comes from computing the $N$-dependent number of factorizations of the total correlator into connected parts.
This factorization strongly depends on the structure of the twists involved in the function. Here the non-composite fields are simple twist-2 single-cycle fields, which yield a manageable result. It would be interesting to try to find a way of determining the $N$-dependence in a more general way.
It would also be important to explore the connection of our results with those of \cite{Ceplak:2021wzz}.

After reducing the factorized multi-cycle four-point function into a sum of connected functions with a finite number of cycles (in our example, the remaining composite field has at most two cycles), we can use covering surfaces methods.
The full $S_N$-invariant correlator is a sum of `Hurwitz blocks', each associated with one of the ${\bf H}$ allowed topologies of covering surfaces, where ${\bf H}$ is a Hurwitz number.
Different types of coalescences of ramification points in these surfaces dictate the resulting twists of operators that appear in the OPE channels of the four-point function. 
Twists configurations can restricts the correlators to such an extent that, for special classes of functions subject to other restraints, e.g.~the ring of NS chiral fields in the D1-D5 CFT, Hurwitz theory may suffice to fix the correlators completely \cite{Pakman:2009ab}.
We would like to explore the structure of Hurwitz blocks in more generality, as well as their connection with conformal blocks.

Since many four-point functions involving untwisted light fields are already known, let us mention some uses of the functions with twisted light NS fields we have calculated. 
One possible application is in the reconstruction of S-matrix elements of a process of absorption and emission of light (or massless) quanta from the heavy object in the bulk, as suggested in \cite{Lunin:2012gz}.
Also, our correlators can be used for deriving functions with $\frac18$-BPS operators, relevant for 3-charge microstate solutions \cite{Giusto:2004id,Giusto:2004ip}. These operators are chiral excitations of Ramond ground states, and the corresponding functions can be obtained from derivatives of the functions derived here, using Ward identities. 
Many particular examples of such correlators are known in the context of D1-D5-P superstrata bulk geometries.
For example, in \cite{Bombini:2019vnc} it is shown that the Ward identity for the simplest Virasoro excitation $L_{-1}$
amounts to applying a differential operator $D_v$ to the function of unexcited fields,%
	\footnote{%
	See Eqs.(5.2)-(5.4) of Ref.\cite{Bombini:2019vnc}; their variable $z$ corresponds to our $v$.
	}
\be
D_v = (1-v)^2 \frac{\pa}{\pa v}  \left( v \frac {\pa}{\pa v} \right) +1 .
\ee
As our four-point functions are known in closed form only in terms of the covering-surface variables $x$, $\bar x$, the question arises of whether we could translate this Ward identity to a differential operator in terms of $x$ instead of the base-sphere anharmonic ratio $v$. The answer is rather simple: since we do know the mapping function $v(x)$ explicitly, we can rewrite $D_v$ as an operator $\tilde D_x $ acting on our functions $A (x,\bar x)$, 
\be
\tilde {D}_x A (x,\bar x)
= 
\left[ \frac{ \{ 1-v(x) \}^2}{v'(x)} \frac{\pa}{\pa x} \left( \frac{v(x)}{v'(x)} \frac{\pa}{\pa x} \right) + 1 \right]  A(x,\bar x) ,
\ee
where $v'(x) = d v/dx$.
Therefore the problem of reconstructing four-point functions with excited states from our correlators --- even in more complicated cases involving also other generators, say $J^+_{-1}$ and integer powers of it --- is rather straightforward. 
Let us note, as a last comment, that once these functions are known, the methods of \cite{Lima:2020kek,Lima:2021wrz} can be used: one computes integrals of the four-point functions with the deformation $\Oint$ to find the anomalous dimension of the heavy fields at second order in conformal perturbation theory. 
Thus we may assess the renormalization or the protection of the excited states.

\bigskip

\noindent
{\bf Acknowledgments}
\\
The work of M.S.~is partially supported by the Bulgarian NSF grant KP-06-H28/5 and that of M.S.~and G.S.~by the Bulgarian NSF grant KP-06-H38/11.
M.S. is grateful for the kind hospitality of the Federal University of Esp\'irito Santo, Vit\'oria, Brazil, where part of his work was done.
We would like to kindly thank an anonymous referee for comments leading to the improvement of the text, in particular the addition of a discussion about spectral flow.

\appendix

\section{Conventions for the D1-D5 CFT}	\label{SectN44CFT}

Here we gather definitions and notations for the seed ${\cal N} = (4,4)$ CFT. In general, we follow \cite{Lima:2021wrz}.
The R-symmetry group is $\rm{SU}(2)_L \times \rm{SU}(2)_R$. 
We work with $(\bbT^4)^N/S_N$, and there is an additional global group  $\rm{SU}(2)_1 \times \rm{SU}(2)_2$.
In the superalgebra,
the  R-currents $J^a_I(z)$, $\tilde J^{\dot a}_I(\bar z)$, and supercurrents $G^{\a A}_I (z)$, $\tilde G^{\dot \a \dot A}_I(\bar z)$ have indices in the SU(2) groups:
$a = 1,2,3$ and $\dot a = \dot 1,\dot2,\dot3$ are triplets of SU(2)$_L$ and SU(2)$_R$;
$\a = \pm$ and $\dot \a = \dot \pm$ doublets of SU(2)$_L$ and SU(2)$_R$;  $A=1,2$ and $\dot A=\dot1,\dot2$ doublets of SU(2)$_1$ and SU(2)$_2$, respectively. The index $I = 1, \cdots, N$ distinguishes the $N$ identical copies of the seed SCFT.
Each copy can be realized in terms of four real bosons plus four real holomorphic and four real anti-holomorphic fermions. They are written in complexified form as $X^{\dot AA}_I(z,\bar z)$, $\psi^{\a \dot A}_I(z)$ and $\tilde \psi^{\dot \a\dot A}_I(\bar z)$, respectively.
Fermions can be conveniently bozonized by chiral bosons $\phi_r(z)$ and $\tilde \phi_r(\bar z)$, 
\be
	\begin{bmatrix}
	\psi^{+ \dot 1}_I (z) \\ \psi^{- \dot 1}_I (z) 
	\end{bmatrix}
	=
	\begin{bmatrix}
	 e^{- i \phi_{2,I}(z)} \\ e^{- i \phi_{1,I}(z)}  
	\end{bmatrix} \ ,
\qquad
	\begin{bmatrix}
	\psi^{+ \dot 2}_I (z) \\ \psi^{- \dot 2}_I (z) 
	\end{bmatrix}
	=
 	\begin{bmatrix}
	e^{ i \phi_{1,I}(z)} \\  - e^{i \phi_{2,I}(z)}  
	\end{bmatrix} 
\label{Bosnpsi2real}
\ee
and similarly for $\tilde \psi^{\dot\a\dot A}_I(\bar z)$.
Exponentials are \emph{always} normal-ordered throughout the paper. See \cite{Burrington:2012yq,Burrington:2015mfa} for cocycles that we ignore. 
The non-vanishing two-point functions are
\begin{align}
\langle \pa X^{\dot A A}_I (z) \pa X^{\dot B B}_I (z') \rangle 
	&=  \frac{2 \e^{\dot A \dot B} \e^{A B}}{(z - z')^2} ,	\label{2ptFuncpaXAA}
\\
\langle \psi^{\a \dot A}_I (z) \psi^{\b \dot B}_I (z') \rangle 
	&= - \frac{\e^{\a\b} \e^{\dot A \dot B}}{z - z'} 	, 
	\label{2ptFuncpsipsi}
\quad \text{or} \quad
\langle \pa \phi_{r,I}(z) \pa \phi_{s,I}(z') \rangle = - \frac{\delta_{rs}}{(z - z')^2}  
\end{align}
Two-point functions between fields on different copies vanish.
The ``magnetic-components'' $J^3$ of the R-current and ${\frak J}^3$ of the SU(2)$_2$ current
can be most conveniently written in bosonized form,
\be	\label{J3curr}
J^3_I(z) =  \frac{i}{2} \big[  \pa \phi_{1,I} (z)  -   \pa \phi_{2,I} (z)  \big] 	,
\qquad
{\frak J}^3_I(z) =  \frac{i}{2} \big[  \pa \phi_{1,I} (z)  +   \pa \phi_{2,I} (z)  \big] 
\ee
We denote the respective eigenvalues as
$j, {\frak j}$, and the ones in the anti-holomorphic sector as $\tilde \jmath, \tilde {\frak j}$.
	Note that these are ``magnetic'', not ``azimuthal'' quantum numbers.

\section{Derivation of the $N$-dependence of twisted $Q$-point functions}	\label{SectCountinFact}

We now derive the key formulas of Sect.\ref{SectMultiCyclCorr} in detail. As mentioned in the text, we use the technology of \cite{Pakman:2009zz}, but generalized for generic, multi-cycle permutations, and without recurring to diagrams.

\subsection{Two-point functions}	\label{SectCountinFactTwp}

First, we derive the normalization factor ${\scr S}_{[g]}$ of the $S_N$-invariant twist $\s_{[g]}$ in Eq.(\ref{Sninvg}).
We want to compute the two-point function
\be	\label{dobsumqwq}
\big\langle \s_{[g]}(0) \s_{[g']}(1) \big\rangle
\equiv
\big\langle \s_{[g]} \s_{[g']} \big\rangle
\equiv 
	\frac1{ {\scr S}_{[g]}^2}
	\sum_{h\in S_N} \sum_{h' \in S_N} \langle \s_{h g h^{-1}} \ \s_{h' g' h'^{-1}}  \rangle
\ee
We omit the arguments $z = 0$, $z'=1$ for economy of space.
The functions inside the sum, which contain individual elements of $S_N$, vanish unless 
\be	\label{hghghco1}
(h g h^{-1}) \ (h' g' h'^{-1}) = \id .
\ee
The class $[g]$ consists of all permutations with the same cycle type of $g$, including its inverse $g^{-1}$.
So 
$\langle \s_{[g]} \s_{[g']} \rangle = 0$ if $g$ and $g'$ have different cycle structures, i.e.~if $[g] \neq [g']$, hence we take $[g'] = [g]$.
Due to symmetry, all terms in the sum are equal, so we need \emph{the number of non-vanishing terms}, i.e.~terms that satisfy Eq.(\ref{hghghco1}). 

For a fixed element $h$ in the first sum in (\ref{dobsumqwq}), we count the non-vanishing terms in the sum over $h'$.
This is the number of elements $h' \in S_N$ that solve the equation
\be	\label{hghqeq}
 h' g h'^{-1} = q \quad \text{for fixed $g$ and fixed $q = (h  g h^{-1})^{-1} \in S_N$}
\ee
Note that $q \in  [g]$, hence $\exists \ k \in S_N$ such that $q = k g k^{-1}$, and 
\be	\label{pgpeq}
p^{-1} g p = g \quad \text{where $p = h'k$, with fixed $k \in S_N$}
\ee
The number of elements $h'$ which solve (\ref{hghqeq}) is the same as the number of elements $p$ which solve (\ref{pgpeq}). The latter are elements elements of the \emph{centralizer} of $g$
\be	\label{CentDef}
\Cent[g] = \{ p \in S_N \ | \ p g p^{-1} = g \} ,
\ee
whose order
(see e.g.~\cite{sagan2001symmetric}),
\be	\label{OrdCentgAp}
\big| \Cent[g] \big| = \prod_n N_n! \ n^{N_n} 
\quad
\text{for}
\quad
g =\prod_n (n)^{N_n} , \quad \sum_n N_n n = N 
\ee
only depends on the cycle structure of $[g]$.
Thus (\ref{dobsumqwq}) reduces to 
\be	\label{dobsumqwqB}
\begin{split}
\big\langle \s_{[g]} \s_{[g']} \big\rangle
	&=  \frac{| \Cent[g] |}{ {\scr S}_{[g]}^2} \sum_{h\in S_N} \langle \s_{h g h^{-1}} \ \s_{(h g h^{-1})^{-1}}  \rangle
	=  \frac{| \Cent[g] | \, | S_N | }{ {\scr S}_{[g]}^2} \ \langle \s_{g } \ \s_{g^{-1}}  \rangle
\end{split}
\ee
By construction, all the terms in this last sum over $h$ are non-vanishing as they trivially satisfy (\ref{hghghco1}), resulting in the factor $|S_N| = N!$.
This gives the normalization factor ${\scr S}_{[g]} = \sqrt{N! | \Cent[g]}$ in Eq.(\ref{Sninvg}).

\subsection{$Q$-point functions}	\label{SectQpointApp}

The $Q$-point function of $S_N$-invariant fields is a multiple sum 
\begin{align}	\label{pdsqapp}
\Big\langle \prod_{i=1}^Q \s_{[g_i]} (z_i) \Big\rangle
	=
	\frac{1}{\prod_i {\scr S}_{[g_i]}}
	\sum_{\substack{h_1 \in S_N \\ \cdots \\ h_Q\in S_N}}
	\Big\langle \s_{h_1 g_1 h_1^{-1}} (z_1) \cdots \s_{h_Q g_Q h_Q^{-1}}(z_Q) \Big\rangle
\end{align}
We will now follow \cite{Pakman:2009zz}, but with some differences:
we do not rely on the existence of diagrams;
we do \emph{not} assume that the $g_i$ are single cycles;
we do  \emph{not} assume (for now) that the functions are connected.
Our goal is to extract the $N$-dependence of the function (\ref{pdsqapp}) which comes from the multiplicity of equivalent terms.
In the r.h.s.~of Eq.(\ref{pdsqapp}), the correlation functions' twists are  individual representatives elements $p_i = h_i g_i h_i^{-1} \in S_N$ within the conjugacy classes $[g_i]$ in the l.h.s.
The non-vanishing correlators are those for which
$\prod_{i=1}^Q p_i = \id$. 
A non-vanishing function
$\langle \s_{p_1} (z_1) \cdots \s_{p_Q}(z_Q) \rangle$
will depend on how the copies inside the permutations interact.
All functions whose sets $\{p_i\} = \{p_1, \cdots, p_Q\}$ are related by a \emph{global} permutation must be equal, as that amounts to an overall relabeling of all copies, and the CFT copies are identical --- only their relative positions within the cycles matter. Thus we have equivalence classes, denoted by $\a$, of the ordered list of permutations $\{p_i\}$,
\be	\label{equiva}
\a : \quad
\{p_1, p_2 , \cdots, p_Q\} \sim \{ k p_1 k^{-1} , k p_2 k^{-1} , \cdots , k p_Q k^{-1} \} , \quad \text{for} \quad k \in S_N
\ee
and functions with $\{p_i\}$ in the same equivalence class $\a$ are equal by symmetry:
\be	\label{eqfuncclas}
\langle \s_{p_1} (z_1) \s_{p_2}(z_2) \cdots \s_{p_Q}(z_Q) \rangle 
	=
\langle \s_{k p_1 k^{-1}} (z_1) \s_{k p_2 k^{-1}}(z_2) \cdots \s_{k p_Q k^{-1}}(z_Q) \rangle	
\ee
	We emphasize the difference between the functions in the r.h.s.~of Eq.(\ref{pdsqapp}) and that in the r.h.s.~of Eq.(\ref{eqfuncclas}): in the former case, each twist has been conjugated by a different $h_i$, and in the latter all twists have undergone a global conjugation by the same element $k$.
Let us call a representative of class $\a$ by 
$\{ p_i^\a \} \in \a$, and the set of different equivalence classes by $\Class \ni \a$.

It is very instructive to look at concrete examples.
Take the $S_N$-invariant four-point function	
\be	\label{forExas}
\Big\langle \s_{[(3)(2)]} (z_1) \s_{[(2)]} (z_2) \s_{[(2)]} (z_3) \s_{[(3)(2)]} (z_4) \Big\rangle
\ee
and consider 
\bsub	\label{alclassecxapl}
\begin{align}
\begin{aligned}	\label{alclassecxapla1}
\a_{\bs 5}^1 \ni \, \{p_1^{\a_{\bs 5}^1}, p_2^{\a_{\bs 5}^1}, p_3^{\a_{\bs 5}^1}, p_4^{\a_{\bs 5}^1} \} 
	&\sim  
	\{
	(\textcolor{PastelBlue}{\bs 1}, 
	\textcolor{PastelGreen}{\bs2}, 
	\textcolor{PastelRed}{\bs3})
	(\textcolor{PastelPurple}{\bs4},
	\textcolor{PastelOrange}{\bs5}) , 
	\  
	(\textcolor{PastelBlue}{\bs1},
	\textcolor{PastelPurple}{\bs4}) , 
	\
	 (\textcolor{PastelBlue}{\bs1},
	 \textcolor{PastelPurple}{\bs4}) , 
	 \  
	 (\textcolor{PastelBlue}{\bs1},
	 \textcolor{PastelRed}{\bs3},
	 \textcolor{PastelGreen}{\bs2})
	 (\textcolor{PastelBlue}{\bs4},
	 \textcolor{PastelOrange}{\bs5}) 
	 \}
\\
	&\sim  \{
		(\textcolor{PastelBlue}{\bs1}, 
	\textcolor{PastelGreen}{\bs7}, 
	\textcolor{PastelRed}{\bs3})
	(\textcolor{PastelPurple}{\bs2},
	\textcolor{PastelOrange}{\bs5}) , 
	\  
	(\textcolor{PastelBlue}{\bs1},
	\textcolor{PastelPurple}{\bs2}) , 
	\
	 (\textcolor{PastelBlue}{\bs1},
	 \textcolor{PastelPurple}{\bs2}) , 
	 \  
	 (\textcolor{PastelBlue}{\bs1},
	 \textcolor{PastelRed}{\bs3},
	 \textcolor{PastelGreen}{\bs7})
	 (\textcolor{PastelBlue}{\bs4},
	 \textcolor{PastelOrange}{\bs5}) \}
\\
&
\end{aligned}
\\
\begin{aligned}	\label{alclassecxapla2}
\a_{\bs 5}^2 \ni \, \{p_1^{\a_{\bs 5}^2}, p_2^{\a_{\bs 5}^2}, p_3^{\a_{\bs 5}^2}, p_4^{\a_{\bs 5}^2} \} 
	&\sim  
	\{
	(\textcolor{PastelBlue}{\bs1}, 
	\textcolor{PastelGreen}{\bs2}, 
	\textcolor{PastelRed}{\bs3})
	(\textcolor{PastelPurple}{\bs4},
	\textcolor{PastelOrange}{\bs5}) , 
	\  
	(\textcolor{PastelBlue}{\bs1},
	\textcolor{PastelPurple}{\bs4}) , 
	\ 
	(\textcolor{PastelBlue}{\bs1},
	\textcolor{PastelRed}{\bs3}) , 
	\  
	(\textcolor{PastelRed}{\bs3},
	\textcolor{PastelOrange}{\bs5},
	\textcolor{PastelPurple}{\bs4})
	(\textcolor{PastelBlue}{\bs1},
	\textcolor{PastelGreen}{\bs2}) 
	\}
\\
	&\sim  
	\{
	(\textcolor{PastelBlue}{\bs1}, 
	\textcolor{PastelGreen}{\bs7}, 
	\textcolor{PastelRed}{\bs3})
	(\textcolor{PastelPurple}{\bs2},
	\textcolor{PastelOrange}{\bs5}) , 
	\  
	(\textcolor{PastelBlue}{\bs1},
	\textcolor{PastelPurple}{\bs2}) , 
	\ 
	(\textcolor{PastelBlue}{\bs1},
	\textcolor{PastelRed}{\bs3}) , 
	\  
	(\textcolor{PastelRed}{\bs3},
	\textcolor{PastelOrange}{\bs5},
	\textcolor{PastelPurple}{\bs2})
	(\textcolor{PastelBlue}{\bs1},
	\textcolor{PastelGreen}{\bs7}) 
	\}
\\
&
\end{aligned}
\\
\begin{aligned}	\label{alclassecxapla3}
\a_{\bs 5}^3 \ni \, \{p_1^{\a_{\bs 5}^3}, p_2^{\a_{\bs 5}^3}, p_3^{\a_{\bs 5}^3}, p_4^{\a_{\bs 5}^3} \} 
	&\sim  
	\{
	(\textcolor{PastelBlue}{\bs1}, 
	\textcolor{PastelGreen}{\bs2}, 
	\textcolor{PastelRed}{\bs3})
	(\textcolor{PastelPurple}{\bs4},
	\textcolor{PastelOrange}{\bs5}) , 
	\  
	(\textcolor{PastelBlue}{\bs1},
	\textcolor{PastelGreen}{\bs2}) , 
	\ 
	(\textcolor{PastelBlue}{\bs1},
	\textcolor{PastelRed}{\bs3}) , 
	\  
	(\textcolor{PastelGreen}{\bs2},
	\textcolor{PastelRed}{\bs3},
	\textcolor{PastelBlue}{\bs1})
	(\textcolor{PastelPurple}{\bs4},
	\textcolor{PastelOrange}{\bs5}) 
	\}
\\
	&\sim
	\{
	(\textcolor{PastelBlue}{\bs1}, 
	\textcolor{PastelGreen}{\bs7}, 
	\textcolor{PastelRed}{\bs3})
	(\textcolor{PastelPurple}{\bs2},
	\textcolor{PastelOrange}{\bs5}) , 
	\  
	(\textcolor{PastelBlue}{\bs1},
	\textcolor{PastelGreen}{\bs7}) , 
	\ 
	(\textcolor{PastelBlue}{\bs1},
	\textcolor{PastelRed}{\bs3}) , 
	\  
	(\textcolor{PastelGreen}{\bs7},
	\textcolor{PastelRed}{\bs3},
	\textcolor{PastelBlue}{\bs1})
	(\textcolor{PastelPurple}{\bs2},
	\textcolor{PastelOrange}{\bs5}) 
	\}
\\
&
\end{aligned}
\\
\begin{aligned}
\a_{\bs 6}^4 \ni \, \{p_1^{\a_{\bs 6}^4}, p_2^{\a_{\bs 6}^4}, p_3^{\a_{\bs 6}^4}, p_4^{\a_{\bs 6}^4} \} 
	&\sim  
	\{
	(\textcolor{PastelBlue}{\bs1}, 
	\textcolor{PastelGreen}{\bs2}, 
	\textcolor{PastelRed}{\bs3})
	(\textcolor{PastelPurple}{\bs4},
	\textcolor{PastelOrange}{\bs5}) , 
	\  
	(\textcolor{PastelBlue}{\bs1},
	\textcolor{Brown}{\bs 6}) , 
	\ 
	(\textcolor{PastelRed}{\bs3},
	\textcolor{Brown}{\bs 6}) , 
	\  
	(\textcolor{Brown}{\bs 6},
	\textcolor{PastelGreen}{\bs2},
	\textcolor{PastelBlue}{\bs1})
	(\textcolor{PastelPurple}{\bs4},
	\textcolor{PastelOrange}{\bs5}) \}
\\
	&\sim 
	\{
	(\textcolor{PastelBlue}{\bs1}, 
	\textcolor{PastelGreen}{\bs7}, 
	\textcolor{PastelRed}{\bs3})
	(\textcolor{PastelPurple}{\bs2},
	\textcolor{PastelOrange}{\bs5}) , 
	\  
	(\textcolor{PastelBlue}{\bs1},
	\textcolor{Brown}{\bs 6}) , 
	\ 
	(\textcolor{PastelRed}{\bs3},
	\textcolor{Brown}{\bs 6}) , 
	\  
	(\textcolor{Brown}{\bs6},
	\textcolor{PastelGreen}{\bs7},
	\textcolor{PastelBlue}{\bs1})
	(\textcolor{PastelPurple}{\bs2},
	\textcolor{PastelOrange}{\bs5}) \}
\end{aligned}
\end{align} \esub
Here we show four different classes $\a_{\bs5}^1 , \a_{\bs5}^2 , \a_{\bs5}^3 ,\a_{\bs6}^4 \in \Class$ contributing to (\ref{forExas}), and two representatives of each class.%
	\footnote{%
	In all four examples, the representatives are related by $\{p_i^\a\} \sim \{k p_i^{\a} k^{-1}\}$ with $k = (2,4,7)$.}
The boldface numbers ${\bf c}$ in $\a_{\bf c}$ indicate the number of  \emph{distinct copies} entering the permutations non-trivially.
Distinct copies are painted with distinct colors.
The coloring emphasizes that a class is determined not by the specific copies (i.e.~the algarisms) that enter the cycles, but by their \emph{relative positions} within the cycles --- that is, different orderings of the colors. \emph{Different representatives of the same class} are different ways of filling one arrangement of relative positions with algarisms $I = 1,\cdots, N$.
\emph{Different classes} are different arrangements of the relative positions, i.e.~different orderings of the colors.
We can see that $\a_{\bs5}^1 \neq \a_{\bs5}^2$, because $p_2^{\a_{\bs5}^1} = (p_3^{\a_{\bs5}^1})^{-1}$, while  $p_2^{\a_{\bs5}^2} \neq (p_3^{\a_{\bs5}^2})^{-1}$.
In classes $\a_{\bs5}^3$ and $\a_{\bs6}^4$  the double cycles $p_1$ and $p_4$ \emph{factorize}, because they contain copies that do not appear in the other permutations.
Note that the permutations in these examples might be elements of $S_N$ with $N \gg 6$, but we omit the \emph{trivial} cycles (of length one). 

\bigskip

We can organize the sum in (\ref{pdsqapp}) as a sum over the different classes $\a \in \Class$ (we are leaving the normalization factors ${\scr S}_{[g_i]}$ behind for a while),
\begin{align}	\label{pdsqapp2}
\begin{aligned}[b]
	\sum_{h_i \in S_N}
	\Big\langle \s_{h_1 g_1 h_1^{-1}} (z_1) \cdots \s_{h_Q g_Q h_Q^{-1}}(z_Q) \Big\rangle
	&=
	\sum_{\a \in \Class}
	{\cal N}_\a 
	\Big\langle \s_{p_1^\a} (z_1) \cdots \s_{p_Q^\a}(z_Q) \Big\rangle
\\
	&=
	\sum_{\bf c}
	\sum_{\a_{\bf c} \in \Class_{\bf c}}
	{\cal N}_{\a_{\bf c}}
	\Big\langle \s_{p_1^{\a_{\bf c}}} (z_1) \cdots \s_{p_Q^{\a_{\bf c}}}(z_Q) \Big\rangle
\end{aligned}
\end{align}
In the first line, $\{p_i^\a \}$ is an arbitrary representative of class $\a$ and ${\cal N}_\a$ is the number of collections $\{p_i\} \in \a$. 
In the second line, we decompose the sum further, 
by cataloguing the classes $\a \in \Class$ 
into subsets $\Class_{\bf c} \subseteq \Class$,
according to the number ${\bf c}$ of distinct copies entering the \emph{non-trivial} cycles of $\{p_i\}$.
By construction, $\cup_{\bf c} \Class_{\bf c} = \Class$.
For fixed ${\bf c}$, there is a collection of different classes $\a_{\bf c} \in \Class_{\bf c}$, with a number ${\cal N}_{\a_{\bf c}}$ of representatives $\{ p_i^{\a_{\bf c}}\}$, one of which is chosen to appear in the correlation function. 

Let us determine ${\cal N}_{\a_{\bf c}}$.
In the l.h.s.~of (\ref{pdsqapp2}), the sum runs over configurations of the individual permutations $g_i$, while in the r.h.s.~there is a sum over different \emph{collections of permutations} $\{p_i^{\a_{\bf c}} \}$ (not individual permutations).
In summing over orbits of $g_i$ in the l.h.s.,  
whenever $h_1 \in \Cent[p_1^{\a}]$ we get the same \emph{collection} $\{p_i^\a\}$;%
	\footnote{%
	Not just a different collection $\{p'^{\a}_i\} \sim \{p^\a_i\} \in \a$ within the same equivalence class $\a$, but \emph{exactly the same collection} $\{p^\a_i\}$.} 
	whenever $h_2 \in \Cent[p_2^{\a}]$ we again get the same collection $\{p_i^\a\}$, and so on, up to $h_Q \in \Cent[p_Q^{\a}]$.
Since in the r.h.s.~of Eq.(\ref{pdsqapp2}) the sums over $h_i$ are independent, the number of repeated occurrences of the collection 
$\{ p_1^{\a} , \cdots , p_Q^{\a} \}$
is
\be	\label{ccenthag12}
| \Cent [p_1^{\a}]| \times \cdots \times |\Cent [p_Q^{\a}]| 
=
| \Cent [ g_1]| \times \cdots \times |\Cent [g_Q]| 
\ee
since $|\Cent [p_i^\a]|$, only depends on the cycle structure of $p_i^a$, which is the same as that of $g_i$.
The same is true for the classes $\a_{\bf c}$, which are special types of $\a$, hence
\be
{\cal N}_{\a_{\bf c}}
	= \left( \prod_{i=1}^Q \big| \Cent[g_i] \big| \right) \times {\cal W}_{\a_{\bf c}}
\ee

The remaining factor ${\cal W}_{\a_{\bf c}}$ counts the number of different sequences $\{ p^{\a_{\bf c}}_i \}$ which are \emph{not identical but still belong to the same equivalence class} $\a_{\bf c}$.
For example, in (\ref{alclassecxapla2}) we can see two sequences $\{ p^{\a_{\bs5}^2}_i\}$ with different individual permutations (compare each cycle in the first line with the one immediately below it in the second line) that belong to the same conjugacy class $\a_{\bs5}^2$ (compare the relative positions of repeated copies, i.e.~the order of the colors). 
To find ${\cal W}_{\a_{\bf c}}$, we proceed in two steps.
First, we must choose the ${\bf c}$ copies that will enter the non-trivial permutations out of the $N$ copies available,
\be	\label{calWas}
{\cal W}_{\a_{\bf c}} = {N \choose {\bf c}} \times w_{\a_{\bf c}}
	= \frac{N!}{(N-{\bf c})! {\bf c}!} \ w_{\a_{\bf c}} .
\ee
The final remaining factor $w_{\a_{\bf c}}$ counts the number of ways we can arrange the ${\bf c}$ copies and still find collections $\{p_i\}$ within the same class.
This number $w_{\a_{\bf c}}$, which we will determine shortly, can depend on ${\bf c}$ and on the cycle structure of the $[g_i]$, but it clearly cannot depend on $N$. So \emph{we have already completely determined the $N$-scaling dependence of the $Q$-point function}.

It turns out that $w_{\a_{\bf c}}$ has a subtle dependence on the factorization of the functions in class $\a_{\bf c}$.
As we can see from the examples (\ref{alclassecxapl}), some classes will have disconnected correlators, and note that connectedness is indeed a class property: all functions 
$\langle \s_{p_1^\a} (z_1) \cdots \s_{p_Q^\a}(z_Q) \rangle$
in the same class $\a$ factorize the same way.
So the r.h.s.~of (\ref{pdsqapp2}) decomposes further,
\begin{align}	\label{pdsqappCon}
\begin{aligned}[b]
\sum_{\a_{\bf c} \in \Class_{\bf c}}
	{\cal N}_{\a_{\bf c}}
	\Big\langle \s_{p_1^{\a_{\bf c}}} (z_1) \cdots \s_{p_Q^{\a_{\bf c}}}(z_Q) \Big\rangle
\\
	=
\left( \prod_{i=1}^Q \big| \Cent[g_i] \big| \right) \frac{N!}{(N-{\bf c})! {\bf c}!}
	&
	\Bigg[
	\sum_{\a_{\bf c} \in \left[\substack{\text{fully} \\ \text{connected}\\\text{classes}} \right]}
	w_{\a_{\bf c}}
	\Big\langle \cdots \Big\rangle 
\\
&
	+
	\sum_{\a_{\bf c} \in \left[ \substack{\text{once} \\ \text{disconnected}\\\text{classes}} \right]}
	w_{\a_{\bf c}}
	\Big\langle \cdots \Big\rangle
	\Big\langle \cdots \Big\rangle
\\
&
	+
	\sum_{\a_{\bf c} \in \left[ \substack{\text{twice} \\ \text{disconnected}\\\text{classes}} \right]}
	w_{\a_{\bf c}}
	\Big\langle \cdots \Big\rangle
	\Big\langle \cdots \Big\rangle
	\Big\langle \cdots \Big\rangle
+ \cdots 
	\Bigg]
\end{aligned}
\end{align}
The possible types of factorizations of the initial correlator  will depend on the original cycle structure of the $[g_i]$, and also on ${\bf c}$.
For example, it is possible that the cycle structure of the $[g_i]$ be incompatible with fully connected classes --- this is what happens with the functions considered in Sect.\ref{SectCompFunct} --- and in this case, the first sum in the r.h.s.~above is void.
The factor $w_{\a_{\bf c}}$ is given by
\be	\label{formwas}
w_{\a_{\bf c}} = {\bf c}! \,  \nu_{\a_{\bf c}} ,
\quad
	\nu_{\a_{\bf c}}
	=
	\begin{cases}
	1
		\ &\text{if no two-point function factorizes}
	\\
	1 / [\prod n_j]_{\a_{\bf c}} 
		\ &\text{if one or more two-point function factorizes}
	\end{cases}
\ee
where the $n_j$ are the lengths of cycles in eventual \emph{two}-point functions that factorize in the class ${\a_{\bf c}}$.

Formula (\ref{formwas}) can be obtained as follows.
In Eq.(\ref{calWas}), we had chosen ${\bf c}$ copies to enter the non-trivial cycles of the set $\{p^{\a_{\bf c}}_1, \cdots, p^{\a_{\bf c}}_Q\}$.
Now we start to fill the cycles of the $p^{\a_{\bf c}}_i$ with these copies. 
We choose an ordering of the copies to fill the first non-repeated slots, and then the copies in the repeated slots are fixed by the structure of the class $\a_{\bf c}$. So we have the freedom of choosing all ${\bf c}!$ orderings of ${\bf c}$ elements to get different collections $\{p^{\a_{\bf c}}_i\}$ belonging to the same class $\a_{\bf c}$.
If no two-point functions factorize, the collections $\{p^{\a_{\bf c}}_i\}$ obtained in this way will all be different, yielding the first line of (\ref{formwas}).
But whenever the class $\a_{\bf c}$ has factorized two-point functions as in 
$
\langle \s_{p_1^{\a_{\bf c}}} (z_1) \cdots \s_{p_Q^{\a_{\bf c}}}(z_Q) \rangle
	= 
	\langle 
	\cdots
	\rangle
	\
	\prod_j
	\langle \s_{(n_j)} \s_{(n_j)^{-1}} \rangle ,
$
the orderings that differ only by cyclic reorderings of the copies in the factorized cycles $(n_j)$ will actually give the same collections $\{  p^{\a_{\bf c}}_i\}$. There are $n_j$ cyclic arrangements of $n_j$ objects, and taking these into account we get the second line of (\ref{formwas}).

\begin{figure}
\centering
\includegraphics[scale=0.2]{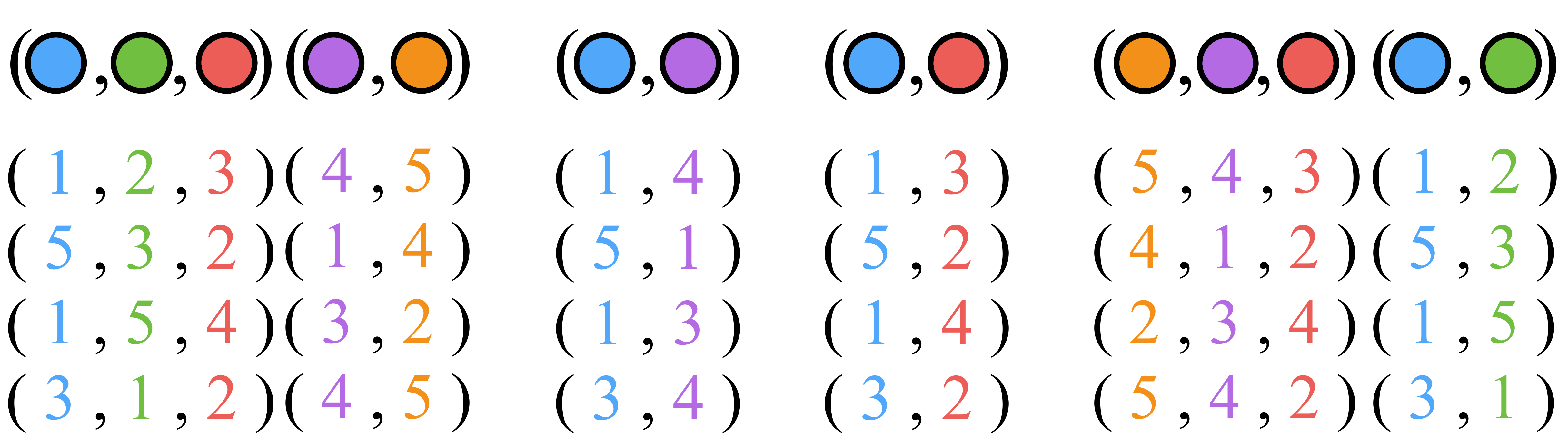}
\caption{%
Four different representatives of class $\a_{\bs5}^2$ (the same featured in example (\ref{alclassecxapla2}). 
All representatives have the same active copies 1, 2, 3, 4, 5 appearing in different orders, but preserving the relative positions of repeated copies.
}
\label{cycles_ordering}
\end{figure}

Again, this is best understood by looking at an example. Consider class $\a_{\bs5}^2$ in (\ref{alclassecxapla2}). 
In Fig.\ref{cycles_ordering} we show different collections $\{ p^{\a_{\bs5}^2}_i\}$, all made with the \emph{same} copies $I =1, 2, 3, 4, 5$.
In each collection (each line), we fill the first five positions with the copies in a given ordering, and then the copies in the remaining positions cannot be chosen: they are fixed by the structure of the class, here highlighted by the coloring (positions with the same color must have the same copy).
Compare the first and the last lines of Fig.\ref{cycles_ordering}, where 
the choices of copies differ only by a cyclic reordering of the first cycle, $(1,2,3) = (3,1,2)$. Hence, \emph{in these two orderings, the first two cycles are the same}.
Nevertheless, \emph{the collections as a whole are not the same} because their remaining cycles are \emph{not} equivalent.
This is a consequence of the connectedness of this class. In accordance with formula (\ref{formwas}), in this case there are $w_{\a_{\bs5}^2} = {\bf c}! = 5! = 120$ ways of ordering the copies $I = 1,2,3,4,5$, all of them giving different sequences $\{ p^{\a_{\bs5}^2}_i\}$ in the same class.

\begin{figure}
\centering
\includegraphics[scale=0.2]{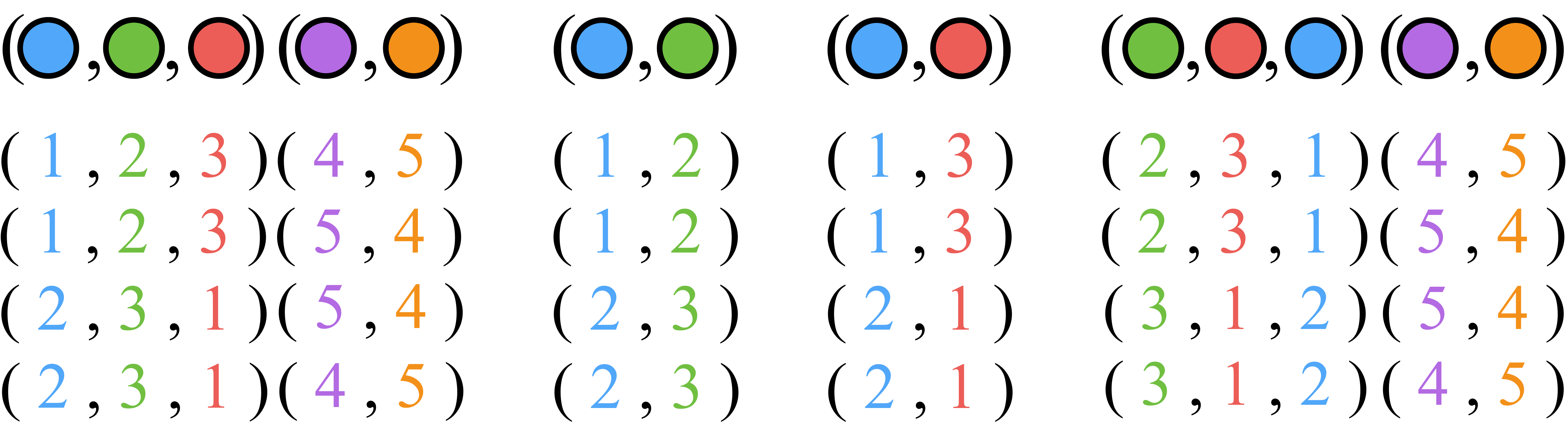}
\caption{%
Different representatives of class $\a_{\bs5}^3$, the same featured in example (\ref{alclassecxapla3}). 
All representatives have the same active copies 1, 2, 3, 4, 5 appearing in different orders, preserving the relative positions of repeated copies.
}
\label{cycles_ordering_fact}
\end{figure}

Now consider the analogous examination of class $\a_{\bs5}^3$ in (\ref{alclassecxapla3}), shown in Fig.\ref{cycles_ordering_fact}.
In this class, there is a factorization of the two-cycles of $p_1$ and $p_4$, making a two-point function:
\be
\begin{split}
&
\Big\langle 
	\s_{
	(\textcolor{PastelBlue}{\bullet}, 
	\textcolor{PastelGreen}{\bullet}, 
	\textcolor{PastelRed}{\bullet} 
	)(
	\textcolor{PastelPurple}{\bullet}, 
	\textcolor{PastelOrange}{\bullet}
	)
	} (z_1) 
	\s_{
	(\textcolor{PastelBlue}{\bullet}, 
	\textcolor{PastelPurple}{\bullet})
	} (z_2) 
	\s_{
	(\textcolor{PastelBlue}{\bullet}, 
	\textcolor{PastelRed}{\bullet})
	} (z_3) 
	\s_{
	(\textcolor{PastelGreen}{\bullet}, 
	\textcolor{PastelRed}{\bullet}, 
	\textcolor{PastelBlue}{\bullet} 
	)(
	\textcolor{PastelPurple}{\bullet}, 
	\textcolor{PastelOrange}{\bullet}
	)
	} (z_4) \Big\rangle
\\
&\qquad
=
\Big\langle 
	\s_{
	(\textcolor{PastelBlue}{\bullet}, 
	\textcolor{PastelGreen}{\bullet}, 
	\textcolor{PastelRed}{\bullet} 
	)
	} (z_1) 
	\s_{
	(\textcolor{PastelBlue}{\bullet}, 
	\textcolor{PastelPurple}{\bullet})
	} (z_2) 
	\s_{
	(\textcolor{PastelBlue}{\bullet}, 
	\textcolor{PastelRed}{\bullet})
	} (z_3) 
	\s_{
	(\textcolor{PastelGreen}{\bullet}, 
	\textcolor{PastelRed}{\bullet}, 
	\textcolor{PastelBlue}{\bullet} )	
	} (z_4) 
\Big\rangle
\
\Big\langle 
	\s_{(\textcolor{PastelPurple}{\bullet}, 
	\textcolor{PastelOrange}{\bullet})
	} (z_1) 
	\s_{
	(\textcolor{PastelPurple}{\bullet}, 
	\textcolor{PastelOrange}{\bullet})
	} (z_4) \Big\rangle
\end{split}	
\ee
Now the choices of $I =1,2,3,4,5$ that only differ by cyclic reordering of the copies entering the factorized cycles do result in \emph{identical} collections $\{p^{\a_{\bs5}^3}_i\}$. As an example, consider the first two lines of Fig.\ref{cycles_ordering_fact}: they differ by a cyclic reordering of copies 4 and 5, entering the factorized cycles, and the configuration of the remaining, non-factorized cycles is left invariant. So the two ordered choices of copies, $1,2,3,4,5$ and $1,2,3,5,4$, must not be counted as different collections $\{p^{\a_{\bs5}^3}_i\}$. The same happens for all other choices, for instance the ones in the third and fourth lines of Fig.\ref{cycles_ordering_fact}.
Thus, in accordance with formula (\ref{formwas}), we must divide the total number ${\bf c}! = 5!$ of ordered choices of copies by the number $n = 2$ of cyclic reorderings of the copies entering the factorized cicles.
Finally, compare the first and the last lines of Fig.\ref{cycles_ordering_fact}, which differ by a cyclic reordering of the first, \emph{non}-factorized cycle $(1,2,3) = (2,3,1)$. These two collections are \emph{not} identical, as you can see that the permutations $p^{\a_{\bs5}^3}_2 = ( \textcolor{PastelBlue}{\bullet}, \textcolor{PastelGreen}{\bullet})$
are not the same in the two collections, 
and neither are the permutations $p^{\a_{\bs5}^3}_3 = ( \textcolor{PastelBlue}{\bullet}, \textcolor{PastelRed}{\bullet})$.

\subsection{Restrictions on the possible factorizations} 
\label{SectQpointAppFact}

The restricted possibilities of factorization of the function (\ref{TermXXZZ}) is due to the fact that the connected functions (\ref{TermXXZZf1}) and (\ref{TermXXZZf0}) are implicitly multiplied by a product of factorized two-point functions
\be	\label{towXXfac}
\big\langle \bar X^{\zeta_i}_{(n_i)}(\infty)  X^{\zeta_i}_{(n_i)}(0)  \big\rangle = 1,
\ee
where the fields $X^{\zeta}_{(n)}$ and $\bar X^{\zeta}_{(n)}$ are the ones whose cycles do not overlap with $ Z_{(2)}$ nor $\bar  Z_{(2)}$.
For a hypothetical connected function with components $\bar X^{\zeta}_{(n)}$ and $X^{\zeta}_{(n)}$ of the original multi-cycle fields in (\ref{TermXXZZ}), such that the remaining components do not \emph{all} match into pairs like (\ref{towXXfac}), the factorization is forbidden.
For example, a factorizations leading to the connected function
\be
\Big\langle
[ \bar X^{\zeta_1}_{(n_1)} \bar X^{\zeta_2}_{(n_2)} ]  \bar Z_{(2)}	 Z_{(2)} X^{\zeta_3}_{(n_1+n_2-1)}
\Big\rangle ,
\ee
seems possible because it satisfies (\ref{compto1}),
but is  forbidden because the factorized operators have different cycles so the analogous to (\ref{towXXfac}) vanishes.

A more subtle case is a factorization leading to a product of three-point functions:
\begin{align}	\label{TermXXZZ3}
\begin{aligned}
&
\Big\langle
\big[ \prod_{\zeta,n} ( \bar X^\zeta_{(n)} )^\Nzn \big] 
\	\bar Z_{(2)}
\	Z_{(2)} 
\big[ \prod_{\zeta,n} ( X^\zeta_{(n)} )^\Nzn \big] 
\Big\rangle 
\\
&
\qquad\qquad
=
\Big\langle
\big[ \bar X^{\zeta_1}_{(n_1)} \bar X^{\zeta_2}_{(n_2)} \big] 
\	\bar Z_{(2)}
\
 X^{\zeta_3}_{(n_1+n_2)} 
\Big\rangle 
\
\Big\langle
  \bar X^{\zeta_3}_{(n_1+n_2)} 
\	 Z_{(2)}
\
\big[  X^{\zeta_1}_{(n_1)} X^{\zeta_2}_{(n_2)} \big] 
\Big\rangle .
\end{aligned}
\end{align}
This is, in general, \emph{not} forbidden.
But for this factorization to occur, the original composite field 
must, necessarily, have at least one component with cycle length equal to the sum of other two components, i.e.
\begin{align*}
\Big[ \prod_{\zeta,n} ( X^\zeta_{(n)} )^\Nzn \Big] 
=
\Big[ X^{\zeta_1}_{(n_1)}  X^{\zeta_2}_{(n_2)}  X^{\zeta_3}_{(n_3)} \cdots  \Big] 
\qquad
\text{with} \qquad n_3 = n_1 + n_2 .
\end{align*}
So functions with fields like
${\scr X} = [ (X^\zeta_{[n]})^{N/n} ]$, for example, 
never factorize as (\ref{TermXXZZ3}). 

Furthermore, in the concrete case of the D1-D5 CFT, the fields carry SU(2) charges --- in particular, the R-charge --- which must add to zero inside a non-vanishing correlation function. This imposes more selection rules. 
For example, if $Z_{[2]}$ is R-neutral, e.g.~$Z_{[2]} = \Oincov_{(2)}$, and the $X^\zeta_{(n)} = R^\zeta_{(n)}$ are Ramond fields (\ref{RamondFields}), all possible three-point functions like (\ref{TermXXZZ3}) vanish.
Meanwhile, for specific configurations, e.g.~
\begin{align*}
\Big[ \prod_{\zeta,n} ( X^\zeta_{(n)} )^\Nzn \Big] 
=
\Big[ R^+_{(n_1)}  R^+_{(n_2)}  R^+_{(n_1+n_2)} \cdots  \Big] 
\quad
\text{and}
\quad
Z_{(2)} = \bar O^{(0,0)}_{(2)}
\end{align*}
(note that $Z_{(2)}$ is an anti-chiral field, with $j = - \frac12$),
the three-point functions in (\ref{TermXXZZ3}) do have zero R-charges.
If all fields in (\ref{TermXXZZ3}) are (possibly different types of) NS chirals, there may also be non-vanishing three-point functions.
For example, if
\begin{align*}
\Big[ \prod_{\zeta,n} ( X^\zeta_{(n)} )^\Nzn \Big] 
=
\Big[ O^{(0,0)}_{(n_1)}  O^{(0,0)}_{(n_2)}  O^{(0,0)}_{(n_1+n_2)} \cdots  \Big] 
\quad
\text{and}
\quad
Z_{(2)} = \bar O^{(0,0)}_{(2)}
\end{align*}
the factorization (\ref{TermXXZZ3}) becomes
\begin{align}	
\Big\langle
\big[\bar O^{(0,0)}_{(n_1)}  \bar O^{(0,0)}_{(n_2)} \big] 
\	 O^{(0,0)}_{(2)}
\
O^{(0,0)}_{(n_1+n_2)}
\Big\rangle 
\
\Big\langle
 \bar O^{(0,0)}_{(n_1+n_2)}
\	\bar O^{(0,0)}_{(2)}
\
\big[  O^{(0,0)}_{(n_1)} O^{(0,0)}_{(n_2)} \big] 
\Big\rangle ,
\end{align}
which is non-vanishing: the R-charges in the first correlator are
$$
- \tfrac12 (n_1 -1 ) - \tfrac12( n_2 - 1) - \tfrac12 (2-1) + \tfrac12 (n_1+n_2 -1) = 0
$$
and likewise for the second correlator.

As mentioned in the main text below (\ref{TermXXZZB}), \emph{if} the factorization into three-point functions occurs, it can be dealt with using basically the same procedure we have used for the factorizations (\ref{TermXXZZB}). 
There will be an additional contribution to the r.h.s.~of Eq.(\ref{facAhlltaa}), with a sum over classes that factorize as (\ref{TermXXZZ3}). The factorized terms can then be reduced to (products of) $S_N$-invariant three-point functions, that will appear in the r.h.s.~of Eq.(\ref{Axxzz}) multiplied by ``symmetry factors''. 
These factors are what connects the $S_N$-invariant three-point functions to the full four-point function with the multi-cycle composite fields.
They can be found following the same type of combinatoric analysis used to derive 
e.g.~${\scr P}$, but are highly ``example-sensitive'': 
given two cycles of length $n_1$ and $n_2$ in the composite field,
the symmetry factors will depend on how many components with cycle length $n_3 = n_1 + n_2$ there are in the field, on the corresponding R-charges etc. Hence it would be cumbersome to try to find a more or less generic formula --- which is yet another reason why we omit these cases in the paper.
Going forward, once the symmetry factors in a given case are known, there still remains to compute the $S_N$-invariant connected three-point functions, as we do in the paper for the connected four-point functions (\ref{An1n2zz}). But, compared with four-point functions, the analysis of twisted three-point functions is much better known. General formulas can be found in the work of Lunin and Mathur \cite{Lunin:2001pw} and, for functions involving only NS chiral fields, many structure constants are known since they form the NS chiral ring \cite{Tormo:2018fnt,Dabholkar:2007ey,Pakman:2009ab}.
We should note, yet, that the functions in (\ref{TermXXZZ3}) include one composite double-cycle field, and these are less studied in the literature. But this kind of structure constant is just what appears in the OPEs we have studied here, see Eq.(\ref{OPEWns}). Hence our own results can also be used to complete the computation of the special cases where the factorization (\ref{TermXXZZ3}) exists.

\subsection{Untwisted composite fields}	\label{SectNormofUntw}

Let us compute the normalization factor ${\scr S}$ of a generic untwisted field (\ref{untwscomp}).
The symmetrization 
\be	\label{untwscompAPp}
	 \rm{Sym}[ \otimes_{i=1}^f  \big( X^{s_i}_{I_1^{(i)}} \otimes \cdots \otimes  X^{s_i}_{I_{p_i}^{(i)}} \big) ] 
\ee
 is a sum of all possible configurations of the copies. A generic term in the sum is
 \be	\label{termXXXX}
 \big( X^{\zeta_1}_{I_1^{(1)}} \otimes \cdots \otimes  X^{\zeta_1}_{I_{p_1}^{(1)}} \big) 
 \otimes
	 \big( 
	 X^{\zeta_2}_{I_2^{(2)}} 
	 \otimes \cdots \otimes  
	 X^{\zeta_2}_{I_{p_2}^{(2)}} \big) 
\otimes
\cdots
 \otimes
	 \big( 
	 X^{s_f}_{I_f^{(f)}} 
	 \otimes \cdots \otimes  
	 X^{s_f}_{I_{p_f}^{(f)}} \big) 
\ee
where the copies $I^{(i)}_{p_i}$ are all distinct. How many equivalent such terms are there?
One must choose $p_1$ copies out of $N$ to enter the first parenthesis (inside of which all fields are equivalent, i.e.~have the same $s$), so there are ${N \choose p_1}$ options.
Then one must choose $p_2$ copies out of the remaining $N - p_1$ copies to enter the second parenthesis, and there are ${ N - p_1 \choose p_2}$ options. And so on. 
The total number of equivalent terms is therefore
\begin{align}	\label{prooSunt}
{\scr S}
=
{N \choose p_1} \times { N - p_1 \choose p_2}
	 \times \cdots \times
	 { N - \sum_{j=1}^{f-1} p_j  \choose p_f}
=
	\frac{N!}{(N - \sum_{i=1}^{f} p_i)! \prod_{i = 1}^f (p_i!)}	
\end{align}
which is the result appearing in (\ref{untwscomp}).
Note that we have not required that every one of the $N$ copies appear in each term (\ref{termXXXX}), that is, we have not required that 
$\sum_i p_i = N$. If this is the case, then the expression in the last line of (\ref{prooSunt}) simplifies
\begin{align}	
	\frac{N!}{(N - \sum_{i=1}^{f} p_i)! \prod_{i = 1}^f (p_i!)}
	=
	\frac{N!}{\prod_{i = 1}^f (p_i!)}
	\quad \text{for} \quad
	\sum_{i=1}^f p_i = N.
\end{align}
If there are only two powers, $p_1 = q$ and $p_2 = N-q$, this formula reduces to (\ref{scrSnpqq}).

\subsubsection*{Factorization of four-point functions}

For untwisted composite fields, since there is no sum over orbits of trivial cycles, we must redo our computations.
For definiteness, consider the operator in (\ref{scrSnpqq}), and the two-point function
\be
\Big\langle
\big[ X_{[1]}^p Y_{[1]}^q \big]^\dagger(\infty) \
Z^\dagger_{[2]}(1) \
Z_{[2]}(v, \bar v) \
\big[ X_{[1]}^p Y_{[1]}^q \big] (0) \
\Big\rangle
 , \quad q = N-p .
\ee
There are two sums over orbits of the 2-cycles, and symmetrization of the copies in the composite fields. Leaving the normalization factors, a generic term in the sum has the following permutation structure
(coordinates omitted for economy of space)
\be	\label{genrtermXYZ}
\Big\langle
\big[ X_{I_1} \cdots X_{I_p} Y_{I_{p+1}} \cdots Y_{I_{p+q}} \big]^\dagger  \
Z^\dagger_{h_1 (2) h_1^{-1}} \
Z_{h_v(2) h_v^{-1}}  \
\big[ X_{J_1} \cdots X_{J_p} Y_{J_{p+1}} \cdots Y_{J_{p+q}} \big] \
\Big\rangle .
\ee
The function can factorize in three ways, depending on the interaction of the cycles in the middle. If the cycles are disjoint, then the factorization is
\be
\Big\langle
\cdots  \
Z^\dagger_{h_1 (2) h_1^{-1}} \
\cdots
\Big\rangle 
\
\Big\langle
\cdots
Z_{h_v(2) h_v^{-1}}  \
\cdots
\Big\rangle 
=0
\ee
which vanishes because the remaining correlators do not satisfy the fundamental condition (\ref{compto1}).
If the cycles are not disjoint, they can either compose to a three cycle, or be the inverse of each other. In the former case, if
$h_1 (2) h_1^{-1} h_v (2) h_v^{-1} = (3)$, then the correlator also vanishes because, again, it fails to satisfy (\ref{compto1}).
The final remaining possibility is that the cycles are the inverses of each other; then the factorized function does satisfy (\ref{compto1}), so this is the only non-vanishing factorization.

\section{Derivation of the master formula}	\label{General4ptNS}

Here we give details of the derivation of the four-point function (\ref{AZv}), namely
\be	\label{GxAp}
	\Big\langle
	\Big[
	X_{[n_1]}^{\{\hat \sigma , \hat \varrho \}} 
	X_{[n_2]}^{\{ \check \sigma , \check \varrho \}} 
	\Big]^{\dagger} (\infty) 
	\
	Z_{[2]}^{\{\a,\b\}\dagger} (1) 
	\
	Z_{[2]}^{\{\a,\b\}} (v)
	\
	\Big[
	X_{[n_1]}^{\{\hat \sigma , \hat \varrho \}}
	X_{[n_2]}^{\{ \check \sigma , \check \varrho \}} 
	\Big](0) 
	\Big\rangle ,
\ee
as parameterized by the pre-image $x$ of $u$ on the covering surface.
Near a ramification point $z_*$, where $z(t) \approx z_* + b_* (t-t_*)^n$, the bosonized fermionic exponentials lift to 
(the lifted field is in the r.h.s.)
\cite{Lunin:2001pw}
\bsub \label{rim}
\begin{align}
\exp \Bigg[ \frac{i}{2n} \sum_{I=1}^n \Big[ \s \phi_{1,I}(z_*) + \varrho \phi_{2,I}(z_*)  \Big] \Bigg]
\s_{(n)}(z_*)
	&\mapsfrom b_*^{- \frac{\s^2 + \varrho^2}{8n} }
	\exp \Bigg[ \frac{i\s}{2} \phi_{1}(t_*) + \frac{i\varrho}{2} \phi_{2}(t_*)   \Bigg]
\end{align}
When inserted at $z = \infty$, the exponential lifts with a \emph{positive} power of $b_*$, 
\begin{align}
\exp \Bigg[ \frac{i}{2n}  \sum_{I=1}^n \Big[ \s \phi_{1,I}(\infty) + \varrho \phi_{2,I}(\infty)  \Big] \Bigg]
\s_{(n)}(\infty)
	&\mapsfrom 
	b_*^{+ \frac{\s^2 + \varrho^2}{8n} }
	\exp \Bigg[ \frac{i\s}{2} \phi_{1}(t_*) + \frac{i\varrho}{2} \phi_{2}(t_*)   \Bigg]
\end{align}
\esub
The coefficients at the branching points of (\ref{coverm}) are
\bsub	\label{bofx}
\begin{align}
b_0 &= 
		x^{-n_2}
		(x-1)^{-n_1}
		(x + \tfrac{n_1}{n_2})^{n_1 + n_2}
		(x + \tfrac{n_1}{n_2} - 1)^{-n_1}
\\
b_{t_0} &=
		(- \tfrac{n_1}{n_2} )^{-n_2}
		(x-1)^{-n_2}
		(x + \tfrac{n_1}{n_2})^{n_1 + n_2}
		(x + \tfrac{n_1}{n_2} - 1)^{n_2-n_1}
\\
b_{t_1}
	&= 
		- n_1
		(x-1)^{-2}
		(x + \tfrac{n_1}{n_2})^{2}
		(x + \tfrac{n_1}{n_2} - 1)^{-2}
		(x + \tfrac{n_1 - n_2}{2n_2})
\\
b_{x}
	&= 
		n_1
		x^{n_1-n_2-2}
		(x-1)^{-(n_1+n_2)}
		(x + \tfrac{n_1}{n_2})^{n_1+n_2}
		(x + \tfrac{n_1}{n_2} - 1)^{n_2-n_1}
		(x + \tfrac{n_1 - n_2}{2n_2})
\\
b_{t_\infty} &=
		(\tfrac{n_1}{n_2} )^{n_2}
		x^{n_1}
		(x-1)^{-(n_1+n_2)}
		(x + \tfrac{n_1}{n_2})^{-n_2}
		(x + \tfrac{n_1}{n_2} - 1)^{n_2}
\\
b_{\infty} &=
		(-1)^{n_2}
		(x-1)^{-(n_1+n_2)}
		(x + \tfrac{n_1}{n_2})^{n_1}
		(x + \tfrac{n_1}{n_2} - 1)^{n_2-n_1}
\end{align}
\esub
and the fields that we  use are  lifted to
\bsub	\label{XXZZbs}
\begin{align}
X^{\{\hat \sigma , \hat \varrho \}\dagger}(\infty) 
X^{\{ \check \sigma , \check \varrho \}\dagger} (t_\infty) 
&=
	b_\infty^{\frac{\hat\s^2+\hat\varrho^2}{8n_1}}
	b_{t_\infty}^{\frac{\check\s^2+\check\varrho^2}{8n_2}}
	e^{
	- \frac{i}{2}
	[ \hat \sigma \phi_{1} (\infty) + \hat \varrho \phi_{2} (\infty) ] 
	}
	\
	e^{
	- \frac{i}{2}
	[ \check \sigma \phi_{1}(t_\infty) + \check \varrho \phi_{2}(t_\infty) ]
	} 
\\
X^{\{\hat \sigma , \hat \varrho \}}(0) 
X^{\{ \check \sigma , \check \varrho \}} (t_0) 
&=
	b_0^{-\frac{\hat\s^2+\hat\varrho^2}{8n_1}}
	b_{t_0}^{-\frac{\check\s^2+\check\varrho^2}{8n_2}}
	e^{
	\frac{i}{2}
	[ \hat \sigma \phi_{1} (0) + \hat \varrho \phi_{2} (0) ] 
	}
	\
	e^{
	\frac{i}{2}
	[ \check \sigma \phi_{1}(t_0) + \check \varrho \phi_{2}(t_0) ]
	} 
\\
Z^{\{ \a,\b \}\dagger} (t_1) 
&=
	b_{t_1}^{-\frac{\a^2 + \b^2}{16}}
	e^{ - \frac{i}{2} [ \a \phi_{1} (t_1) + \b \phi_{2} (t_1) ] }
\\
Z^{\{ \a,\b \}} (x) 
&=
	b_{x}^{-\frac{\a^2 + \b^2}{16}}
	e^{ \frac{i}{2} [ \a \phi_{1} (x) + \b \phi_{2} (x) ] }
\end{align}\esub
Lifted fields carry no twist indices because they are untwisted.
Apart from the $b_*$ factors, the function (\ref{Gcoverx}) is a six-point function of exponentials only, whose computation is immediate.
Therefore, 
\be	\label{GbosnAp}
\begin{split}
&\AmZ{x}_\cover
	= 
	b_\infty^{\frac{1}{4n_1}}	
	b_{t_\infty}^{\frac{1}{4n_2}}
	b_{t_1}^{- \frac{1}{8}}
	b_{x}^{- \frac{1}{8}}
	b_0^{-\frac{1}{4n_1}}	
	b_{t_0}^{-\frac{1}{4n_2} }
\\
&\times
	\left[
	\frac{ 
	(t_\infty - t_1) (t_0 - x) 
	}{
	(t_\infty - x) (t_0 - t_1) 
	} 
	\right]^{\frac{ \a \check\sigma + \b \check\varrho }{4} }
	\left( \frac{x}{t_1} \right)^{ \frac{\a \hat\sigma + \b \hat\varrho}{4}}
	\left( \frac{t_0}{t_\infty} 
	\right)^{
	\frac{\check \sigma \hat \sigma + \hat \varrho \check \varrho }{4} 
			}
	( t_\infty - t_0 )^{- \frac{\check\s^2 + \check\varrho^2}{4} }
	(t_1 - x )^{-\frac{\a^2 + \b^2}{4} }
\end{split}	
\ee
where we need to express $t_1, t_0, t_\infty$ all in terms of $x$ via (\ref{tim}).
The final result for the base sphere four-point function, parameterized by $x$, is (\ref{GeSLcover}), i.e.~$\AmZ{x} = e^{S_L} \AmZ{x}_\cover$,
where $e^{S_L}$ is the bare-twist correlation function; using either the results in \cite{Avery:2010qw} or the stress-tensor method of \cite{Lima:2020kek,Lima:2021wrz}, we can find  
\begin{align}	\label{SLofx}
\begin{split}
S_L(x)    &=
	- 
	\frac{
	2 n_2^2 + (2+3 n_2) (n_1 - n_2) n_1
	}{
	8 n_1 n_2}
	\log x
\\
&\quad
	+
	\frac{
	2 n_2^2 + (2+3 n_2) (n_1 + n_2) n_1
	}{
	8 n_1 n_2}
	\log ( x-1 )
\\
&\quad
	+
	\frac{
	2 n_2^2 + (2-3 n_2) (n_1 + n_2) n_1
	}{
	8 n_1 n_2}
	\log ( x + \tfrac{n_1}{n_2} )
\\
&\quad
	-
	\frac{
	2 n_2^2 + (2-3 n_2) (n_1 - n_2) n_1
	}{
	8 n_1 n_2}
	\log ( x + \tfrac{n_1}{n_2} -1 )
\\
&\quad
	-
	\frac{1}{4}
	\log ( x + \tfrac{n_1-n_2}{2n_2}  )
\end{split}
\end{align}
apart from a constant which we fix later by looking at OPE limits of the final correlator.
Combining (\ref{GbosnAp}) with (\ref{GeSLcover})-(\ref{SLofx}), we obtain the final result (\ref{GenricAapp}).

\section{List of double-cycle four-point functions}\label{AppListofFun}

Here we list a collection of selected examples of functions $\AmZ{x}$ and $\Amint{x}$.
For economy of space, we omit the arguments of the fields.

\subsubsection*{Functions with NS chirals}

\begin{align}	\label{NSfunctionappD}
\begin{aligned}
\Big\langle 
	\Big[ \bar O^{(0,0)}_{[n_1]} \bar O^{(0,0)}_{[n_2]} \Big] 
	\bar O^{(0,0)}_{[2]}
	O^{(0,0)}_{[2]}
	\Big[ O^{(0,0)}_{[n_1]} O^{(0,0)}_{[n_2]} \Big] 
\Big\rangle
	&=
		- \frac{1}{4 n_1}
	\frac{ ( 1 - x)^2}{x + \frac{n_1 - n_2}{2n_2}}
\\
\Big\langle 
	\Big[ \bar O^{(2,2)}_{[n_1]} \bar O^{(2,2)}_{[n_2]} \Big]
	\bar O^{(0,0)}_{[2]}
	O^{(0,0)}_{[2]}
	\Big[ O^{(2,2)}_{[n_1]} O^{(2,2)}_{[n_2]} \Big]
\Big\rangle
	&=
		- \frac{1}{4 n_1}
	\frac{ (x + \frac{n_1}{n_2})^2}{x + \frac{n_1 - n_2}{2n_2}}
\\
\Big\langle 
	\Big[ \bar O^{(1\pm,1\pm)}_{[n_1]} \bar O^{(1\pm,1\pm)}_{[n_2]} \Big]
	\bar O^{(0,0)}_{[2]}
	O^{(0,0)}_{[2]}
	\Big[ O^{(1\pm,1\pm)}_{[n_1]} O^{(1\pm,1\pm)}_{[n_2]} \Big]
\Big\rangle
	&=
		- \frac{1}{4 n_1}
	\frac{ (x + \frac{n_1}{n_2}) (x-1)}{x + \frac{n_1 - n_2}{2n_2}}
%
%
\\
\Big\langle 
	\Big[ \bar O^{(2,2)}_{[n_1]} \bar O^{(0,0)}_{[n_2]} \Big]
	\bar O^{(0,0)}_{[2]}
	O^{(0,0)}_{[2]}
	\Big[ O^{(2,2)}_{[n_1]} O^{(0,0)}_{[n_2]} \Big]
\Big\rangle
	&=
		- \frac{1}{4 n_1}
	\frac{ x^2}{x + \frac{n_1 - n_2}{2n_2}}
%
%
%
\\
\Big\langle 
	\Big[ \bar O^{(1\pm,1\pm)}_{[n_1]} \bar O^{(1\mp,1\mp)}_{[n_2]} \Big]
	\bar O^{(0,0)}_{[2]}
	O^{(0,0)}_{[2]}
	\Big[ O^{(1\pm,1\pm)}_{[n_1]} O^{(1\mp,1\mp)}_{[n_2]} \Big]
\Big\rangle
	&=
		- \frac{1}{4 n_1}
	\frac{ x (x-1 + \frac{n_1}{n_2})}{x + \frac{n_1 - n_2}{2n_2}}
\\
\Big\langle 
	\Big[ \bar O^{(0,0)}_{[n_1]} \bar O^{(0,0)}_{[n_2]} \Big] 
	\bar O^{(1+,1+)}_{[2]}
	O^{(1+,1+)}_{[2]}
	\Big[ O^{(0,0)}_{[n_1]} O^{(0,0)}_{[n_2]} \Big] 
\Big\rangle
	&=
	\frac1{16n_1^2}
	\frac{
	x
	(x -1)^{4}
	(x + \frac{n_1 - n_2}{n_2})
	}{ (x + \frac{n_1-n_2}{2n_2})^4}
\\
\Big\langle 
	\Big[ \bar O^{(2,2)}_{[n_1]}  \bar O^{(2,2)}_{[n_2]} \Big] 
	\bar O^{(1+,1+)}_{[2]}
	O^{(1+,1+)}_{[2]}
	\Big[ O^{(2,2)}_{[n_1]} O^{(2,2)}_{[n_2]} \Big] 
\Big\rangle
	&=
	\frac1{16n_1^2}
	\frac{
	x
	(x + \frac{n_1}{n_2})^{4}
	(x + \frac{n_1 - n_2}{n_2})
	}{ (x + \frac{n_1-n_2}{2n_2})^4}
\end{aligned}
\end{align}

\begin{table}
\begin{center}
\begin{tabular}{r|| c c c c  }
&$O^{(0)}_{[2]}$
&$O^{(1+)}_{[2]}$
&$O^{(1-)}_{[2]}$
&$O^{(2)}_{[2]}$
\\
\hline
$\{r ,s \}$
& {\footnotesize $\{0 , 0 \}$}
& {\footnotesize $\{1 , 0 \}$}
& {\footnotesize $\{0 , 1 \}$}
& {\footnotesize $\{1 , 1 \}$}
\end{tabular}
\caption{Parameters  $r,s$  for different NS chirals.}
\label{ONSrs}
\end{center}
\end{table}

\subsubsection*{Functions with Ramond ground states and NS chirals.}

With parameters $(r,s)$ such that the NS chirals are given by the choices in Table \ref{ONSrs}, 
\be
A^{\zeta_1\zeta_2 ; (p)}_{n_1,n_2} (v,\bar v) 
	= 
	\Big\langle 
	\big[ R^{\zeta_1}_{[n_1]} R^{\zeta_2}_{[n_2]}]^\dagger (\infty) 
	O^{(p,p)\dagger}_{[2]}(1)
	O^{(p,p)}_{[2]}(v,\bar v) 
	\big[ R^{\zeta_1}_{[n_1]} R^{\zeta_2}_{[n_2]}] (0) 
	\Big\rangle
\ee
is given by formula (\ref{GenricAapp}) as
\begin{flalign}
&&\begin{split}
A^{++(p)}_{n_1,n_2}(x) 
	&= C \,
		x^{ - \frac{n_1-n_2}{2} }
		(x-1)^{ \frac{n_1 + n_2 - 2(r+s)}{2} }
		(x+ \tfrac{n_1}{n_2} )^{ - \frac{n_1+n_2 - 4 - 2s+s}{2} }
	\\
	&\qquad\quad
	\times
		(x- 1+ \tfrac{n_1}{n_2})^{ \frac{n_1-n_2}{2} }
		( x + \tfrac{n_1-n_2}{2})^{-1 - r(r+1) - s(s+1)}
\end{split}&&
\end{flalign}
\begin{flalign}
&&\begin{split}
A^{+-(p)}_{n_1,n_2}(x) 
	&= C \,
		x^{ - \frac{n_1-n_2-4 - 2(r+s)}{2} }
		(x-1)^{ \frac{n_1+n_2}{2} }
		(x+ \tfrac{n_1}{n_2} )^{ - \frac{n_1+n_2 - 2r(r+1) - 2s(s+1)}{2} }
	\\
	&\qquad\quad
	\times
		(x- 1+ \tfrac{n_1}{n_2})^{ \frac{n_1-n_2 -2(r+s)}{2} }
		( x + \tfrac{n_1-n_2}{2})^{-1 - r(r+1) - s(s+1)}
\end{split}&&
\end{flalign}
\begin{flalign}
&&\begin{split}
A^{\dot1+(p)}_{n_1,n_2}(x) 
	&= C \,
		x^{- \frac{n_1-n_2 - 2r}{2} }
		(x-1)^{ \frac{n_1+n_2 - 2s}{2} }
		(x+ \tfrac{n_1}{n_2} )^{ - \frac{n_1+n_2 -2r(r+1) - 2s(s+1)}{2} }
	\\
	&\qquad\quad
	\times
		(x- 1+ \tfrac{n_1}{n_2})^{ \frac{n_1-n_2+2 + 2r}{2} }
		( x + \tfrac{n_1-n_2}{2})^{-1 - r(r+1) - s(s+1)}
\end{split}&&
\end{flalign}
\begin{flalign}
&&\begin{split}
A^{\dot1-(0)}_{n_1,n_2}(x) 
	&= C \,
		x^{ - \frac{n_1-n_2 - 2-2s}{2} }
		(x-1)^{ \frac{n_1+n_2+2}{2} }
		(x+ \tfrac{n_1}{n_2} )^{ - \frac{n_1+n_2 -2r(r+1) -2s(s+1)}{2} }
	\\
	&\qquad\quad
	\times
		(x- 1+ \tfrac{n_1}{n_2})^{ \frac{n_1-n_2 - 2s}{2} }
		( x + \tfrac{n_1-n_2}{2})^{-1 - r(r+1) - s(s+1)}
\end{split}&&
\end{flalign}
\begin{flalign}
&&\begin{split}
A^{\dot1\dot1(0)}_{n_1,n_2}(x) 
	&= C \,
		x^{ - \frac{n_1-n_2}{2} }
		(x-1)^{ \frac{n_1 + n_2+2 + 2(r-s)}{2} }
		(x+ \tfrac{n_1}{n_2} )^{ - \frac{n_1+n_2 -2}{2} }
	\\
	&\qquad\quad
	\times
		(x- 1+ \tfrac{n_1}{n_2})^{ \frac{n_1-n_2}{2} }
		( x + \tfrac{n_1-n_2}{2})^{-1 - r(r+1) - s(s+1)}
\end{split}&&
\end{flalign}
\begin{flalign}
&&\begin{split}
A^{\dot1\dot2(0)}_{n_1,n_2}(x) 
	&= C \,
		x^{ - \frac{n_1-n_2-2}{2} }
		(x-1)^{ \frac{n_1 + n_2}{2} }
		(x+ \tfrac{n_1}{n_2} )^{ - \frac{n_1+n_2 }{2} }
	\\
	&\qquad\quad
	\times
		(x- 1+ \tfrac{n_1}{n_2})^{ \frac{n_1-n_2+2}{2} }
		( x + \tfrac{n_1-n_2}{2})^{-1}
\end{split}&&
\end{flalign}

\bibliographystyle{utphys}

\bibliography{JHEP_279P_0222_Resubmission_2b} 
\end{document}